\documentclass[11pt]{article}
\pdfoutput=1

\usepackage{jcappub}
\usepackage{lineno}
\usepackage{color, xcolor, appendix,latexsym,amsmath,amssymb,graphicx,booktabs,epsfig,hyperref,url,nicefrac,diagbox,relsize,longtable,comment,makecell,multirow,enumitem,siunitx,placeins,orcidlink}

\usepackage[normalem]{ulem} 
\newcommand\redout{\bgroup\markoverwith
{\textcolor{red}{\rule[0.5ex]{2pt}{0.8pt}}}\ULon}

\usepackage{dsfont}
\usepackage[english]{babel}
\usepackage[utf8]{inputenc}
\usepackage{letltxmacro}
\LetLtxMacro{\originaleqref}{\eqref}
\renewcommand{\eqref}{Eq.~\originaleqref}
\addto\extrasenglish{%
  \renewcommand{\figurename}{Figure}%
  \renewcommand{\tablename}{Table}%
}

\newcommand{\ssim}{\mathchar"5218\relax\,}

\numberwithin{equation}{section}

\definecolor{MyBlue}{rgb}{0.15,0.15,0.70}
\hypersetup{
colorlinks=true,
citecolor=MyBlue,
linkcolor=MyBlue,
urlcolor=MyBlue
}
\definecolor{lightgray}{gray}{0.9}

\newcommand{\mrs}{Aix-Marseille Univ., Universit\'e de Toulon, CNRS, CPT, Marseille, France}

\newcommand{\geneva}{D\'epartement de Physique Th\'eorique, Universit\'e de Gen\`eve, 24 quai Ernest Ansermet, 1211 Gen\`eve 4, Switzerland}
\newcommand{\gwsc}{Gravitational Wave Science Center (GWSC), Universit\'e de Gen\`eve, CH-1211 Geneva, Switzerland}

\graphicspath{figures/} 

\title{Combining underground and on-surface third-generation gravitational-wave interferometers}

\author[1,2]{Francesco Iacovelli\,\orcidlink{0000-0002-4875-5862},}
\author[1,2]{Enis Belgacem\,\orcidlink{0000-0003-4920-0911},}
\author[1,2]{Michele Maggiore\,\orcidlink{0000-0001-7348-047X},}
\author[3]{Michele Mancarella\,\orcidlink{0000-0002-0675-508X},}
\author[1,2]{Niccol\`o Muttoni\,\orcidlink{0000-0002-4214-2344}\,}

\affiliation[1]{\geneva}
\affiliation[2]{\gwsc}
\affiliation[3]{\mrs}

\emailAdd{Francesco.Iacovelli@unige.ch}
\emailAdd{Enis.Belgacem@unige.ch}
\emailAdd{Michele.Maggiore@unige.ch}
\emailAdd{mancarella@cpt.univ-mrs.fr}
\emailAdd{Niccolo.Muttoni@unige.ch}

\abstract{
Recently, detailed studies have been made to compare the performance of the European next generation GW observatory Einstein Telescope (ET) in a single-site triangular configuration with the performance of a configuration featuring two L-shaped detectors in different sites, still taken to have all other ET characteristics except for the geometry, in particular, underground and composed of a low-frequency interferometer working at cryogenic temperatures and a high-frequency interferometer working at room temperature.
Here we study a further possibility for a European network, made by a single L-shaped underground detector, like  one of the detectors considered for the 2L version of ET,  and a single third-generation 20-km L-shaped interferometer on the surface. 
We compare the performances of such a network to those of the triangle and of the 2L-underground ET configurations. We then examine the performance of an intercontinental network made by a 40-km CE in the US, together with any of these European networks.
}

\begin{document}


\newcommand{\mmin}{m_{\rm min}}
\newcommand{\mmax}{m_{\rm max}}

\newcommand{\dgw}{d_L^{\,\rm gw}}
\newcommand{\dem}{d_L^{\,\rm em}}
\newcommand{\dcom}{d_{\rm com}}
\newcommand{\dmax}{d_{\rm max}}
\newcommand{\hatO}{\hat{\Omega}}
\newcommand{\ngw}{n_{\rm GW}}
\newcommand{\ngal}{n_{\rm gal}}

\newcommand{\red}{\textcolor{red}} 

\newcommand{\blue}{\textcolor{blue}}
\newcommand{\green}{\textcolor{green}}
\newcommand{\cyan}{\textcolor{cyan}}
\newcommand{\magenta}{\textcolor{magenta}}
\newcommand{\yellow}{\textcolor{yellow}}

\newcommand{\hc}{{\cal H}}

\newcommand{\nn}{\nonumber}

\newcommand{\Lrr}{\Lambda_{\rm\scriptscriptstyle RR}}
\newcommand{\scDE}{{\textsc{DE}}}
\newcommand{\scR}{{\textsc{R}}}
\newcommand{\scM}{{\textsc{M}}}

\newcommand{\hla}{\hat{\lambda}}
\newcommand{\iBox}{\Box^{-1}}
\newcommand{\Stu}{St\"uckelberg }
\newcommand{\phib}{\bar{\phi}}

\newcommand{\Fmn}{F_{\mu\nu}}
\newcommand{\FMN}{F^{\mu\nu}}
\newcommand{\Am}{A_{\mu}}
\newcommand{\An}{A_{\nu}}
\newcommand{\Amu}{A_{\mu}}
\newcommand{\Anu}{A_{\nu}}
\newcommand{\AMU}{A^{\mu}}
\newcommand{\AN}{A^{\nu}}
\newcommand{\ANU}{A^{\nu}}

\renewcommand\({\left(}
\renewcommand\){\right)}
\renewcommand\[{\left[}
\renewcommand\]{\right]}
\newcommand\del{{\mbox {\boldmath $\nabla$}}}
\newcommand\n{{\mbox {\boldmath $\nabla$}}}
\newcommand{\ra}{\rightarrow}

\def\lsim{\raise 0.4ex\hbox{$<$}\kern -0.8em\lower 0.62
ex\hbox{$\sim$}}

\def\gsim{\raise 0.4ex\hbox{$>$}\kern -0.7em\lower 0.62
ex\hbox{$\sim$}}

\def\lbar{{\hbox{$\lambda$}\kern -0.7em\raise 0.6ex
\hbox{$-$}}}

\newcommand\eq[1]{eq.~(\ref{#1})}
\newcommand\eqs[2]{eqs.~(\ref{#1}) and (\ref{#2})}
\newcommand\Eq[1]{Equation~(\ref{#1})}
\newcommand\Eqs[2]{Equations~(\ref{#1}) and (\ref{#2})}
\newcommand\Eqss[3]{Equations~(\ref{#1}), (\ref{#2}) and (\ref{#3})}
\newcommand\eqss[3]{eqs.~(\ref{#1}), (\ref{#2}) and (\ref{#3})}
\newcommand\eqsss[4]{eqs.~(\ref{#1}), (\ref{#2}), (\ref{#3})
and (\ref{#4})}
\newcommand\eqssss[5]{eqs.~(\ref{#1}), (\ref{#2}), (\ref{#3}),
(\ref{#4}) and (\ref{#5})}
\newcommand\eqst[2]{eqs.~(\ref{#1})--(\ref{#2})}
\newcommand\Eqst[2]{Eqs.~(\ref{#1})--(\ref{#2})}
\newcommand\pa{\partial}
\newcommand\p{\partial}
\newcommand\pdif[2]{\frac{\pa #1}{\pa #2}}
\newcommand\pfun[2]{\frac{\delta #1}{\delta #2}}

\newcommand\ee{\end{equation}}
\newcommand\be{\begin{equation}}
\def\bea{\begin{array}}
\def\eea{\end{array}}\def\ea{\end{array}}
\newcommand\ees{\end{eqnarray}}
\newcommand\bees{\begin{eqnarray}}
\def\nn{\nonumber}
\newcommand\sub[1]{_{\rm #1}}
\newcommand\su[1]{^{\rm #1}}

\def\v#1{\hbox{\boldmath$#1$}}
\def\vepsilon{\v{\epsilon}}
\def\vPhi{\v{\Phi}}
\def\vomega{\v{\omega}}
\def\vsigma{\v{\sigma}}
\def\vmu{\v{\mu}}
\def\vxi{\v{\xi}}
\def\vpsi{\v{\psi}}
\def\vth{\v{\theta}}
\def\vphi{\v{\phi}}
\def\vchi{\v{\chi}}

\newcommand{\om}{\omega}
\newcommand{\Om}{\Omega}

\def\f{\phi}
\def\D{\Delta}
\def\a{\alpha}
\def\b{\beta}
\def\ab{\alpha\beta}

\def\s{\sigma}
\def\g{\gamma}
\def\G{\Gamma}
\def\d{\delta}
\def\Si{\Sigma}
\def\eps{\epsilon}
\def\veps{\varepsilon}
\def\Ups{\Upsilon}
\def\Upsun{{\Upsilon}_{\odot}}

\def\dslash{\hspace{-1mm}\not{\hbox{\kern-2pt $\partial$}}}
\def\Dslash{\not{\hbox{\kern-2pt $D$}}}
\def\pslash{\not{\hbox{\kern-2.1pt $p$}}}
\def\kslash{\not{\hbox{\kern-2.3pt $k$}}}
\def\qslash{\not{\hbox{\kern-2.3pt $q$}}}


\newcommand{\vac}{|0\rangle}
\newcommand{\cav}{\langle 0|}
\newcommand{\hint}{H_{\rm int}}
\newcommand{\va}{{\bf a}}
\newcommand{\vb}{{\bf b}}
\newcommand{\vp}{{\bf p}}
\newcommand{\vq}{{\bf q}}
\newcommand{\vk}{{\bf k}}
\newcommand{\vx}{{\bf x}}
\newcommand{\xp}{{\bf x}_{\perp}}
\newcommand{\vy}{{\bf y}}
\newcommand{\vz}{{\bf z}}
\newcommand{\vu}{{\bf u}}

\def\p1{{\bf p}_1}
\def\p2{{\bf p}_2}
\def\k1{{\bf k}_1}
\def\k2{{\bf k}_2}

\newcommand{\emn}{\eta_{\mu\nu}}
\newcommand{\ers}{\eta_{\rho\sigma}}
\newcommand{\emr}{\eta_{\mu\rho}}
\newcommand{\ens}{\eta_{\nu\sigma}}
\newcommand{\ems}{\eta_{\mu\sigma}}
\newcommand{\enr}{\eta_{\nu\rho}}
\newcommand{\eMN}{\eta^{\mu\nu}}
\newcommand{\eRS}{\eta^{\rho\sigma}}
\newcommand{\eMR}{\eta^{\mu\rho}}
\newcommand{\eNS}{\eta^{\nu\sigma}}
\newcommand{\eMS}{\eta^{\mu\sigma}}
\newcommand{\eNR}{\eta^{\nu\rho}}
\newcommand{\ema}{\eta_{\mu\alpha}}
\newcommand{\emb}{\eta_{\mu\beta}}
\newcommand{\ena}{\eta_{\nu\alpha}}
\newcommand{\enb}{\eta_{\nu\beta}}
\newcommand{\eab}{\eta_{\alpha\beta}}
\newcommand{\eAB}{\eta^{\alpha\beta}}

\newcommand{\gmn}{g_{\mu\nu}}
\newcommand{\grs}{g_{\rho\sigma}}
\newcommand{\gmr}{g_{\mu\rho}}
\newcommand{\gns}{g_{\nu\sigma}}
\newcommand{\gms}{g_{\mu\sigma}}
\newcommand{\gnr}{g_{\nu\rho}}
\newcommand{\gsn}{g_{\sigma\nu}}
\newcommand{\gsm}{g_{\sigma\mu}}
\newcommand{\gMN}{g^{\mu\nu}}
\newcommand{\gRS}{g^{\rho\sigma}}
\newcommand{\gMR}{g^{\mu\rho}}
\newcommand{\gNS}{g^{\nu\sigma}}
\newcommand{\gMS}{g^{\mu\sigma}}
\newcommand{\gNR}{g^{\nu\rho}}
\newcommand{\gLR}{g^{\lambda\rho}}
\newcommand{\gSN}{g^{\sigma\nu}}
\newcommand{\gSM}{g^{\sigma\mu}}
\newcommand{\gAB}{g^{\alpha\beta}}
\newcommand{\gab}{g_{\alpha\beta}}

\newcommand{\gBmn}{\bar{g}_{\mu\nu}}
\newcommand{\gBrs}{\bar{g}_{\rho\sigma}}
\newcommand{\gBMN}{\bar{g}^{\mu\nu}}
\newcommand{\gBRS}{\bar{g}^{\rho\sigma}}
\newcommand{\gBMS}{\bar{g}^{\mu\sigma}}
\newcommand{\gBAB}{\bar{g}^{\alpha\beta}}
\newcommand{\gBma}{\bar{g}_{\mu\alpha}}
\newcommand{\gBnb}{\bar{g}_{\nu\beta}}
\newcommand{\gBab}{\bar{g}_{\a\b}}
\newcommand{\gbmn}{\bar{g}_{\mu\nu}}
\newcommand{\gbrs}{\bar{g}_{\rho\sigma}}
\newcommand{\gbMN}{\bar{g}^{\mu\nu}}
\newcommand{\gbRS}{\bar{g}^{\rho\sigma}}
\newcommand{\gbMS}{\bar{g}^{\mu\sigma}}
\newcommand{\gbAB}{\bar{g}^{\alpha\beta}}
\newcommand{\gbma}{\bar{g}_{\mu\alpha}}
\newcommand{\gbnb}{\bar{g}_{\nu\beta}}
\newcommand{\gbab}{\bar{g}_{\a\b}}

\newcommand{\hmn}{h_{\mu\nu}}
\newcommand{\hrs}{h_{\rho\sigma}}
\newcommand{\hmr}{h_{\mu\rho}}
\newcommand{\hns}{h_{\nu\sigma}}
\newcommand{\hms}{h_{\mu\sigma}}
\newcommand{\hnr}{h_{\nu\rho}}
\newcommand{\hrn}{h_{\rho\nu}}
\newcommand{\hra}{h_{\rho\alpha}}
\newcommand{\hsb}{h_{\sigma\beta}}
\newcommand{\hma}{h_{\mu\alpha}}
\newcommand{\hna}{h_{\nu\alpha}}
\newcommand{\hmb}{h_{\mu\beta}}
\newcommand{\has}{h_{\alpha\sigma}}
\newcommand{\hab}{h_{\alpha\beta}}
\newcommand{\hnb}{h_{\nu\beta}}
\newcommand{\hcr}{h_{\times}}

\newcommand{\hMN}{h^{\mu\nu}}
\newcommand{\hRS}{h^{\rho\sigma}}
\newcommand{\hMR}{h^{\mu\rho}}
\newcommand{\hRM}{h^{\rho\mu}}
\newcommand{\hRN}{h^{\rho\nu}}
\newcommand{\hNS}{h^{\nu\sigma}}
\newcommand{\hMS}{h^{\mu\sigma}}
\newcommand{\hNR}{h^{\nu\rho}}
\newcommand{\hAB}{h^{\alpha\beta}}
\newcommand{\hij}{h_{ij}}
\newcommand{\hIJ}{h^{ij}}
\newcommand{\hkl}{h_{kl}}
\newcommand{\hTTij}{h_{ij}^{\rm TT}}
\newcommand{\HTTij}{H_{ij}^{\rm TT}}
\newcommand{\dhTTij}{\dot{h}_{ij}^{\rm TT}}
\newcommand{\hTTab}{h_{ab}^{\rm TT}}

\newcommand{\thmn}{\tilde{h}_{\mu\nu}}
\newcommand{\thrs}{\tilde{h}_{\rho\sigma}}
\newcommand{\thmr}{\tilde{h}_{\mu\rho}}
\newcommand{\thns}{\tilde{h}_{\nu\sigma}}
\newcommand{\thms}{\tilde{h}_{\mu\sigma}}
\newcommand{\thnr}{\tilde{h}_{\nu\rho}}
\newcommand{\thrn}{\tilde{h}_{\rho\nu}}
\newcommand{\thab}{\tilde{h}_{\alpha\beta}}
\newcommand{\thMN}{\tilde{h}^{\mu\nu}}
\newcommand{\thRS}{\tilde{h}^{\rho\sigma}}
\newcommand{\thMR}{\tilde{h}^{\mu\rho}}
\newcommand{\thRM}{\tilde{h}^{\rho\mu}}
\newcommand{\thRN}{\tilde{h}^{\rho\nu}}
\newcommand{\thNS}{\tilde{h}^{\nu\sigma}}
\newcommand{\thMS}{\tilde{h}^{\mu\sigma}}
\newcommand{\thNR}{\tilde{h}^{\nu\rho}}
\newcommand{\thAB}{\tilde{h}^{\alpha\beta}}

\newcommand{\vvarphi}{\hat{\varphi}}
\newcommand{\hhmn}{\hat{h}_{\mu\nu}}
\newcommand{\hhrs}{\hat{h}_{\rho\sigma}}
\newcommand{\hhmr}{\hat{h}_{\mu\rho}}
\newcommand{\hhns}{\hat{h}_{\nu\sigma}}
\newcommand{\hhms}{\hat{h}_{\mu\sigma}}
\newcommand{\hhnr}{\hat{h}_{\nu\rho}}
\newcommand{\hhra}{\hat{h}_{\rho\alpha}}

\newcommand{\hhMN}{\hat{h}^{\mu\nu}}
\newcommand{\hhRS}{\hat{h}^{\rho\sigma}}
\newcommand{\hhMR}{\hat{h}^{\mu\rho}}
\newcommand{\hhNS}{\hat{h}^{\nu\sigma}}
\newcommand{\hhMS}{\hat{h}^{\mu\sigma}}
\newcommand{\hhNR}{\hat{h}^{\nu\rho}}
\newcommand{\hhAB}{\hat{h}^{\alpha\beta}}

\newcommand{\sh}{\mathsf{h}}
\newcommand{\shmn}{\mathsf{h}_{\mu\nu}}
\newcommand{\shrs}{\mathsf{h}_{\rho\sigma}}
\newcommand{\shmr}{\mathsf{h}_{\mu\rho}}
\newcommand{\shns}{\mathsf{h}_{\nu\sigma}}
\newcommand{\shms}{\mathsf{h}_{\mu\sigma}}
\newcommand{\shnr}{\mathsf{h}_{\nu\rho}}
\newcommand{\shra}{\mathsf{h}_{\rho\alpha}}
\newcommand{\shsb}{\mathsf{h}_{\sigma\beta}}
\newcommand{\shma}{\mathsf{h}_{\mu\alpha}}
\newcommand{\shna}{\mathsf{h}_{\nu\alpha}}
\newcommand{\shmb}{\mathsf{h}_{\mu\beta}}
\newcommand{\shas}{\mathsf{h}_{\alpha\sigma}}
\newcommand{\shab}{\mathsf{h}_{\alpha\beta}}
\newcommand{\shnb}{\mathsf{h}_{\nu\beta}}
\newcommand{\shcr}{\mathsf{h}_{\times}}
\newcommand{\shMN}{\mathsf{h}^{\mu\nu}}
\newcommand{\shRS}{\mathsf{h}^{\rho\sigma}}
\newcommand{\shMR}{\mathsf{h}^{\mu\rho}}
\newcommand{\shNS}{\mathsf{h}^{\nu\sigma}}
\newcommand{\shMS}{\mathsf{h}^{\mu\sigma}}
\newcommand{\shNR}{\mathsf{h}^{\nu\rho}}
\newcommand{\shAB}{\mathsf{h}^{\alpha\beta}}
\newcommand{\shij}{\mathsf{h}_{ij}}
\newcommand{\shIJ}{\mathsf{h}^{ij}}
\newcommand{\shkl}{\mathsf{h}_{kl}}
\newcommand{\shTTij}{\mathsf{h}_{ij}^{\rm TT}}
\newcommand{\shTTab}{\mathsf{h}_{ab}^{\rm TT}}

\newcommand{\bhmn}{\bar{h}_{\mu\nu}}
\newcommand{\bhrs}{\bar{h}_{\rho\sigma}}
\newcommand{\bhmr}{\bar{h}_{\mu\rho}}
\newcommand{\bhns}{\bar{h}_{\nu\sigma}}
\newcommand{\bhms}{\bar{h}_{\mu\sigma}}
\newcommand{\bhnr}{\bar{h}_{\nu\rho}}
\newcommand{\bhRS}{\bar{h}^{\rho\sigma}}
\newcommand{\bhMN}{\bar{h}^{\mu\nu}}
\newcommand{\bhNR}{\bar{h}^{\nu\rho}}
\newcommand{\bhMR}{\bar{h}^{\mu\rho}}
\newcommand{\bhAB}{\bar{h}^{\alpha\beta}}

\newcommand{\hax}{h^{\rm ax}}
\newcommand{\haxmn}{h^{\rm ax}_{\mu\nu}}
\newcommand{\hpol}{h^{\rm pol}}
\newcommand{\hpolmn}{h^{\rm pol}_{\mu\nu}}

\newcommand{\dgzz}{{^{(2)}g_{00}}}
\newcommand{\qgzz}{{^{(4)}g_{00}}}
\newcommand{\tgzi}{{^{(3)}g_{0i}}}
\newcommand{\dgij}{{^{(2)}g_{ij}}}
\newcommand{\zTzz}{{^{(0)}T^{00}}}
\newcommand{\dTzz}{{^{(2)}T^{00}}}
\newcommand{\dTii}{{^{(2)}T^{ii}}}
\newcommand{\uTzi}{{^{(1)}T^{0i}}}

\newcommand{\xm}{x^{\mu}}
\newcommand{\xn}{x^{\nu}}
\newcommand{\xr}{x^{\rho}}
\newcommand{\xs}{x^{\sigma}}
\newcommand{\xa}{x^{\a}}
\newcommand{\xb}{x^{\b}}

\newcommand{\hatk}{\hat{\bf k}}
\newcommand{\hatn}{\hat{\bf n}}
\newcommand{\hatx}{\hat{\bf x}}
\newcommand{\haty}{\hat{\bf y}}
\newcommand{\hatz}{\hat{\bf z}}
\newcommand{\hatr}{\hat{\bf r}}
\newcommand{\hatu}{\hat{\bf u}}
\newcommand{\hatv}{\hat{\bf v}}
\newcommand{\xim}{\xi_{\mu}}
\newcommand{\xin}{\xi_{\nu}}
\newcommand{\xia}{\xi_{\a}}
\newcommand{\xib}{\xi_{\b}}
\newcommand{\xiM}{\xi^{\mu}}
\newcommand{\xiN}{\xi^{\nu}}

\newcommand{\tA}{\tilde{\bf A} ({\bf k})}

\newcommand{\pam}{\pa_{\mu}}
\newcommand{\pal}{\pa_{\mu}}
\newcommand{\pan}{\pa_{\nu}}
\newcommand{\parho}{\pa_{\rho}}
\newcommand{\pas}{\pa_{\sigma}}
\newcommand{\paM}{\pa^{\mu}}
\newcommand{\paN}{\pa^{\nu}}
\newcommand{\paR}{\pa^{\rho}}
\newcommand{\paS}{\pa^{\sigma}}
\newcommand{\paa}{\pa_{\alpha}}
\newcommand{\pab}{\pa_{\beta}}
\newcommand{\pat}{\pa_{\theta}}
\newcommand{\paf}{\pa_{\phi}}

\newcommand{\Dam}{D_{\mu}}
\newcommand{\Dan}{D_{\nu}}
\newcommand{\Dar}{D_{\rho}}
\newcommand{\Das}{D_{\sigma}}
\newcommand{\DaM}{D^{\mu}}
\newcommand{\DaN}{D^{\nu}}
\newcommand{\DaR}{D^{\rho}}
\newcommand{\DaS}{D^{\sigma}}
\newcommand{\Daa}{D_{\alpha}}
\newcommand{\Dab}{D_{\beta}}

\newcommand{\DBm}{\bar{D}_{\mu}}
\newcommand{\DBn}{\bar{D}_{\nu}}
\newcommand{\DBr}{\bar{D}_{\rho}}
\newcommand{\DBs}{\bar{D}_{\sigma}}
\newcommand{\DBt}{\bar{D}_{\tau}}
\newcommand{\DBa}{\bar{D}_{\alpha}}
\newcommand{\DBb}{\bar{D}_{\beta}}
\newcommand{\DBM}{\bar{D}^{\mu}}
\newcommand{\DBN}{\bar{D}^{\nu}}
\newcommand{\DBR}{\bar{D}^{\rho}}
\newcommand{\DBS}{\bar{D}^{\sigma}}
\newcommand{\DBA}{\bar{D}^{\alpha}}

\newcommand{\GMnr}{{\Gamma}^{\mu}_{\nu\rho}}
\newcommand{\Glmn}{{\Gamma}^{\lambda}_{\mu\nu}}
\newcommand{\barGMnr}{{\bar{\Gamma}}^{\mu}_{\nu\rho}}
\newcommand{\GMns}{{\Gamma}^{\mu}_{\nu\sigma}}
\newcommand{\GInr}{{\Gamma}^{i}_{\nu\rho}}
\newcommand{\Rmn}{R_{\mu\nu}}
\newcommand{\Gmn}{G_{\mu\nu}}
\newcommand{\RMN}{R^{\mu\nu}}
\newcommand{\GMN}{G^{\mu\nu}}
\newcommand{\Rmnrs}{R_{\mu\nu\rho\sigma}}
\newcommand{\RMnrs}{{R^{\mu}}_{\nu\rho\sigma}}
\newcommand{\Tmn}{T_{\mu\nu}}
\newcommand{\Smn}{S_{\mu\nu}}
\newcommand{\Tab}{T_{\a\b}}
\newcommand{\TMN}{T^{\mu\nu}}
\newcommand{\TAB}{T^{\a\b}}
\newcommand{\TBmn}{\bar{T}_{\mu\nu}}
\newcommand{\TBMN}{\bar{T}^{\mu\nu}}
\newcommand{\TRS}{T^{\rho\sigma}}
\newcommand{\tmn}{t_{\mu\nu}}
\newcommand{\tMN}{t^{\mu\nu}}
\newcommand{\RUmn}{R_{\mu\nu}^{(1)}}
\newcommand{\RDmn}{R_{\mu\nu}^{(2)}}
\newcommand{\RTmn}{R_{\mu\nu}^{(3)}}
\newcommand{\RBmn}{\bar{R}_{\mu\nu}}
\newcommand{\RBmr}{\bar{R}_{\mu\rho}}
\newcommand{\RBnr}{\bar{R}_{\nu\rho}}

\newcommand{\dddM}{\kern 0.2em \raise 1.9ex\hbox{$...$}\kern -1.0em \hbox{$M$}}
\newcommand{\dddQ}{\kern 0.2em \raise 1.9ex\hbox{$...$}\kern -1.0em \hbox{$Q$}}
\newcommand{\dddI}{\kern 0.2em \raise 1.9ex\hbox{$...$}\kern -1.0em\hbox{$I$}}
\newcommand{\dddJ}{\kern 0.2em \raise 1.9ex\hbox{$...$}\kern-1.0em
\hbox{$J$}}
\newcommand{\dddcalJ}{\kern 0.2em \raise 1.9ex\hbox{$...$}\kern-1.0em
\hbox{${\cal J}$}}

\newcommand{\dddO}{\kern 0.2em \raise 1.9ex\hbox{$...$}\kern -1.0em
\hbox{${\cal O}$}}
\def\dddz{\raise 1.5ex\hbox{$...$}\kern -0.8em \hbox{$z$}}
\def\dddd{\raise 1.8ex\hbox{$...$}\kern -0.8em \hbox{$d$}}
\def\dddbd{\raise 1.8ex\hbox{$...$}\kern -0.8em \hbox{${\bf d}$}}
\def\ddbd{\raise 1.8ex\hbox{$..$}\kern -0.8em \hbox{${\bf d}$}}
\def\dddx{\raise 1.6ex\hbox{$...$}\kern -0.8em \hbox{$x$}}

\newcommand{\hti}{\tilde{h}}
\newcommand{\hf}{\tilde{h}_{ab}(f)}
\newcommand{\Hti}{\tilde{H}}
\newcommand{\fmin}{f_{\rm min}}
\newcommand{\fmax}{f_{\rm max}}
\newcommand{\frot}{f_{\rm rot}}
\newcommand{\fpol}{f_{\rm pole}}
\newcommand{\omax}{\o_{\rm max}}
\newcommand{\orot}{\o_{\rm rot}}
\newcommand{\op}{\o_{\rm p}}
\newcommand{\tmax}{t_{\rm max}}
\newcommand{\tobs}{t_{\rm obs}}
\newcommand{\fobs}{f_{\rm obs}}
\newcommand{\temis}{t_{\rm emis}}
\newcommand{\DE}{\D E_{\rm rad}}
\newcommand{\DEm}{\D E_{\rm min}}
\newcommand{\msun}{~{\rm M}_{\odot}}
\newcommand{\mtot}{M_{\rm tot}}
\newcommand{\rsun}{R_{\odot}}
\newcommand{\ogw}{\omega_{\rm gw}}
\newcommand{\fgw}{f_{\rm gw}}
\newcommand{\oL}{\omega_{\rm L}}
\newcommand{\kL}{k_{\rm L}}
\newcommand{\lL}{\l_{\rm L}}
\newcommand{\mns}{M_{\rm NS}}
\newcommand{\rns}{R_{\rm NS}}
\newcommand{\tret}{t_{\rm ret}}
\newcommand{\Sch}{Schwarzschild }
\newcommand{\rtid}{r_{\rm tidal}}

\newcommand{\ot}{\o_{\rm t}}
\newcommand{\mt}{m_{\rm t}}
\newcommand{\gt}{\g_{\rm t}}
\newcommand{\xit}{\tilde{\xi}}
\newcommand{\xtr}{\xi_{\rm t}}
\newcommand{\xtj}{\xi_{{\rm t},j}}
\newcommand{\dxtj}{\dot{\xi}_{{\rm t},j}}
\newcommand{\ddxtj}{\ddot{\xi}_{{\rm t},j}}
\newcommand{\teff}{T_{\rm eff}}
\newcommand{\samp}{S_{\xi_{\rm t}}^{\rm ampl}}

\newcommand{\mpl}{M_{\rm Pl}}
\newcommand{\mplr}{m_{\rm Pl}}
\newcommand{\mgut}{M_{\rm GUT}}
\newcommand{\lpl}{l_{\rm Pl}}
\newcommand{\tpl}{t_{\rm Pl}}
\newcommand{\ls}{\lambda_{\rm s}}
\newcommand{\Ogw}{\Omega_{\rm gw}}
\newcommand{\hogw}{h_0^2\Omega_{\rm gw}}
\newcommand{\hn}{h_n(f)}

\newcommand{\sinc}{{\rm sinc}\, }
\newcommand{\Ein}{E_{\rm in}}
\newcommand{\Eout}{E_{\rm out}}
\newcommand{\Et}{E_{\rm t}}
\newcommand{\Er}{E_{\rm refl}}
\newcommand{\lm}{\l_{\rm mod}}

\newcommand{\mrI}{\mathrm{I}}
\newcommand{\mrJ}{\mathrm{J}}
\newcommand{\mrW}{\mathrm{W}}
\newcommand{\mrX}{\mathrm{X}}
\newcommand{\mrY}{\mathrm{Y}}
\newcommand{\mrZ}{\mathrm{Z}}
\newcommand{\mrM}{\mathrm{M}}
\newcommand{\mrS}{\mathrm{S}}

\newcommand{\rmI}{\mathrm{I}}
\newcommand{\rmJ}{\mathrm{J}}
\newcommand{\rmW}{\mathrm{W}}
\newcommand{\rmX}{\mathrm{X}}
\newcommand{\rmY}{\mathrm{Y}}
\newcommand{\rmZ}{\mathrm{Z}}
\newcommand{\rmM}{\mathrm{M}}
\newcommand{\rmS}{\mathrm{S}}
\newcommand{\rmU}{\mathrm{U}}
\newcommand{\rmV}{\mathrm{V}}


\newcommand{\et}{{{\bf e}^t}}
\newcommand{\etm}{{\bf e}^t_{\mu}}
\newcommand{\etn}{{\bf e}^t_{\nu}}

\newcommand{\er}{{{\bf e}^r}}
\newcommand{\erm}{{\bf e}^r_{\mu}}
\newcommand{\ern}{{\bf e}^r_{\nu}}

\newcommand{\hz}{H^{(0)}}
\newcommand{\hu}{H^{(1)}}
\newcommand{\hd}{H^{(2)}}
\newcommand{\thz}{\tilde{H}^{(0)}}
\newcommand{\thu}{\tilde{H}^{(1)}}
\newcommand{\thd}{\tilde{H}^{(2)}}
\newcommand{\tK}{\tilde{K}}
\newcommand{\tZ}{\tilde{Z}}
\newcommand{\tQ}{\tilde{Q}}

\newcommand{\inT}{\int_{-\infty}^{\infty}}
\newcommand{\intz}{\int_{0}^{\infty}}
\newcommand{\Dl}{\int{\cal D}\lambda}

\newcommand{\fnl}{f_{\rm NL}}

\newcommand{\ode}{\Omega_{\rm DE}}
\newcommand{\oma}{\Omega_{M}}
\newcommand{\ora}{\Omega_{R}}
\newcommand{\ovac}{\Omega_{\rm vac}}
\newcommand{\ola}{\Omega_{\Lambda}}
\newcommand{\oxi}{\Omega_{\xi}}
\newcommand{\oga}{\Omega_{\gamma}}

\newcommand{\lc}{\Lambda_c}
\newcommand{\rde}{\rho_{\rm DE}}
\newcommand{\wde}{w_{\rm DE}}
\newcommand{\rvac}{\rho_{\rm vac}}
\newcommand{\rlam}{\rho_{\Lambda}}

\newcommand{\NM}[1]{{\color{blue}{[\bf{NM}}: #1]}}

\maketitle
\flushbottom

\section{Introduction}

The detection of gravitational waves (GWs) in the last few years demonstrated the possibility of a novel way to observe the Universe by ``listening'' to the ripples of spacetime. The second generation detectors LIGO and Virgo revealed the existence of stellar-scale black holes more massive than expected and neutron-star mergers emitting both GWs and light signals across the entire spectrum, while providing additional confirmation of Einstein's theory of General Relativity~\cite{LIGOScientific:2016aoc,LIGOScientific:2017vwq,LIGOScientific:2017zic,LIGOScientific:2017ync,LIGOScientific:2020ibl,LIGOScientific:2021djp,LIGOScientific:2021psn,LIGOScientific:2021sio,LIGOScientific:2021aug}. 

Building on this success, the GW community is preparing the jump toward third-generation (3G) GW detectors, new observatories that overcome the limitations imposed by existing second-generation detector infrastructures and are designed to detect GW sources along the cosmic history up to the early Universe. The Einstein Telescope (ET)  \cite{Punturo:2010zz,Hild:2010id,Maggiore:2019uih} is the 3G European observatory project, while the US community effort is represented by the Cosmic Explorer (CE) project~
\cite{Reitze:2019iox,Evans:2021gyd,Evans:2023euw}.
These detectors will provide an improvement in sensitivity by one order of magnitude and a significant enlargement of the bandwidth, both toward low and high frequencies, and will have extraordinary potential for discoveries in astrophysics, cosmology, and fundamental physics. 
 
In the last few years, comprehensive studies of the scientific potential of various 3G  detector networks have been performed \cite{Borhanian:2022czq,Branchesi:2023mws,Gupta:2023lga}. In particular, in ref.~\cite{Branchesi:2023mws} two different  geometries for ET were compared, a single-site triangular geometry made of  3 nested detectors [each comprising two interferometers, a low frequency (LF) and a high-frequency (HF) one, for a total of six nested interferometers],
and a network made of two identical  L-shaped detectors (``2L'' in the following), again made by a LF and a HF  interferometer each, located in two different sites within Europe (see also \cite{Puecher:2023twf,Bhagwat:2023jwv,Franciolini:2023opt,Iacovelli:2023nbv} for further follow-up studies). 
These different configurations for ET were considered both in a ET-only scenario, and 
in a broader world-wide network including also  a single 40-km Cosmic Explorer (CE) detector, or  two CE detectors with arm-lengths of 20 and 40~km, respectively.

The main aim of the present paper is to investigate the performance of a hypothetical European detector network made by  a single L-shaped underground detector  with the amplitude spectral density (ASD) of ET  and a single 3G L-shaped interferometer on the surface,
and  compare it with the other 3G detector network configurations that have been recently studied. We will first  compare the performances of these configurations taken as European networks in isolation, and we will then further compare their performances when they are part of a broader world-wide network including also  a single 40-km Cosmic Explorer (CE) detector in the US.

\section{Detector configurations}\label{sect:detconfig}

As discussed  in ref.~\cite{Branchesi:2023mws}, a single-L 3G detector, operated in isolation and  not inserted in  a network, would miss many of the science goals expected from the next generation of GW detectors. A network of  at least two L-shaped 3G detectors with the characteristics of ET would instead allow  reaching them \cite{Branchesi:2023mws}, as would also be the case for  two 3G detectors with the characteristics of CE \cite{Evans:2023euw,Gupta:2023lga}. 

Each of these designs, ET or CE, has its own advantages. The ET design allows  reaching a better sensitivity at  low frequencies,  thanks to the cryogenic LF instrument and to the fact that in an underground facility the seismic noise is lower. On the other hand, the fact that  CE is not underground allows one to make longer arms (for a given cost), which provide a better sensitivity above about 10~Hz. \autoref{fig:All_ASDs} shows the ASD  for a  single-L ET (with 10~km arms and  with 15~km arms), for CE (with 20-km arms and with 40-km arms), and, by comparison, with the ASD expected for LIGO and Virgo 
by the end of the O5 run.\footnote{When comparing a triangle to an L-shaped interferometer with the same  ASD, one must take into account that the triangle is made of three nested detectors (a detector being an LF and HF interferometer pair), with an opening angle of $60^{\circ}$. 
For the triangle configuration,  one must then  project the GW tensor of the incoming wave onto each of these three  components [see eqs.~(9)--(11) of \cite{Jaranowski:1998qm} for explicit expressions], and then
combine the results at the level of the SNR and parameter estimation to obtain the ET capabilities. 
 See also the discussion in Section~2 of
ref.~\cite{Branchesi:2023mws}.
\label{foot:ASD-Delta}}

The question that we want to address in this paper is what happens when one combines a single L-shaped   detector with the ASD of ET (therefore underground, and made of a LF and a HF interferometer), with a single 3G L-shaped  detector 
on the surface, again located in Europe. For instance, a very interesting question is whether the better low-frequency sensitivity of a single  L-shaped detector with the ASD of ET becomes of limited value, without a partner with a similar low-frequency sensitivity or, on the contrary,  it is sufficient to combine it with a  3G surface detector to exploit its low-frequency capabilities. For the underground  L-shaped  detector we will use the most recent publicly available ET curve (shown in \autoref{fig:All_ASDs}). This is the same sensitivity curve that was also used in the study~\cite{Branchesi:2023mws}. For the surface L-shaped European detector we will use, as an example, the ASD of a 20-km CE detector. On the basis of the analysis made during the ET conceptual design, one should not expect that the CE sensitivity can be easily realized with a surface detector in Europe due to strong constraints on interferometer length. We use it in our analysis since it is the only readily available model of a sensitivity curve for a next-generation GW detector at the surface.\footnote{The ET sensitivity  curve that we use is available at
\url{https://apps.et-gw.eu/tds/?content=3&r=18213}. For CE--20km and CE--40km  we use the curves available at
\url{https://dcc.cosmicexplorer.org/cgi-bin/DocDB/ShowDocument?.submit=Identifier&docid=T2000017&version=}. Again, these are the same sensitivity curves that have been used in ref.~\cite{Branchesi:2023mws}. }
\\[.15cm]  
More in detail, we consider the following configurations:

\begin{itemize}

\item ET in its standard 10-km triangle configuration. We will refer it as ET-$\Delta$. 

\item ET in the configuration of two L-shaped detectors with 15 km arms, taken to be in the two candidate sites in Sardinia and in the Meuse-Rhine region. We will refer to it as ET-2L (see below for the relative orientation among the two L).

\item A hybrid configuration of two detectors, both taken to be in Europe, one L-shaped, underground and with the ASD of ET with 15 km arms, and the other again L-shaped but on the surface, with the ASD of CE, with 20 km arms. We will refer to it as ``Hybrid''.

\end{itemize}
\begin{figure}[tbp]
    \centering
    \includegraphics[width=.6\textwidth]{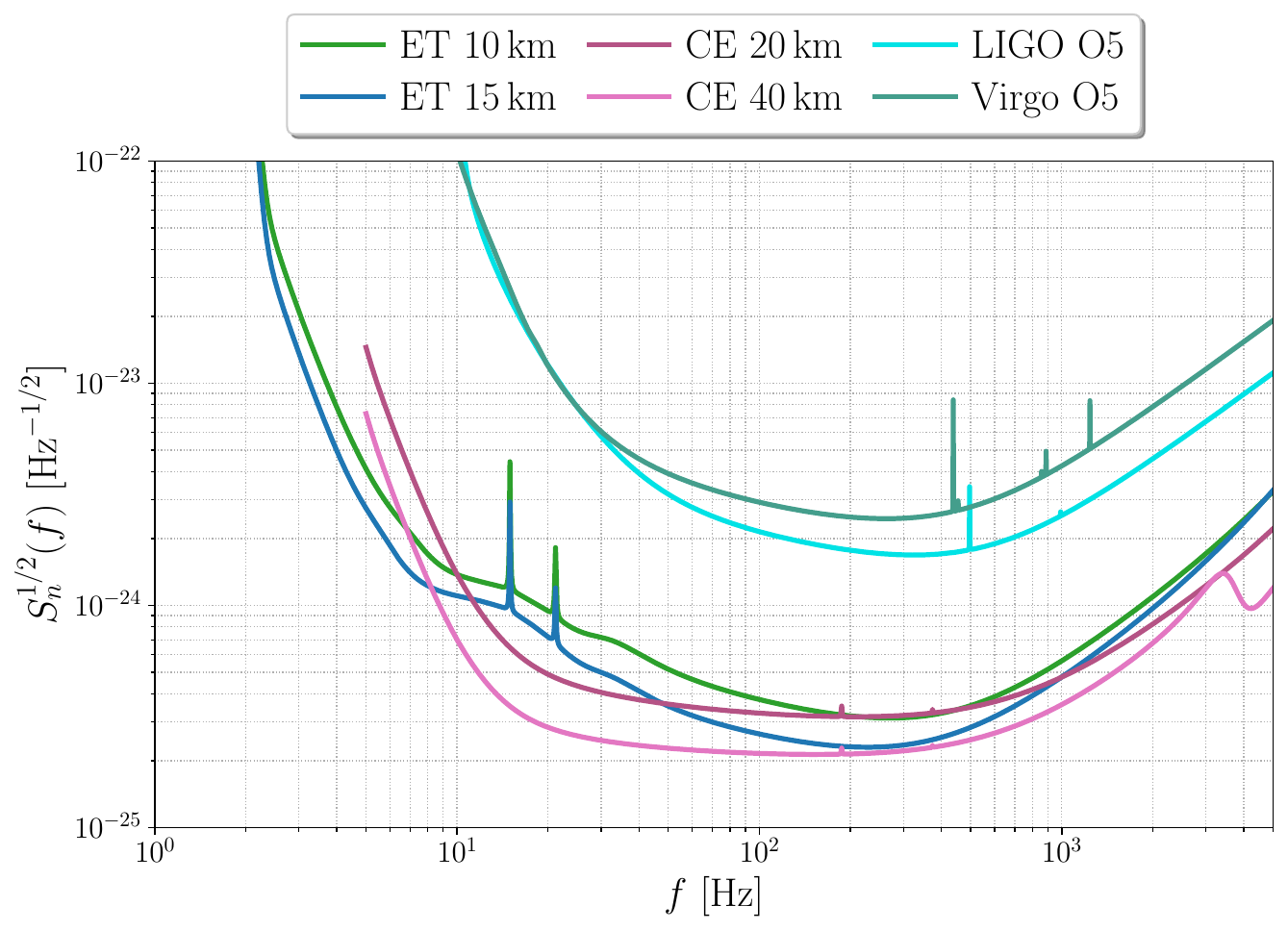}
    \caption{Amplitude spectral densities (ASDs) used in this work. Note that the ASDs are always defined as if we had a single L-shaped detector of the given arm-length. To get the overall sensitivity of the triangle to a given signal, one must then combine the ASD of the three interferometers as discussed in \autoref{foot:ASD-Delta}.}
    \label{fig:All_ASDs}
\end{figure}
Each of these configurations will then be studied also in correlation with a single 40-km CE located in the US.

The first two configurations above are the two main reference configurations that were studied  in \cite{Branchesi:2023mws}, and they are both being actively considered for ET.\footnote{Their different   arm-lengths reflect the fact that the total length of the vacuum pipes to be installed (which is one of the most expensive items) is the same for a 10-km triangle and for a 15-km 2L. Indeed, for the triangle we have 3 tunnels and 4 vacuum tubes per tunnel (since, in a  nested detector configuration, in each tunnel we have one arm of the HF interferometer and one arm of the LF interferometer of one detector,  as well as   one arm of the HF interferometer and one arm of the LF interferometer of another detector), so $10\, {\rm km}\times 3 \times 4=120\, {\rm km}$, while for 2L we have overall 4 arms, but just 2 tubes per arm (the HF and the LF interferometers), and
$15\, {\rm km}\times 4 \times 2=120\, {\rm km}$~\cite{Barsotti:2023}. Furthermore, taking into account that the nested detectors configuration of the triangle requires larger tunnels, also the volume of excavated rocks (another crucial aspects of a cost analysis)  for a 10-km triangle and for a 15-km 2L is roughly comparable, see also footnote~5 in \cite{Branchesi:2023mws}.} The Hybrid configuration, in contrast, has never been considered to date.
Placing a 20-km  interferometer on the surface, as in the Hybrid configuration, requires 
a sufficiently large underpopulated area, both to construct such a long infrastructure and to avoid anthropic noise. In Europe it is more difficult to find viable sites of this type, compared to the US. Here, for the sake of the exercise, we will put  the single-L surface detector in a largely underpopulated area in   Spain\footnote{We use  for definiteness the coordinates ($41^{\circ}21'57''\,{\rm N}, 6^{\circ}06'4''\,{\rm O}$),
that fall inside  a  large flat and desert area  with a population density of 6 ${\rm persons}/{\rm km}^2$, with the closest town having about 1000 inhabitants.} and, for definiteness, in the Hybrid configuration we will locate the single-L
detector with the ET ASD in the Sardinia candidate site, resulting in a cord-distance  of about 1300~km. Similar results would, however,  be obtained placing the 
surface detector elsewhere in Spain, or in fact in any suitable place in Europe, as well as placing the underground detector in any of the two candidate locations in Europe, as long as the two sites have a similar cord distance; in particular, this holds for a surface detector  in this hypothetical site in Spain and an underground detector 
in the Meuse-Rhine candidate site, as the cord distance in this case  is about 1390~km. We stress that the specific choice that we make here for the location of a surface detector as well as, more generally, the choice of studying a configuration of a single L-shaped ET-like underground detector, and a single L-shaped 3G on-surface detector, both located in Europe, does not correspond to any project currently under study from the many points of view (geological, topographical, 
financial, political, etc.) that are necessary for determining viable detector configurations and optimal site selection. At the present stage, this must just  be considered as an exercise, whose aim is to better understand the performances of 3G networks in different settings. 

In a 2L network an important choice is the relative orientation among the two detectors. When taking into account the Earth's curvature, the relative orientation between  two L-shaped detectors is  defined with reference to the great circle that connects them~\cite{Flanagan:1993ix,Christensen:1996da}. We denote by  $\beta$ the angle describing the relative orientation of the two detectors, defined with reference to this great circle, so that  $\beta=0^{\circ}$ corresponds to the case where the arms of the two interferometers make the same angle with respect to the great circle, while $\beta=45^{\circ}$ corresponds to the situation in which  one of the two interferometers is rotated by $45^{\circ}$ from the $\beta=0^{\circ}$ orientation.
For $\beta=45^{\circ}$ the accuracy of the parameter estimation for coalescing binaries is maximized,\footnote{Actually, this is true only for the standard tensorial polarization of General Relativity (GR), and not for extra polarization modes that could appear in extensions of GR.}
while the sensitivity to stochastic backgrounds is minimized (to the extent that it becomes exactly zero in the limit  $d/\lambda\ra 0$, where $\lambda=c/f$ is the  wavelength of a GW with frequency $f$, and $d$ the distance between the two detectors). Conversely, for  parallel detectors ($\beta=0^{\circ}$) the sensitivity to stochastic background is maximized,  while the accuracy of the parameter estimation for coalescing binaries is minimized.\footnote{Note, however, that for parallel detectors the sensitivity to parameter estimation  is of course non-zero and in fact still reasonably close to the optimal value. The two LIGO detectors are set parallel to each other, as was natural for two detectors whose initial target was to get the first detections, in which case one maximizes the chances that the two detectors  both go above detection threshold for the same event.} Here we will follow the same strategy as in  ref.~\cite{Branchesi:2023mws}, considering both the case $\beta=0^{\circ}$, and the case of $\beta$ close, but not exactly equal, to $45^{\circ}$, 
using a value $\beta=45^{\circ}-2.51^{\circ}$. This small misalignment, with respect to $\beta=45^{\circ}$, has essentially no impact on the quality of the reconstruction of the parameters of coalescing binaries, while it allows us to recover an interesting sensitivity to stochastic backgrounds, about $10\%$ of the optimal sensitivity which is obtained for parallel detectors.\footnote{The precise value $2.51^{\circ}$ of the misalignment angle has no special meaning. It was chosen, somewhat arbitrarily, in   ref.~\cite{Branchesi:2023mws} because, for the case of one detector in the Sardinia candidate site and one in the Meuse-Rhine candidate site, it corresponds to 
$\alpha=45^{\circ}$, where $\alpha$ is the relative angle defined
using the local North at the two detector sites; i.e., for these specific locations, $\alpha\simeq \beta+2.51^{\circ}$.\label{foot:alpha}} 
We will denote the corresponding 2L configurations
as 2L-$0^{\circ}$ (i.e. $\beta=0^{\circ}$) and
2L-mis (for misaligned, i.e. $\beta=45^{\circ}-2.51^{\circ}$).\footnote{Actually, the only situation for which the precise value $2.51^{\circ}$ of the misalignment angle  is relevant is for the study of  stochastic backgrounds in the misaligned configuration. For all results on compact binary coalescences, the plots of the results obtained  with 
$\beta=45^{\circ}-2.51^{\circ}$, on the scales that we use,  are  visually indistinguishable from those with  $\beta=45^{\circ}$; similarly for  $\beta=0^{\circ}$ or $\beta=-2.51^{\circ}$.}
So, more precisely, the all-European networks that we will consider are:

\begin{enumerate}

\item ET-$\Delta$: the 10-km triangle, with the ET ASD, therefore underground. We locate it for definiteness in Sardinia.
 
\item ET-2L-$0^{\circ}$: two L-shaped detectors of 15~km arms, with the ET ASD, therefore both underground, taken to be parallel with respect to the great circle that joins them (i.e. $\beta =0^{\circ}$).
We locate them  in Sardinia and in the Meuse-Rhine region.

\item ET-2L-mis: as ET-2L-$0^{\circ}$, but  at $\beta =45^{\circ}-2.51^{\circ}$.

\item Hybrid-$0^{\circ}$: a hybrid configuration, with an underground L-shaped  detector with the ASD of ET with 15~km arms (located for definiteness in Sardinia) and a surface L-shaped detector with the ASD of CE-20km (located as an example in Spain, with a baseline among the two sites of order 1300~km).

\item Hybrid-mis. As Hybrid-$0^{\circ}$, but  at $\beta =45^{\circ}-2.51^{\circ}$.

\end{enumerate}

We will also compare  with the results that could be obtained by the most advanced  2G detector network,  namely  LIGO Hanford, LIGO Livingston, Virgo, KAGRA and LIGO India,  using the publicly available best sensitivities that are planned to be achieved by the end of the O5 run \cite{KAGRA:2020rdx}. We denote this network as LVKI~O5. 

Finally, each of the five European networks above will also be studied in a broader world-wide network, adding further a single-L CE detector of 40~km in the US.\footnote{When performing the correlation with a 40-km CE in the US,  we will place for definiteness the US detector in Idaho using as representative the location and orientation in Table III of \cite{Borhanian:2020ypi}. This results in an alignment of about $191^\circ$ with respect to the Sardinia site, $237^\circ$ with respect to the Meuse-Rhine site (in the situation in which this is misaligned with respect to the Sardinia one) and $227^\circ$ with respect to the Spain site (in the situation in which this is misaligned with respect to the Sardinia one). We stress that the exact choice for the location is not relevant as far as it can provide a long baseline when in a network with European detectors. \label{foot:Idaho}} We observe that the recent report of the NSF MPS AC Subcommittee
on Next-Generation Gravitational-Wave Detector Concepts~\cite{Evans:2023euw}
includes a network of CE-40km and ET (taken in its ET-$\Delta$ configuration) among the recommended world-wide next-generation networks. Our analysis will allow us to compare also with different possible versions of a European project, such as the ET-2L, or the Hybrid configurations.

\section{Methodology}\label{sect:methods}

Our methodology is identical to the one  already followed in Section~3 of ref.~\cite{Branchesi:2023mws}, whom we refer the reader for more details. For compact binary coalescences (CBCs), we perform parameter estimation in the Fisher matrix approximation,  using the 
\texttt{GWFAST} code~\cite{Iacovelli:2022bbs,Iacovelli:2022mbg}.\footnote{The code is publicly available at \url{https://github.com/CosmoStatGW/gwfast}.} \texttt{GWFAST} is a Fisher matrix code tuned toward the needs of 3G detectors. In the context of the activities of the ET  Observational Science Board (OSB),\footnote{See \url{https://www.et-gw.eu/index.php/the-et-collaboration/observational-science-board}.}  extended   cross--checks have performed  between \texttt{GWFAST} and  other recently developed Fisher-matrix codes for 3G detectors, such as \texttt{GWBENCH}~\cite{Borhanian:2020ypi},   \texttt{GWFISH} \cite{Dupletsa:2022wke},  \texttt{TiDoFM} \cite{Chan:2018csa,Li:2021mbo} and the code used in  \cite{Pieroni:2022bbh}. The Fisher matrix formalism has well-known limitations \cite{Vallisneri:2007ev,Rodriguez:2013mla}, but is currently the only computationally  practical way
of dealing with parameter estimation for large populations (see, however, ref.~\cite{Nitz:2021pbr} for progress toward the use of full inference on large populations, and ref.~\cite{Mancarella:2024qle} for a mixed approach in which some parameters are dealt analytically while some are dealt with the Fisher matrix).

We use state--of--the art  waveforms; for binary black holes (BBHs) we use \textsc{IMRPhenomXPHM} (which includes precessing spins and higher-order modes)~\cite{Pratten:2020ceb}. For  binary neutron stars (BNSs) we use \textsc{IMRPhenomD\_NRTidalv2}~\cite{Khan:2015jqa,Dietrich:2019kaq}, which includes tidal effects. 
The parameters of the waveform are 
\be
 \{{\cal M}_c, \eta, d_L, \theta, \phi, \iota, \psi, t_c, \Phi_c, \vchi_1, \vchi_2,  \Lambda_1, \Lambda_2\}\, ,
\ee 
where ${\cal M}_c$ is the detector--frame chirp mass, $\eta$ the symmetric mass ratio, $d_L$ the luminosity distance to the source, $\theta$ and $\phi$  the sky position coordinates, $\iota$ the  angle between the orbital angular momentum of the binary and the line of sight, $\psi$ the polarisation angle, $t_c$ the time of coalescence, $\Phi_c$ the phase at coalescence, $\vchi_{i}$ the dimensionless spin vector of the component $i=\{1,2\}$ of the binary, and $\Lambda_i$ their dimensionless tidal deformabilities. 
Instead of $\Lambda_1, \Lambda_2$, we will actually use the two combinations $\tilde{\Lambda}$ and $\delta\tilde{\Lambda}$ defined in \cite{Wade:2014vqa}. We will show the results for the combination
\be\label{deftildeLambda}
\tilde{\Lambda} = \dfrac{8}{13} \left[(1+7\eta-31\eta^2)(\Lambda_1 + \Lambda_2) + \sqrt{1-4\eta}\, (1+9\eta-11\eta^2)(\Lambda_1 - \Lambda_2)\right]\, ,
\ee
which enters at 5PN order, while $\delta\tilde{\Lambda}$ only enters at 6PN and is more poorly constrained.
For BBHs we will perform the inference on all parameters except, of course, the tidal deformabilities,  that vanish for BHs and, as in LVK parameter estimations, rather than $\iota$, we will use $\theta_{JN}$, defined as  the angle between the total angular momentum and the line of sight; this is the same as $\iota$ only in the absence of precession.
For BNSs, instead, we include tidal deformability but, given the small expected values of their spin magnitudes, we only consider the aligned spin components in the analysis, thus performing estimation on $\chi_{1,z}$ and $\chi_{2,z}$.  The labels `1' and `2' always refer, respectively, to the heaviest and lightest component of the binary system.

We also use state--of--the art population models; in particular, 
we use the same catalog of
BBH and BNS  already used in~\cite{Branchesi:2023mws}.\footnote{These
catalogs are publicly available at  \url{https://apps.et-gw.eu/tds/?content=3&r=18321}.} The catalog for BNSs is based on work developed in refs.~\cite{Mapelli:2017hqk,Giacobbo:2017qhh,Giacobbo:2018hze,Mapelli:2019ipt,Giacobbo:2019fmo,Santoliquido:2020axb}, while the catalog of BBHs is based on \cite{Mapelli:2021syv,Mapelli:2021gyv}.

As in \cite{Branchesi:2023mws}, we assume an uncorrelated 85\% duty cycle in each L-shaped detector, and in each of the three detectors composing the triangle.
All other technical details of the inference process performed below are as described  
in \cite{Iacovelli:2022bbs,Branchesi:2023mws}.

\section{Results}

\subsection{Horizons}

\begin{figure}[tb]
    \centering
    \begin{tabular}{c c}
       \includegraphics[height=.4\textwidth]{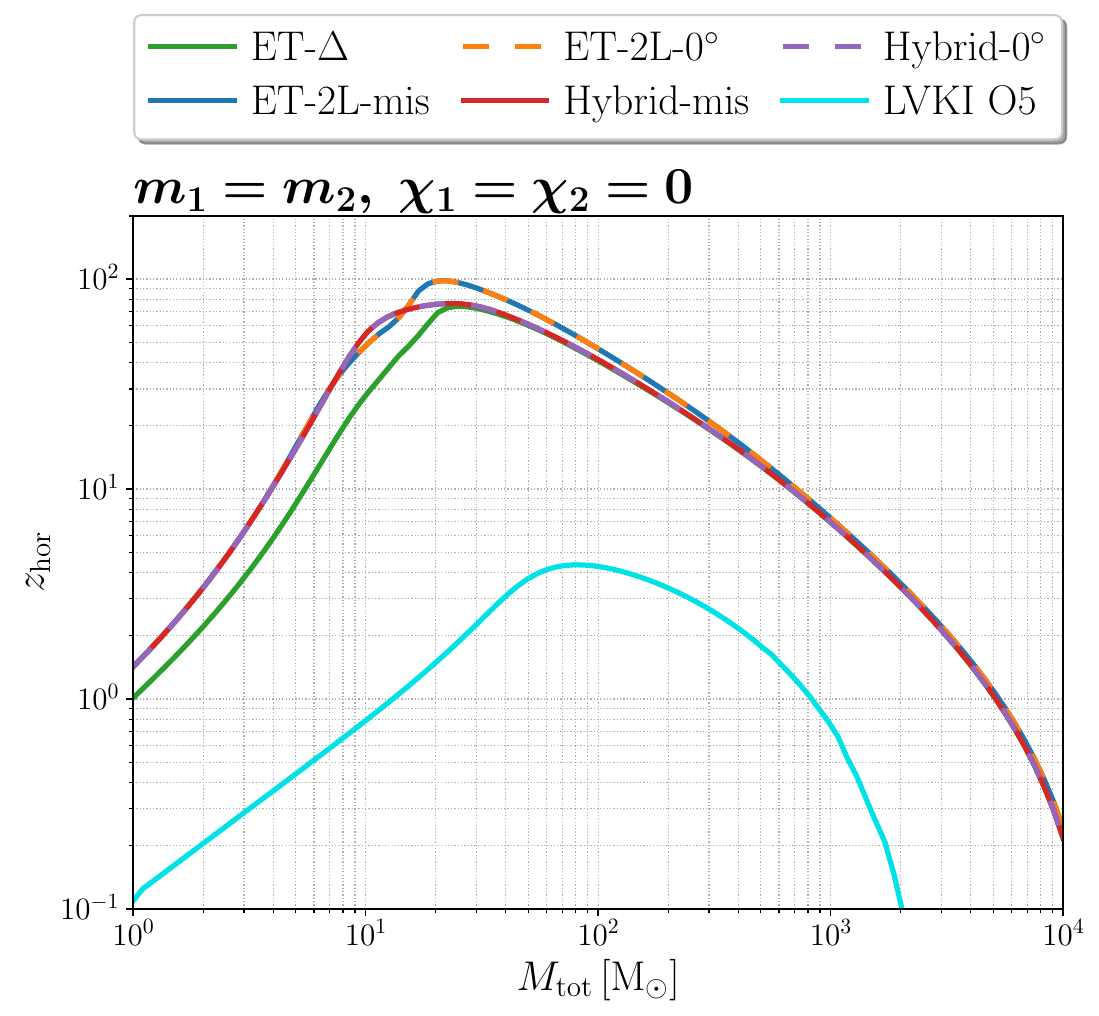}  &  \includegraphics[height=.4\textwidth]{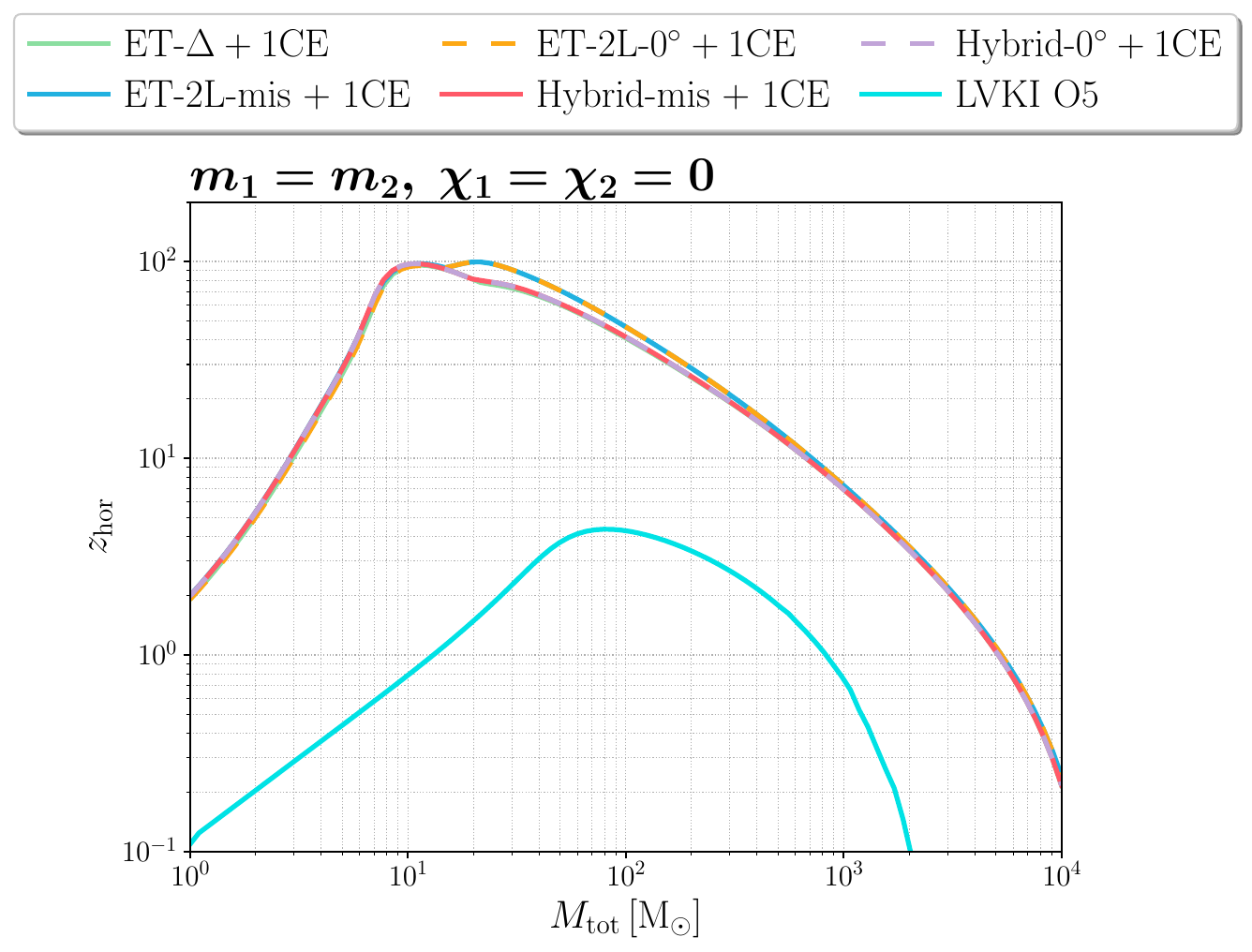}\\
    \end{tabular}
    \caption{Detector horizons for equal mass, non-spinning binaries for the various detector geometries considered for European-only networks (\emph{left panel}), and adding to each 3G network a single 40-km CE detector in the US (\emph{right panel}). In each panel we also add, for comparison, LVKI~O5. On the scale of the figure, ET-2L-$0^{\circ}$ and ET-2L-mis are  indistinguishable, and the same for Hybrid-$0^{\circ}$ and Hybrid-mis.}
    
    \label{fig:All_horizons}
\end{figure}

\autoref{fig:All_horizons} shows the detector horizons  for the  detector networks considered, as a function of the total mass of the binary,  for equal-mass non-spinning coalescing binaries. The left panel gives the results for the 3G European networks that we are considering 
(and, for comparison, also LVKI~O5), while in the right panel each of these 3G
European-only networks  is further enlarged by adding  a 40-km CE located in the US (and we compare again also LVKI~O5). Note that, on this logarithmic scale, the difference between the 2L configurations misaligned or at $0^{\circ}$ are not visible.

First of all,  \autoref{fig:All_horizons} shows that, in terms of detection horizons,  each of these 3G configurations allows us to make a large jump with respect to the best possible 2G detector network. We also see that, above a few hundreds$\msun$, all the 3G networks in the left panel have very similar detection horizons;  below about $20\msun$, however, in the European-only setting some difference appear, and the ET-$\Delta$ configuration is  less performant, while the ET-2L  and Hybrid configurations   have rather similar  horizons. In particular,  for $\mtot=2.7\msun$, a typical value for the total mass of a  BNS, 
the values of the horizons for the 3G configurations in the left panel are given in \autoref{tab:horizon_tabs}, where we see that, for BNS, all networks reach $z_{\rm hor}\simeq 5.3$ except ET-$\Delta$, that reaches $z_{\rm hor}\simeq 3.3$. In contrast, we see from  the right-panel of \autoref{fig:All_horizons} that,  when CE-40km is added, the redshift horizons become closer to each other, almost everywhere in the mass range shown. For $\mtot=2.7\msun$, the horizons are given
in the right panel of \autoref{tab:horizon_tabs}, and they are all between 8.4 and 9.0. Note that, in any case, all these configurations cover the peak of the star formation rate, at $z\sim 2-3$, and therefore the large majority of BNSs.

The detection of subsolar-mass black holes would be a smoking gun signature for their primordial origin. For an equal-mass binary with total mass $\mtot =1.0\msun$ (i.e.,  a $0.5\msun +0.5\msun$ binary), the horizon of all European configurations is $z_{\rm hor}\simeq 1.4$, except for ET-$\Delta$, for which $z_{\rm hor}\simeq 1.0$. Once put into a network with the 40-km CE in the US, however, all these configuration have $z_{\rm hor}$ between 1.9 and 2.0.

Of course, the detection horizons only tell us  a part of the story, and the accuracy of parameter estimation depends significantly also on other aspects (see e.g.~\cite{Vitale:2016icu,Vitale:2018nif,Ng:2021sqn,Ng:2022vbz,Iacovelli:2022bbs,Fairhurst:2023beb,Mancarella:2023ehn} for discussions in the context of 3G detectors). As an  example, the angular resolution of a network of two L-shaped interferometers is very sensitive to the relative orientation between the two Ls, and to the distance between the detectors. In the next subsection we then examine the performances of these configurations for parameter estimation of coalescing binaries.

\begin{table}[tbp]
    \begin{minipage}{.5\linewidth}
      \centering
        \begin{tabular}{!{\vrule width 1pt}l|c!{\vrule width 1pt}}
            \toprule
            \midrule
            \multicolumn{1}{!{\vrule width 1pt}c|}{Detector configuration} & $z_{\rm hor}(2.7\msun)$ \\
            \midrule
            ET-$\Delta$ & 3.3 \\
            ET-2L-mis & 5.3 \\
            ET-2L-$0^\circ$ & 5.3 \\
            Hybrid-mis & 5.3 \\
            Hybrid-$0^\circ$ & 5.3 \\
            \midrule
            LVKI O5 & 0.3 \\
            \midrule
            \bottomrule
            \end{tabular}
    \end{minipage}%
    \begin{minipage}{.5\linewidth}
      \centering
        \begin{tabular}{!{\vrule width 1pt}l|c!{\vrule width 1pt}}
            \toprule
            \midrule
            \multicolumn{1}{!{\vrule width 1pt}c|}{Detector configuration} & $z_{\rm hor}(2.7\msun)$ \\
            \midrule
            ET-$\Delta$ + 1CE & 8.5 \\
            ET-2L-mis + 1CE & 9.0 \\
            ET-2L-$0^\circ$ + 1CE & 8.4 \\
            Hybrid-mis + 1CE & 8.9 \\
            Hybrid-$0^\circ$ + 1CE & 8.9 \\
            \midrule
            LVKI O5 & 0.3 \\
            \midrule
            \bottomrule
        \end{tabular}
    \end{minipage}
    \caption{Horizon redshifts for equal mass non-spinning binaries evaluated at a source-frame total mass $M_{\rm tot} = 2.7\msun$ for the various configurations considered  without (\emph{left table}) and with (\emph{right table}) a single 40-km CE detector in the US.}
    \label{tab:horizon_tabs}
\end{table}

\subsection{Parameter reconstruction of coalescing binaries}\label{sect:PE}

\subsubsection{European networks}

\begin{figure}[!htbp]
    \centering
    \includegraphics[width=.95\textwidth]{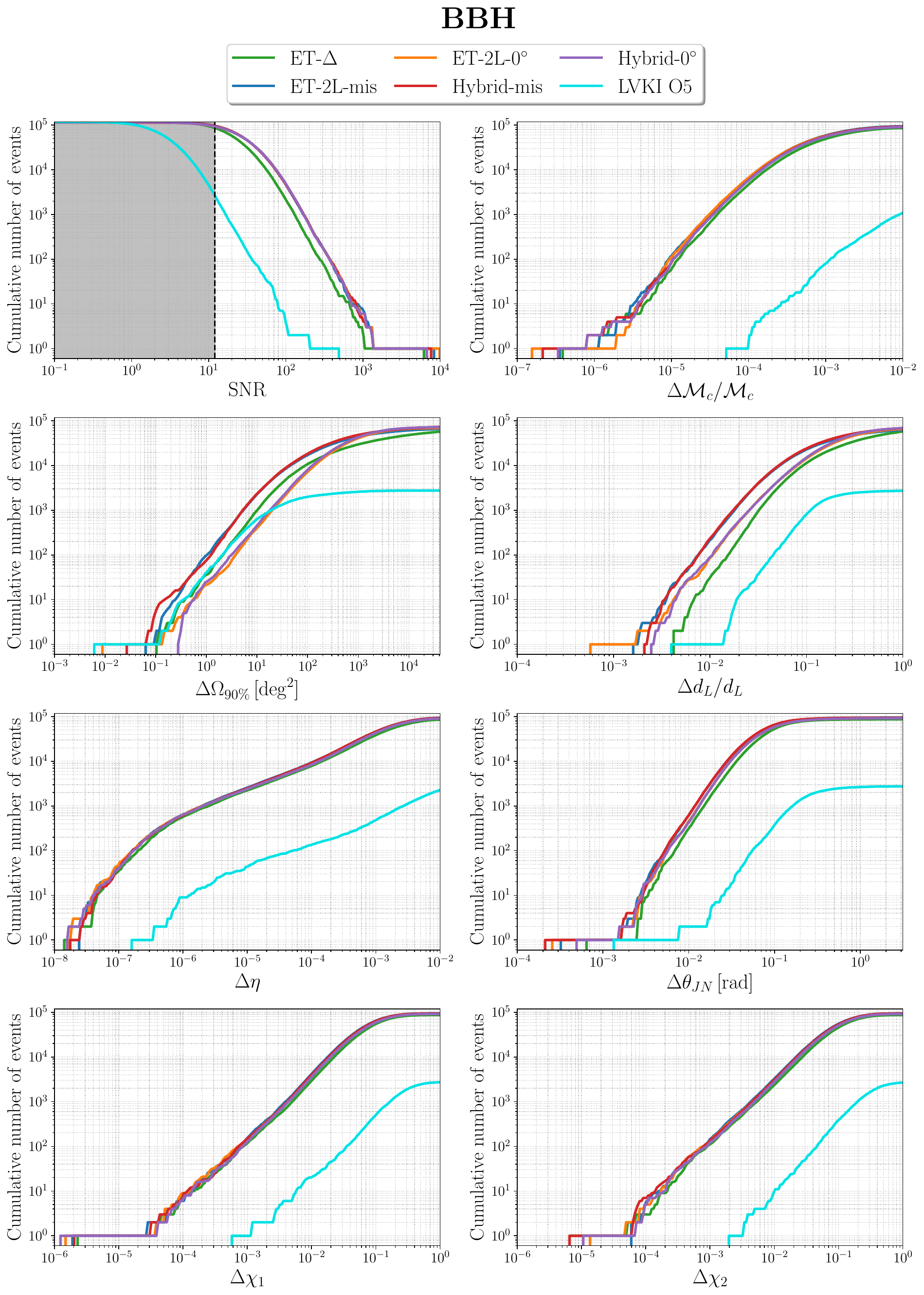}
    \caption{Cumulative distributions of the number of detections per year, for the SNRs and for the error on the parameters, for BBH signals, for the European networks considered.}
    \label{fig:PE_BBH_noCE}
\end{figure}

We begin by considering the networks where all 3G detectors are located in Europe. 
\autoref{fig:PE_BBH_noCE} shows the
cumulative distributions of the number of detections per year, for the network SNRs and for the error on the parameters, for BBH signals.
We see that the most significant differences appear in the SNR distribution, angular localization, and luminosity distance.
For the SNR distributions, among the 3G networks, the ET-$\Delta$  configuration is noticeably less performant, even on this logarithmic scale, while all others 3G configurations are equivalent. For angular resolution, the best results come from ET-2L-mis and Hybrid-mis, which are very similar among them, and clearly better than ET-$\Delta$, which in turn is clearly better than ET-2L-$0^{\circ}$ and Hybrid-$0^{\circ}$. For the accuracy on the luminosity distance, again ET-2L-mis and Hybrid-mis are very similar among them and provide the best results, followed in this case by ET-2L-$0^{\circ}$ and Hybrid-$0^{\circ}$, while ET-$\Delta$ is the less performing configuration. Note that, for angular localization, LVKI at O5 sensitivity is very competitive, given that it is composed by five detectors, with large baselines among them; indeed, for events localized better than a few degrees, only ET-2L-mis and Hybrid-mis perform better than LVKI~O5.
\autoref{tab:BBH_numbers_loc_noCE} gives, for each of these 3G networks,  the number of BBHs with 
angular resolution $\Delta\Omega_{90\%}$ (defined, as in \cite{Branchesi:2023mws}, as the sky localisation area  at $90\%$ c.l.)
smaller than $1\, {\rm deg}^2$, or smaller than $10\, {\rm deg}^2$ (first two column), 
as well as the BBHs that (independently of their angular localization) have
$\Delta d_L/d_L\leq5\times 10^{-3}$, or $\Delta d_L/d_L\leq10^{-2}$ (third and fourth column).\footnote{All these figures are for one year of data taking, with the duty cycles given at the end of \autoref{sect:methods}.} 
Observe, in particular, that for BBHs with $d_L$ measured better than $1\%$, we have 217 events for Hybrid-mis and 28 for ET-$\Delta$, a difference by one order of magnitude.
For all other parameters the differences are not so significant (except, more marginally, for $\theta_{JN}$).

\autoref{fig:histz_BBH_noCE} shows the distribution in redshift of ``golden events'' defined, as in ref.~\cite{Branchesi:2023mws}, as events with especially high SNR (left column), or  especially good reconstruction of luminosity distance (middle column) or of sky location (right column), for BBHs in these European-only networks. We observe that, also on this metric, ET-2L-mis and Hybrid-mis give the best performance in all three panels; in the case of  the SNR, also ET-2L-$0^{\circ}$ and Hybrid-$0^{\circ}$ perform very well;  however, the -$0^{\circ}$ configurations are again the less performant, among the 3G configurations, for  events with especially good angular localizations.
Note, however, that all 3G configurations now perform much better than LVKI~O5  for well-localized events at large redshift. LVKI~O5 has no BBH localized to better than $10\, {\rm deg}^2$ beyond $z=1$, while all other 3G networks detect hundreds of events with this localization at $z>1$, and a few of them even up to $z\sim 3-4$.

\begin{figure}[!tbp]
    \centering
    \begin{tabular}{c c c}
        \includegraphics[width=.31\textwidth]{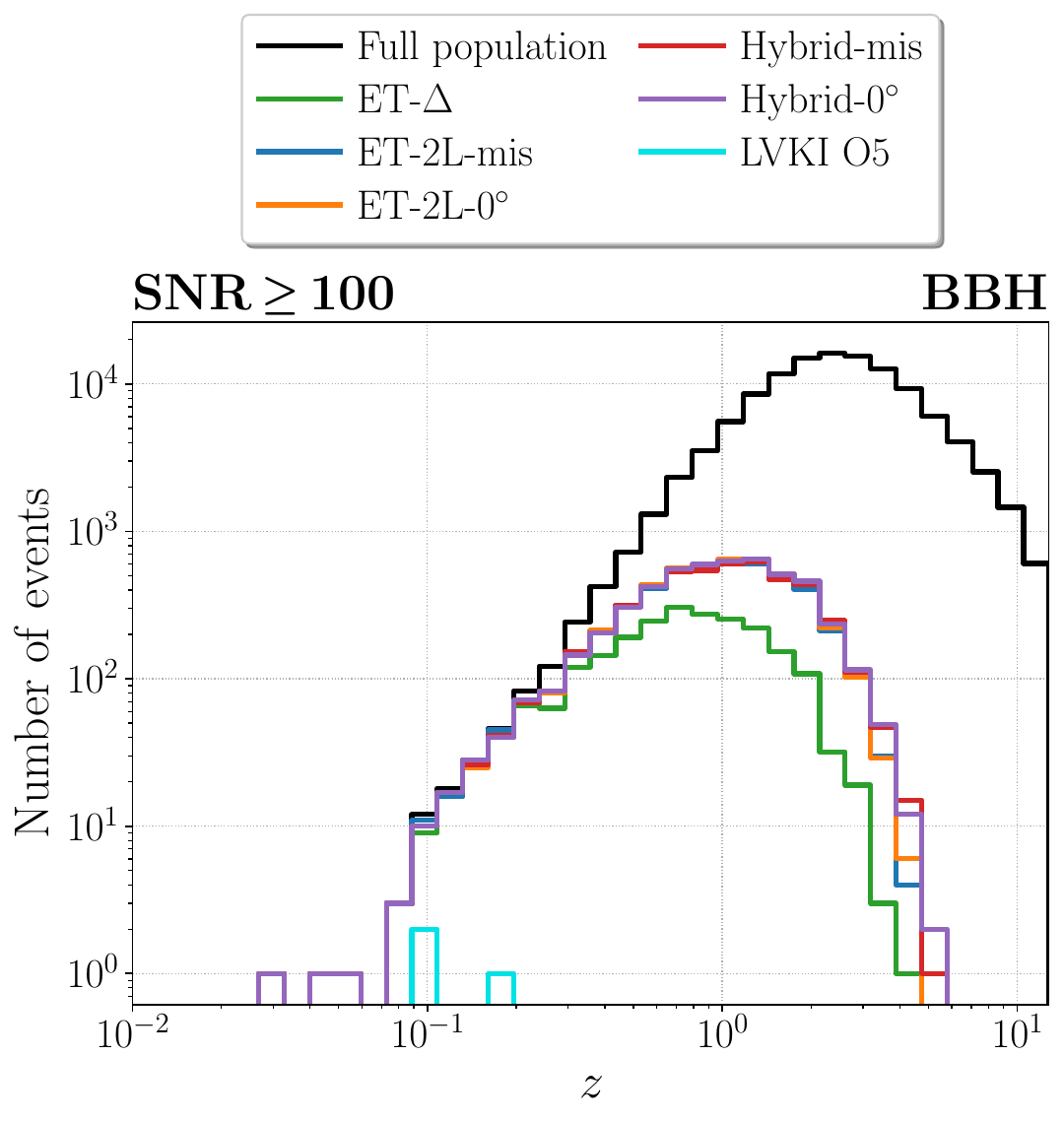}  & \includegraphics[width=.31\textwidth]{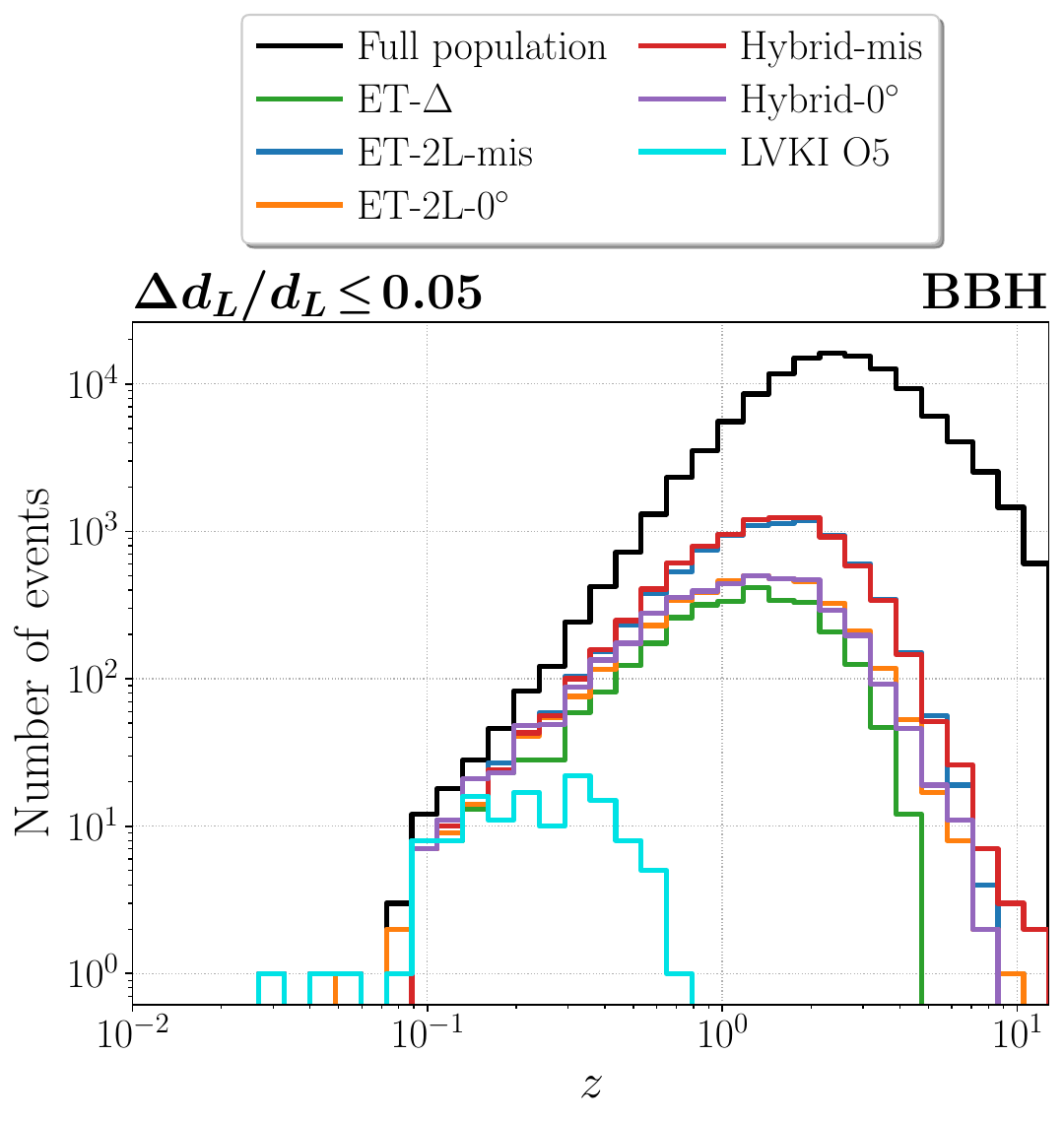} & \includegraphics[width=.31\textwidth]{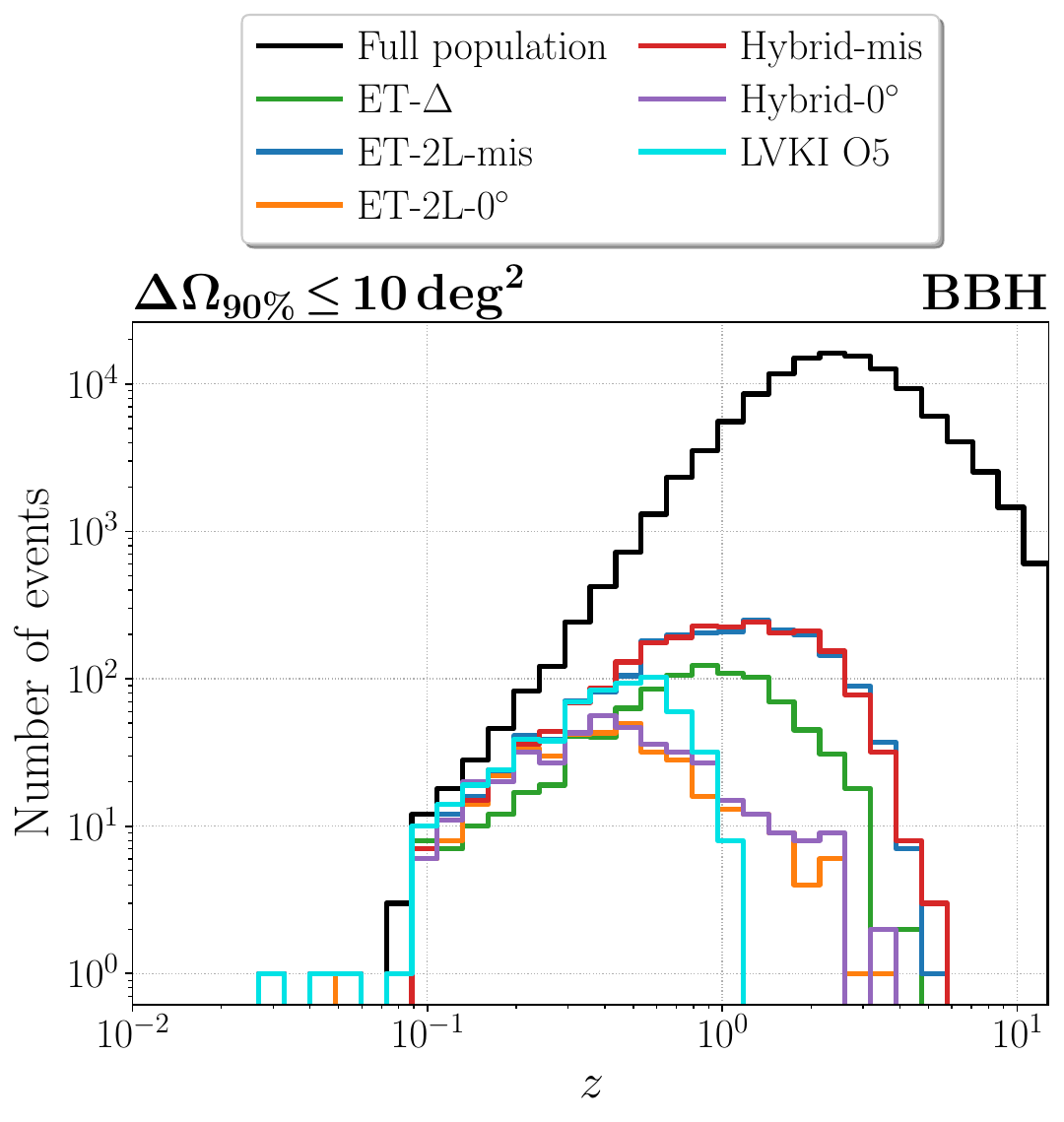} \\
    \end{tabular}
    \caption{Redshift distribution of BBHs detected with ${\rm SNR} \geq 100$ (left column), relative error on the luminosity distance $\Delta d_L/d_L \leq 0.05$ (central column), or sky location $\Delta\Omega_{90\%} \leq 10~{\rm deg}^2$ (right column) for the various detector geometries considered with all detectors located in Europe.}
    \label{fig:histz_BBH_noCE}
\end{figure}
\FloatBarrier
\begin{table}[!htbp]
    \centering
    \begin{tabular}{!{\vrule width 1pt}m{3.1cm}|S[table-format=2.0]|S[table-format=4.0]|S[table-format=2.0]|S[table-format=3.0]!{\vrule width 1pt}}
        \toprule
        \midrule
        \multicolumn{5}{!{\vrule width 1pt}c!{\vrule width 1pt}}{\textbf{BBH}}\\
        \midrule
        \hfil\multirow{3}{*}{\shortstack[c]{Detector\\configuration}}\hfill & \multicolumn{4}{c!{\vrule width 1pt}}{Detections with} \\
        & \multicolumn{2}{c|}{$\Delta\Omega_{90\%}\leq$} & \multicolumn{2}{c!{\vrule width 1pt}}{$\Delta d_L/d_L \leq$}\\
        \cmidrule(lr){2-5}
        & \multicolumn{1}{c|}{$1~{\rm deg}^2$} & \multicolumn{1}{c|}{$10~{\rm deg}^2$}  & \multicolumn{1}{c|}{$5\times10^{-3}$} & \multicolumn{1}{c!{\vrule width 1pt}}{$10^{-2}$}\\
        \midrule
        \textrm{ET-}$\Delta$ & 35 & 914 & 2 & 28 \\
        \textrm{ET-2L-mis} & 92 & 2124 & 29 & 202 \\
        \textrm{ET-2L-}$0^\circ$ & 21 & 374 & 15 & 79 \\
        \textrm{Hybrid-mis} & 70 & 2180 & 32 & 217 \\
        \textrm{Hybrid-}$0^\circ$ & 24 & 416 & 17 & 84 \\
        \midrule
        \bottomrule
    \end{tabular}
    \caption{Number of detected BBH sources at the considered European networks with different cuts on the sky localization or on the relative error on the luminosity distance.}
    \label{tab:BBH_numbers_loc_noCE}
\end{table}

\hyperref[fig:PE_BNS_noCE]{Figures~\ref*{fig:PE_BNS_noCE}} and \ref{fig:histz_BNS_noCE}
show the corresponding results for BNS. From \autoref{fig:PE_BNS_noCE} we see that, for angular localization, ET-2L-mis is  better than the other configurations, which are close among them, especially for localizations better than $100\, {\rm deg}^2$. For luminosity distance, ET-2L-mis is clear the best, followed by Hybrid-mis, which in turn is clearly better than  ET-$\Delta$ and
ET-2L-$0^{\circ}$. \autoref{tab:BNS_numbers_loc_noCE} gives, for each of these networks,  the number of BNSs with $\Delta\Omega_{90\%}\leq10\, {\rm deg}^2$ or with $\Delta\Omega_{90\%}\leq100\, {\rm deg}^2$, and the number of BNSs that (independently of their angular localization) have
$\Delta d_L/d_L\leq5\times 10^{-2}$, or $\Delta d_L/d_L\leq10^{-1}$. 

Observe that the Hybrid-mis configuration actually localizes BNS somewhat better than ET-$\Delta$: for $\Delta\Omega_{90\%}\leq10\, {\rm deg}^2$ it has 12 events against 8, while for $\Delta\Omega_{90\%}\leq100\, {\rm deg}^2$ it has 288 events against 184. It is instructive to understand the mechanism behind this result. In general, a better low-frequency sensitivity allows  BNSs to stay longer in the bandwidth, up to several hours or a day; then, the modulation of the signal due to  Earth's  rotation (which is an effect  taken into account in \texttt{GWFAST}) allows us to improve the angular localization; since ET-$\Delta$ has a better low-frequency sensitivity than Hybrid-mis, this effect favors ET-$\Delta$ over Hybrid-mis. On the other hand, two well-separated detectors can better triangulate the signal, with respect to a single-site detector, and this favors Hybrid-mis over ET-$\Delta$. We see that, overall, in the comparison among these two configurations, the effect of the long baseline dominates. On the other hand, ET-2L-mis benefits of both effects (better low-frequency sensitivity and long baseline) and has clearly the best performances, with 25 BNS localized better than  $10\, {\rm deg}^2$, and 559 better than  $100\, {\rm deg}^2$.

From \autoref{fig:PE_BNS_noCE} we  see that significant differences also show up in the polarization angle $\psi$ and in orbit inclination $\theta_{JN}$, where the parallel configurations (and, for $\psi$, ET-$\Delta$), clearly perform less well. In general, whatever the observable, we always find that, among the 3G networks,  ET-2L-mis and  Hybrid-mis are the two top performers, sometime with clear differences with respect to the other configurations, sometime together with other configurations. The same message emerges from the golden BNS events in 
\autoref{fig:histz_BNS_noCE}. Note, however, that the LVKI-O5 network, while clearly inferior for all other observables, is the one that get the best results for events with angular localization below a few degrees, thanks to the fact of being made by five detectors, with large baseline distances.

\begin{figure}[!htbp]
    \centering
    \includegraphics[width=.95\textwidth]{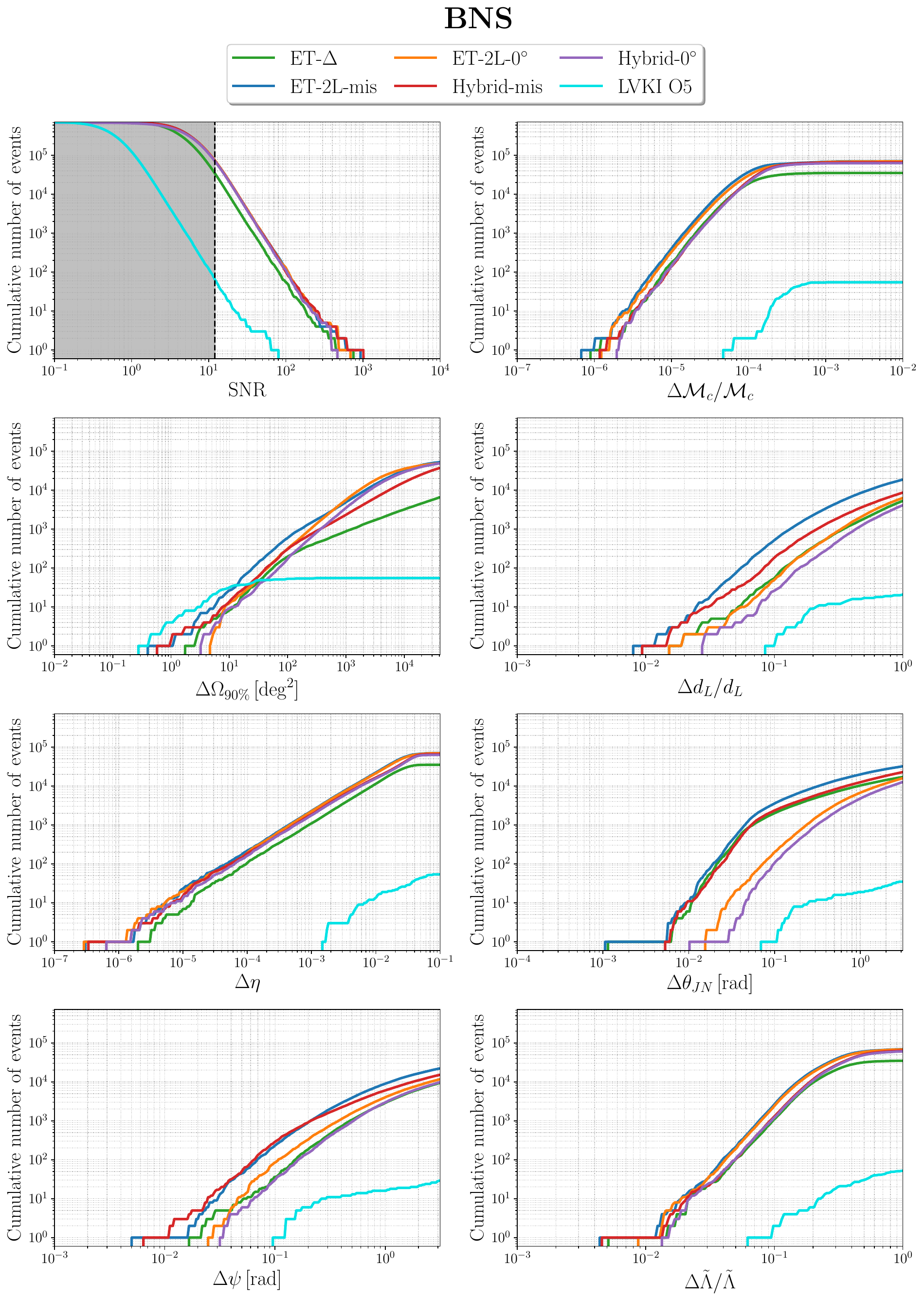}
    \caption{Cumulative distributions of the number of detections per year, for the SNRs and for the error on the parameters, for BNS signals, for the European networks considered.}
    \label{fig:PE_BNS_noCE}
\end{figure}

\begin{figure}[!htbp]
    \centering
    \begin{tabular}{c c c}
       \includegraphics[width=.31\textwidth]{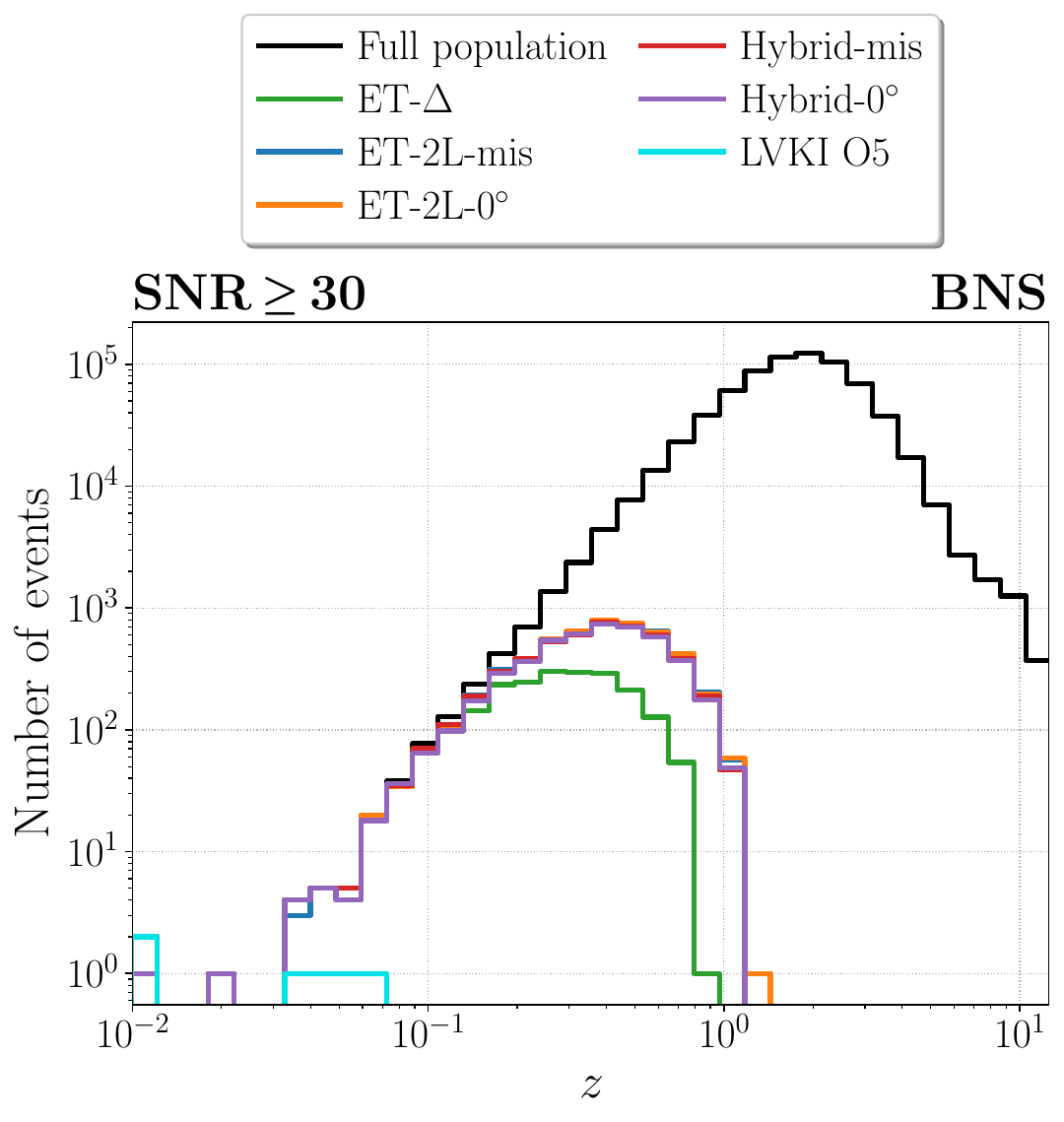}  & \includegraphics[width=.31\textwidth]{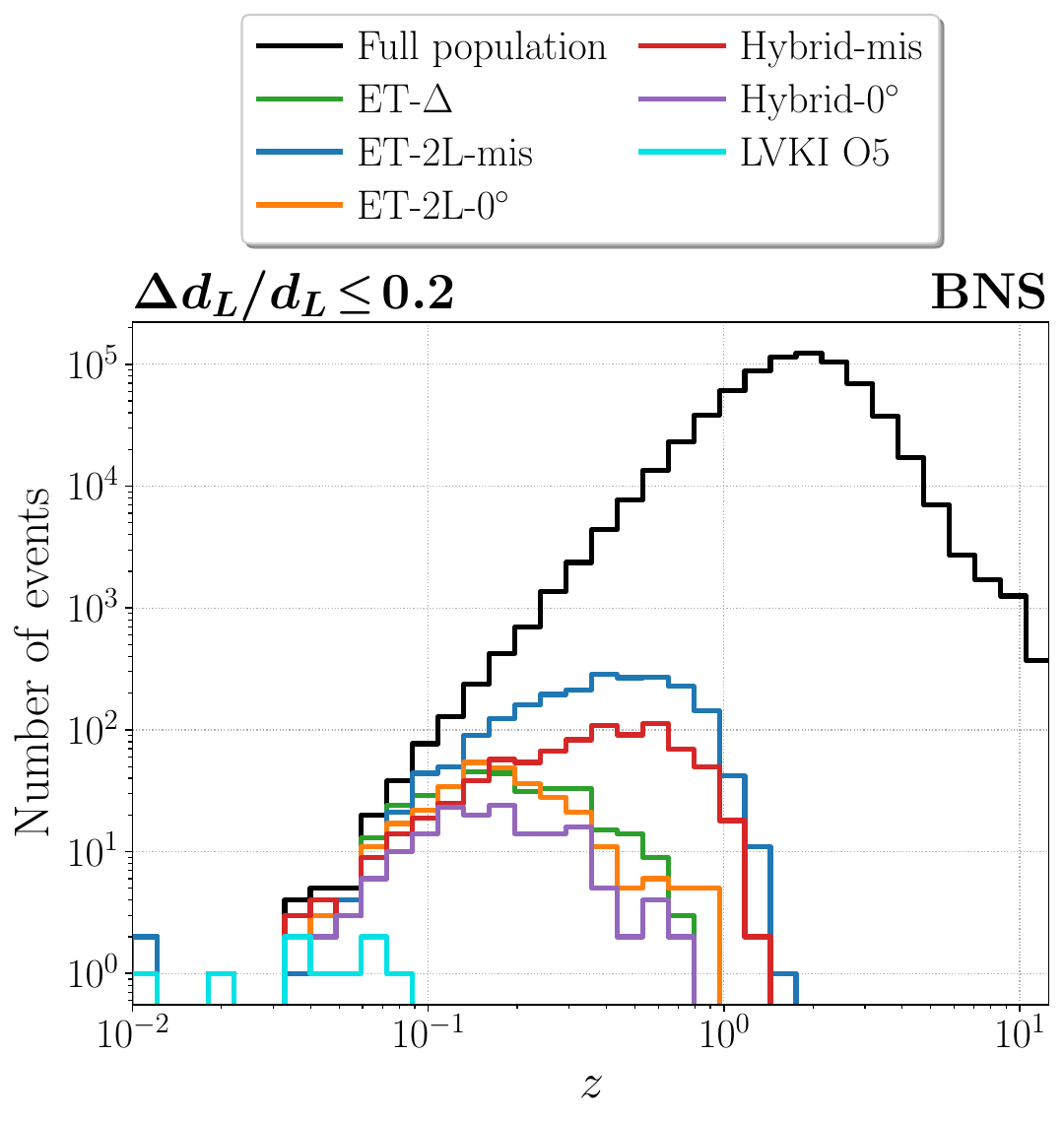} & \includegraphics[width=.31\textwidth]{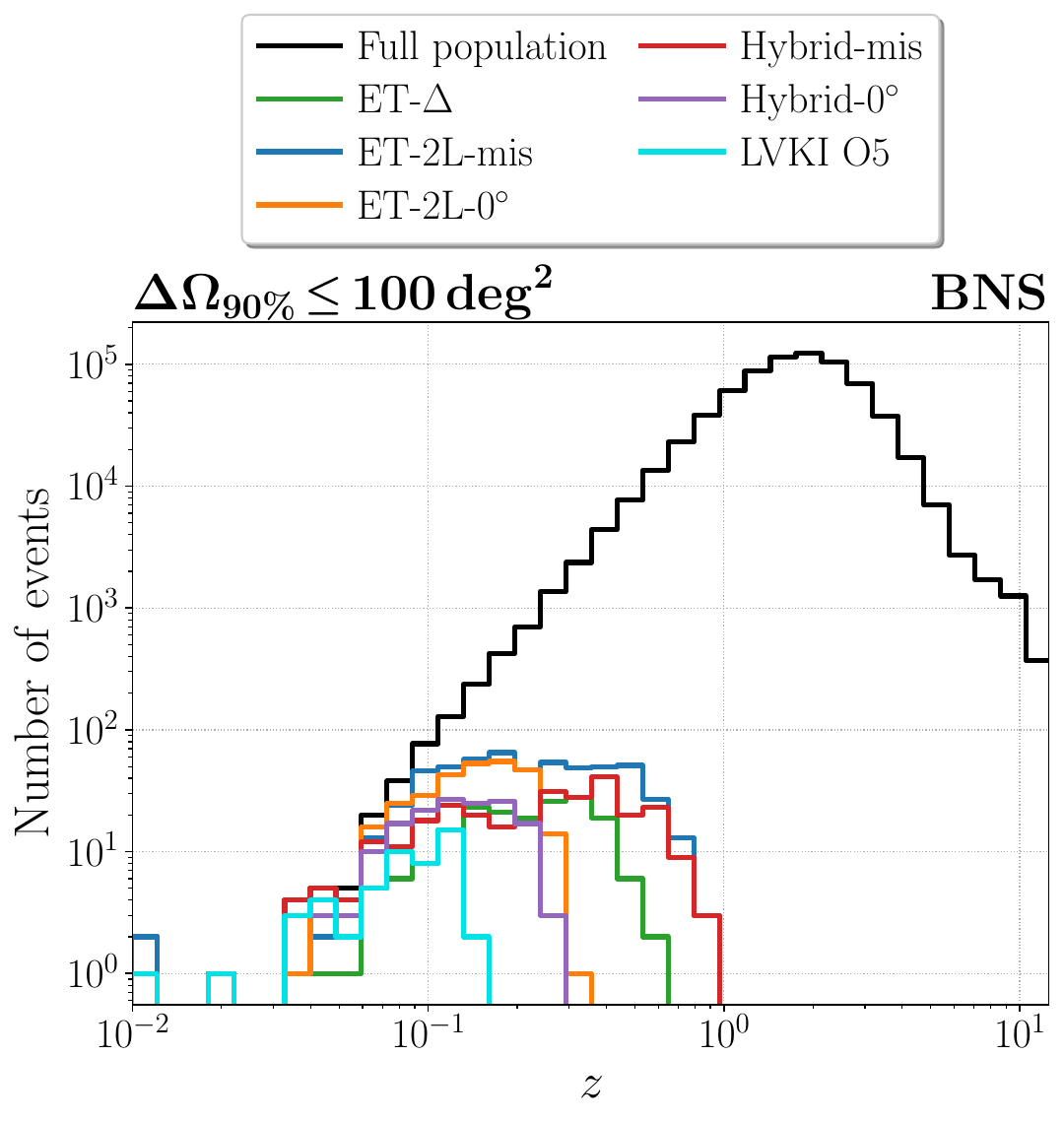} \\

    \end{tabular}
    \caption{Redshift distribution of BNSs detected with ${\rm SNR} \geq 30$ (left column), relative error on the luminosity distance $\Delta d_L/d_L \leq 0.2$ (central column), or sky location $\Delta\Omega_{90\%} \leq 100~{\rm deg}^2$ (right column) for the various detector geometries considered with all detectors located in Europe.}
    \label{fig:histz_BNS_noCE}
\end{figure}
\FloatBarrier
\begin{table}[!htbp]
    \centering
    \begin{tabular}{!{\vrule width 1pt}m{3.1cm}|S[table-format=2.0]|S[table-format=3.0]|S[table-format=2.0]|S[table-format=3.0]!{\vrule width 1pt}}
        \toprule
        \midrule
        \multicolumn{5}{!{\vrule width 1pt}c!{\vrule width 1pt}}{\textbf{BNS}}\\
        \midrule
        \hfil\multirow{3}{*}{\shortstack[c]{Detector\\configuration}}\hfill & \multicolumn{4}{c!{\vrule width 1pt}}{Detections with} \\
        & \multicolumn{2}{c|}{$\Delta\Omega_{90\%}\leq$} & \multicolumn{2}{c!{\vrule width 1pt}}{$\Delta d_L/d_L \leq$}\\
        \cmidrule(lr){2-5}
        & \multicolumn{1}{c|}{$10~{\rm deg}^2$} & \multicolumn{1}{c|}{$100~{\rm deg}^2$}  & \multicolumn{1}{c|}{$5\times10^{-2}$} & \multicolumn{1}{c!{\vrule width 1pt}}{$10^{-1}$}\\
        \midrule
        \textrm{ET-}$\Delta$ & 8 & 184 & 8 & 52 \\
        \textrm{ET-2L-mis} & 25 & 559 & 69 & 479 \\
        \textrm{ET-2L-}$0^\circ$ & 12 & 293 & 7 & 48 \\
        \textrm{Hybrid-mis} & 12 & 288 & 28 & 169 \\
        \textrm{Hybrid-}$0^\circ$ & 9 & 157 & 4 & 25 \\
        \midrule
        \bottomrule
    \end{tabular}
    \caption{Number of detected BNS sources at the considered European networks with different cuts on the sky localization and relative error on the luminosity distance.}
    \label{tab:BNS_numbers_loc_noCE}
\end{table}

\subsubsection{European networks together with a CE-40km in the US}

\begin{figure}[!tbp]
    \centering
    \includegraphics[width=.95\textwidth]{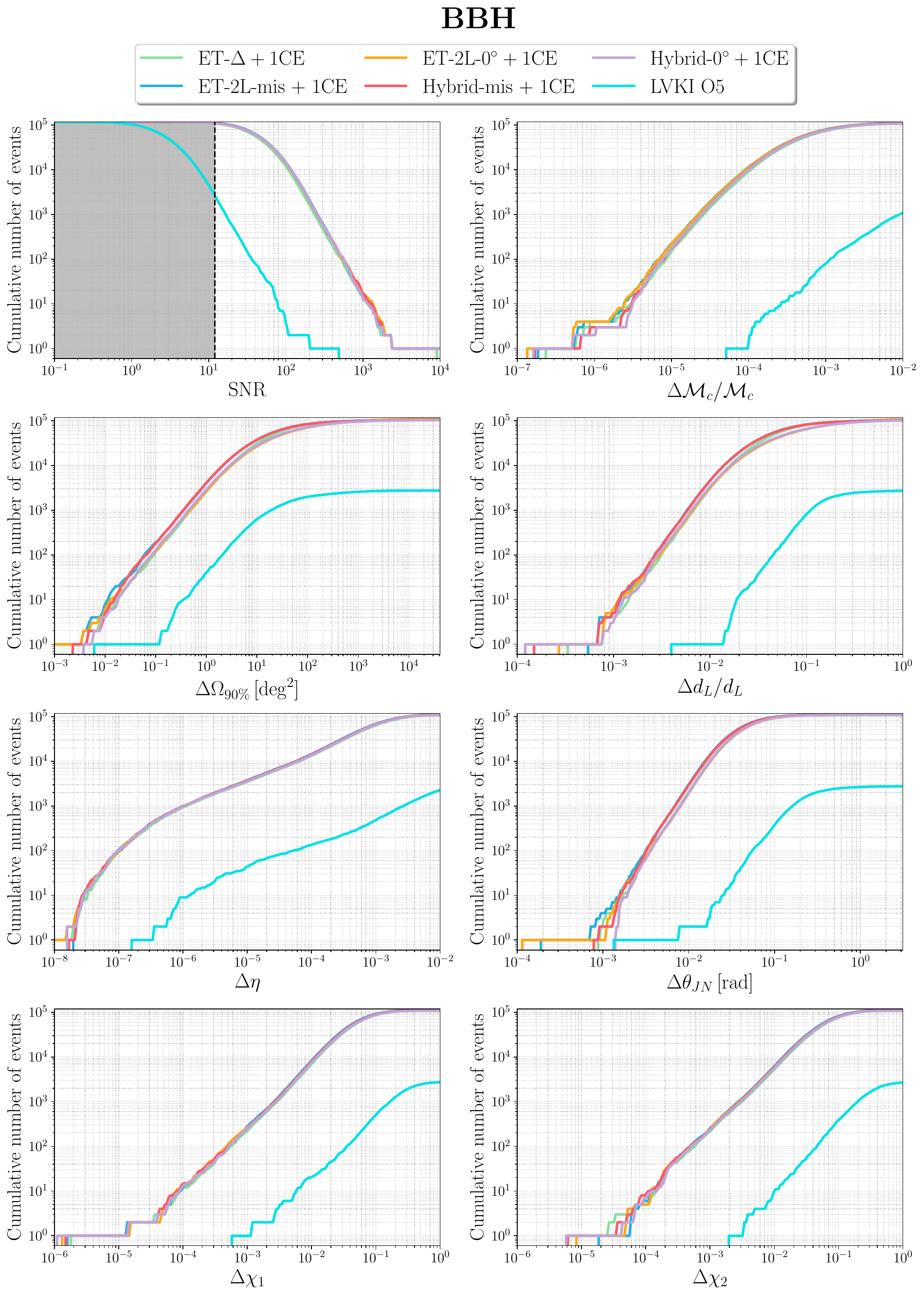}
    \caption{Cumulative distributions of the number of detections per year, for the SNRs and for the error on the parameters, for BBH signals, for the various detector geometries considered, including a single 40-km CE detector in the US.}
    \label{fig:PE_BBH_wCE}
\end{figure}

We now consider a set of broader world-wide networks, in which we add a 40-km CE detector in the US to each of the 3G European configurations studied in the previous subsection. The results are given in \hyperref[fig:PE_BBH_wCE]{Figures~\ref*{fig:PE_BBH_wCE}} and \ref{fig:histz_BBH_wCE} for BBHs, and in 
\hyperref[fig:PE_BNS_wCE]{Figures~\ref*{fig:PE_BNS_wCE}} and \ref{fig:histz_BNS_wCE} for BNSs.

We see that, for BBHs,  all network configurations become essentially equivalent, on all observables.\footnote{The results for ET-$\Delta$, ET-2L-mis and ET-2L-$0^{\circ}$, together with 1CE, are the same already shown in Figure~18 of \cite{Branchesi:2023mws}. In that figure, the spread in the results looked visually larger because, there, were included also the configurations with 2 CE in the US.} 
This is confirmed by the values given in \autoref{tab:BBH_numbers_loc_wCE}.

For BNSs the differences are  larger, in particular for angular localization and luminosity distance, where the best results are obtained by 
(ET-2L-mis + 1CE) and (Hybrid-mis + 1CE), that are typically better by a factor of order 2 than ET-$\Delta$, which in turn is somewhat better than  (ET-2L-$0^{\circ}$ + 1CE) and (Hybrid-$0^{\circ}$ + 1CE). 
As an example, 
\autoref{tab:BNS_numbers_loc_wCE} gives, for each of these networks,  the number of BNSs with $\Delta\Omega_{90\%}\leq10\, {\rm deg}^2$ or with $\Delta\Omega_{90\%}\leq100\, {\rm deg}^2$, as well as the BNSs that (independently of their angular localization) have
$\Delta d_L/d_L\leq5\times 10^{-2}$, or $\Delta d_L/d_L\leq10^{-1}$ (again, for one yr of data and our choice of duty cycle).

Significant differences appear also  for the polarization angle and the orbit inclination, for which again (ET-2L-mis + 1CE) and (Hybrid-mis + 1CE) are the best configurations.

\begin{figure}[!tbp]
    \centering
    \begin{tabular}{c c c}
       \includegraphics[width=.31\textwidth]{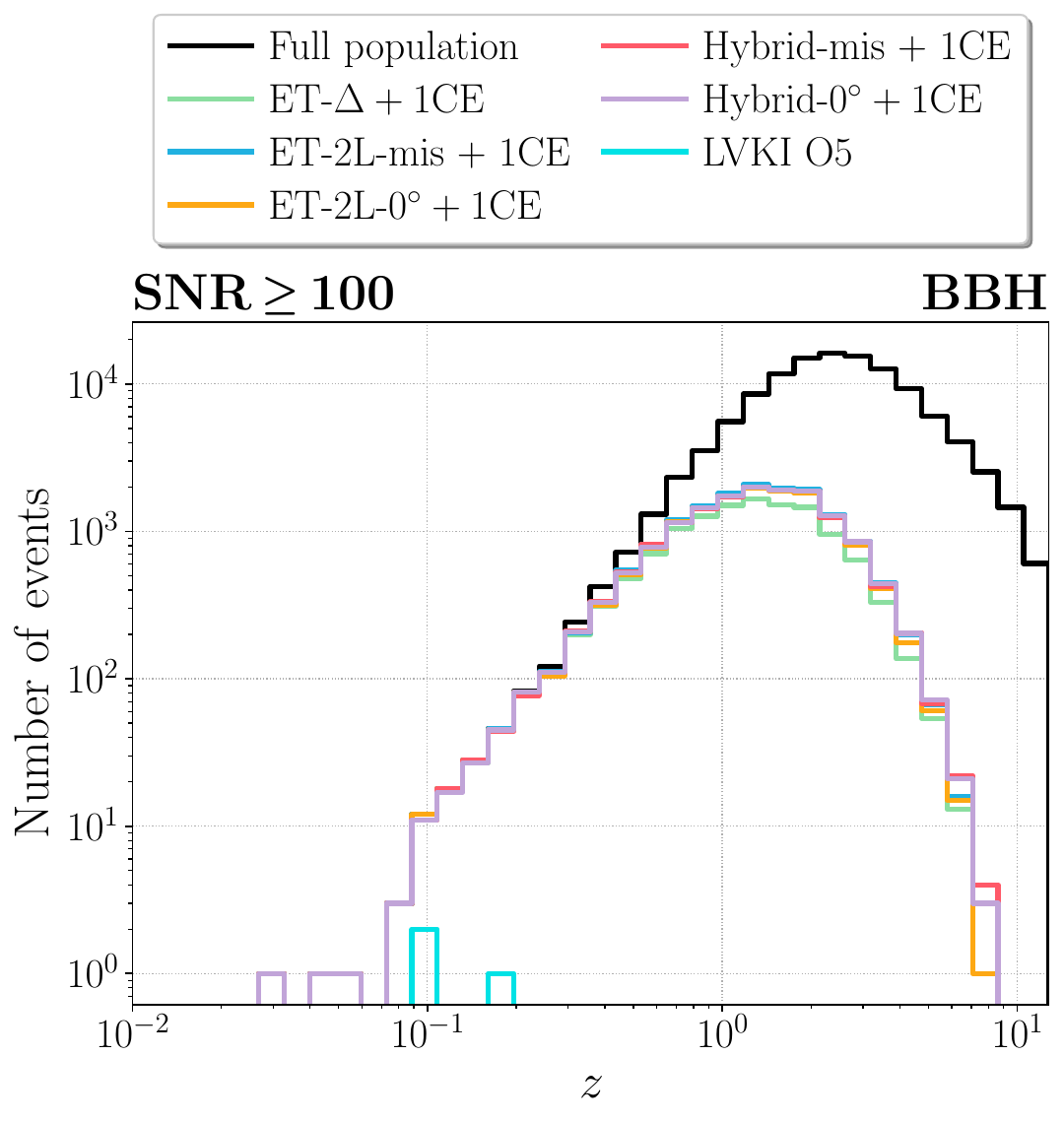}  & \includegraphics[width=.31\textwidth]{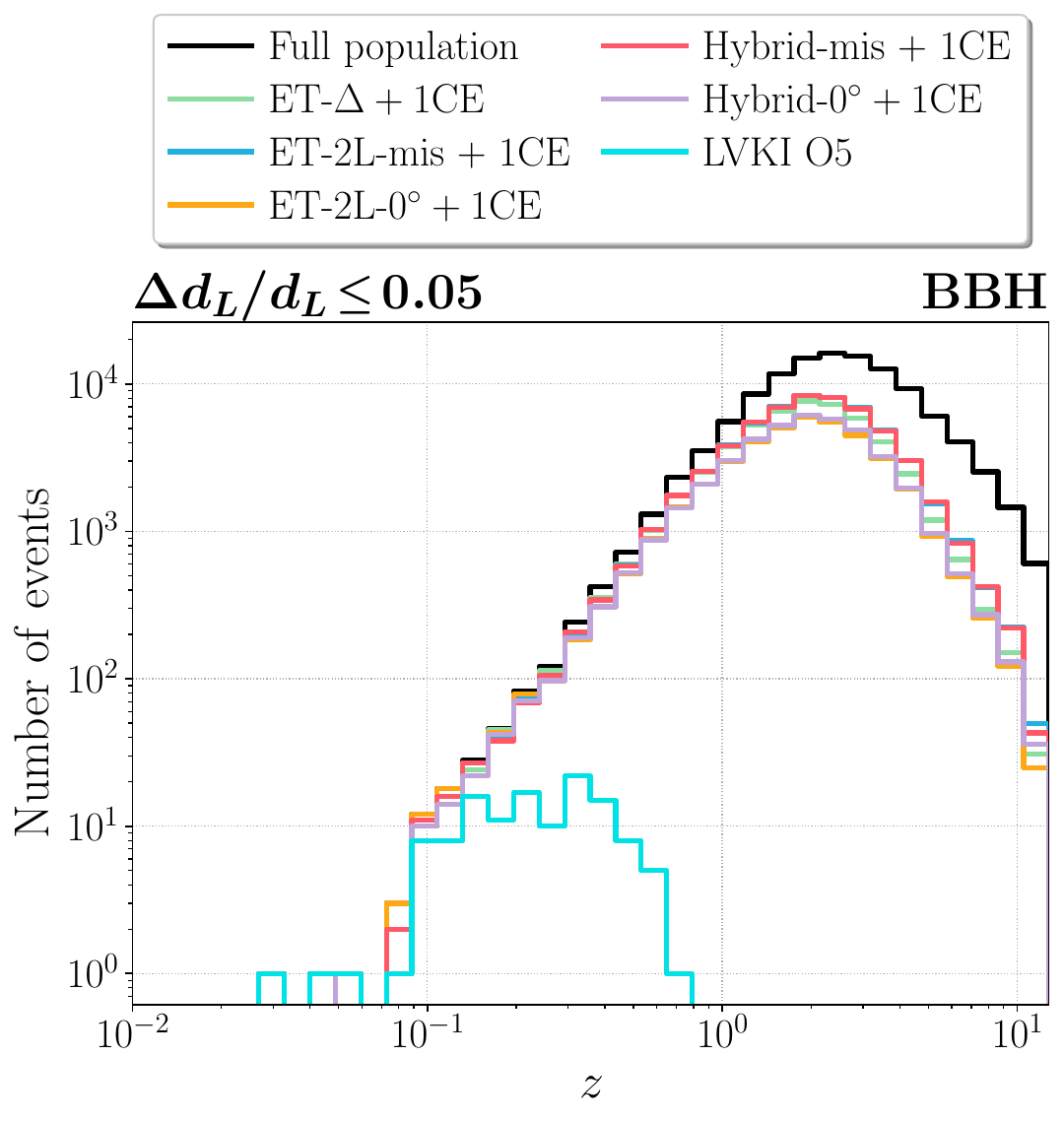} & \includegraphics[width=.31\textwidth]{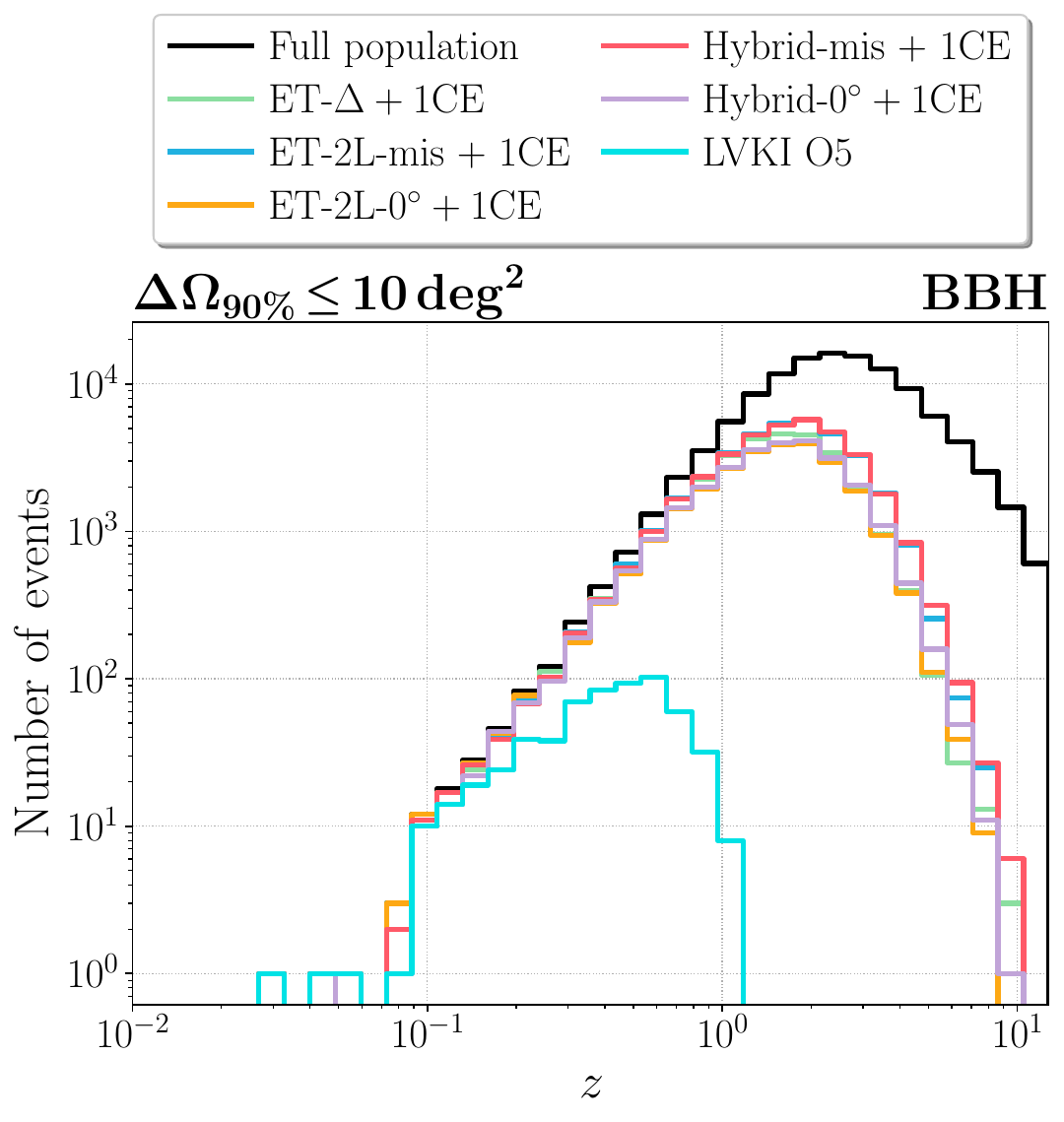} 
    \end{tabular}
    \caption{Redshift distribution of BBHs detected with ${\rm SNR} \geq 100$ (left column), relative error on the luminosity distance $\Delta d_L/d_L \leq 0.05$ (central column), or sky location $\Delta\Omega_{90\%} \leq 10~{\rm deg}^2$ (right column) for the various detector geometries considered, including a single 40-km CE detector in the US.}
    \label{fig:histz_BBH_wCE}
\end{figure}

\begin{table}[!tbp]
    \centering
    \begin{tabular}{!{\vrule width 1pt}m{3.1cm}|S[table-format=4.0]|S[table-format=5.0]|S[table-format=3.0]|S[table-format=4.0]!{\vrule width 1pt}}
        \toprule
        \midrule
        \multicolumn{5}{!{\vrule width 1pt}c!{\vrule width 1pt}}{\textbf{BBH}}\\
        \midrule
        \hfil\multirow{3}{*}{\shortstack[c]{Detector\\configuration}}\hfill & \multicolumn{4}{c!{\vrule width 1pt}}{Detections with} \\
        & \multicolumn{2}{c|}{$\Delta\Omega_{90\%}\leq$} & \multicolumn{2}{c!{\vrule width 1pt}}{$\Delta d_L/d_L \leq$}\\
        \cmidrule(lr){2-5}
        & \multicolumn{1}{c|}{$1~{\rm deg}^2$} & \multicolumn{1}{c|}{$10~{\rm deg}^2$}  & \multicolumn{1}{c|}{$5\times10^{-3}$} & \multicolumn{1}{c!{\vrule width 1pt}}{$10^{-2}$}\\
        \midrule
        \textrm{ET-}$\Delta + {\rm 1CE}$ & 2447 & 29924 & 395 & 2901 \\
        \textrm{ET-2L-mis + 1CE} & 3743 & 36457 & 575 & 4301 \\
        \textrm{ET-2L-}$0^\circ + {\rm 1CE}$ & 2464 & 25782 & 400 & 2995 \\
        \textrm{Hybrid-mis + 1CE} & 3810 & 36344 & 581 & 4276 \\
        \textrm{Hybrid-}$0^\circ + {\rm 1CE}$ & 2704 & 27043 & 433 & 3153 \\
        \midrule
        \bottomrule
    \end{tabular}
    \caption{Number of detected BBH sources at the considered networks including a single 40-km CE detector in the US, with different cuts on the sky localization and relative error on the luminosity distance.}
    \label{tab:BBH_numbers_loc_wCE}
\end{table}

\begin{figure}[!tbp]
    \centering
    \includegraphics[width=.95\textwidth]{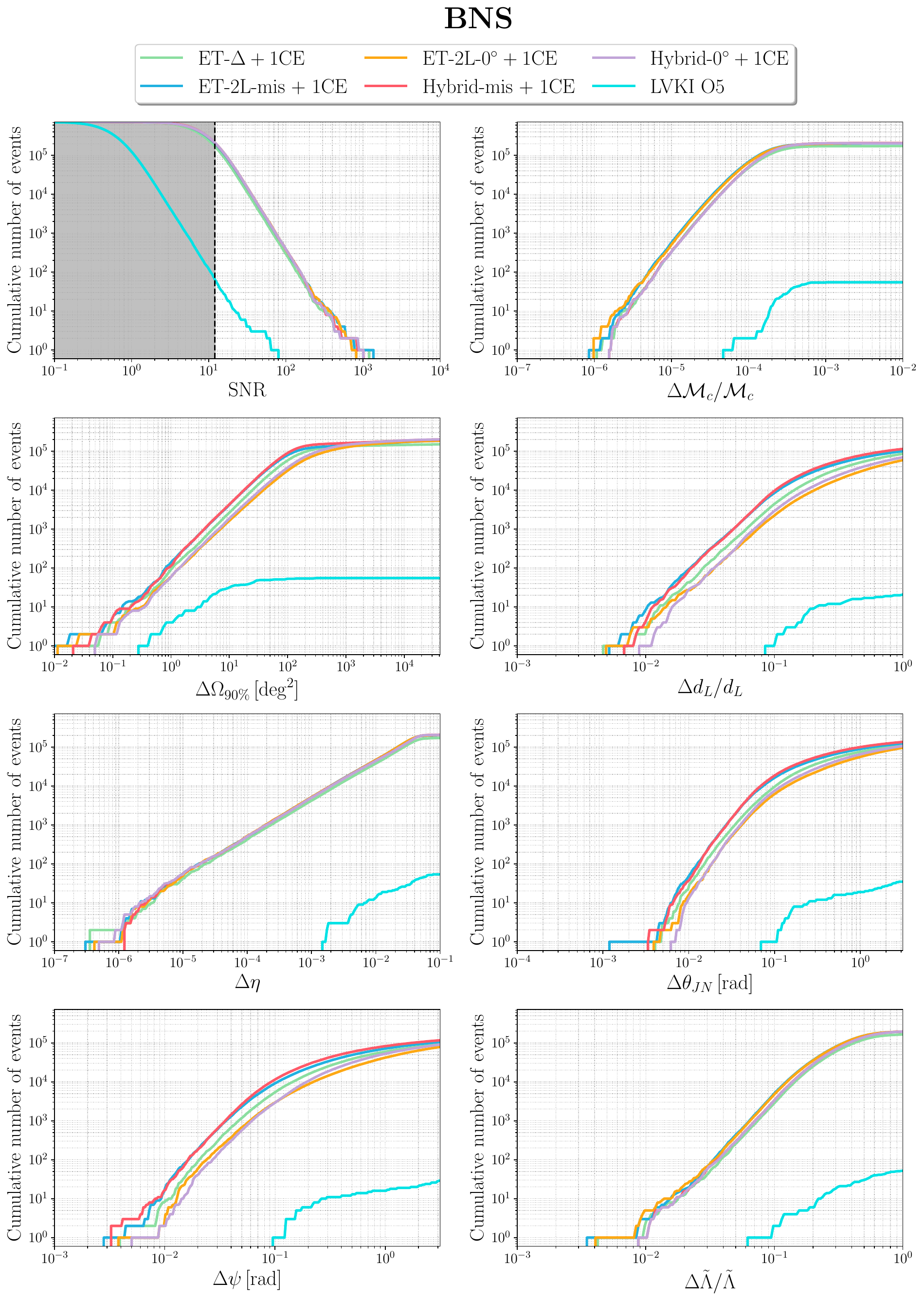}
    \caption{Cumulative distributions of the number of detections per year, for the SNRs and for the error on the parameters, for BNS signals,  for the various detector geometries considered, including a single 40-km CE detector in the US.}
    \label{fig:PE_BNS_wCE}
\end{figure}

\begin{figure}[!tbp]
    \centering
    \begin{tabular}{c c c}
       \includegraphics[width=.31\textwidth]{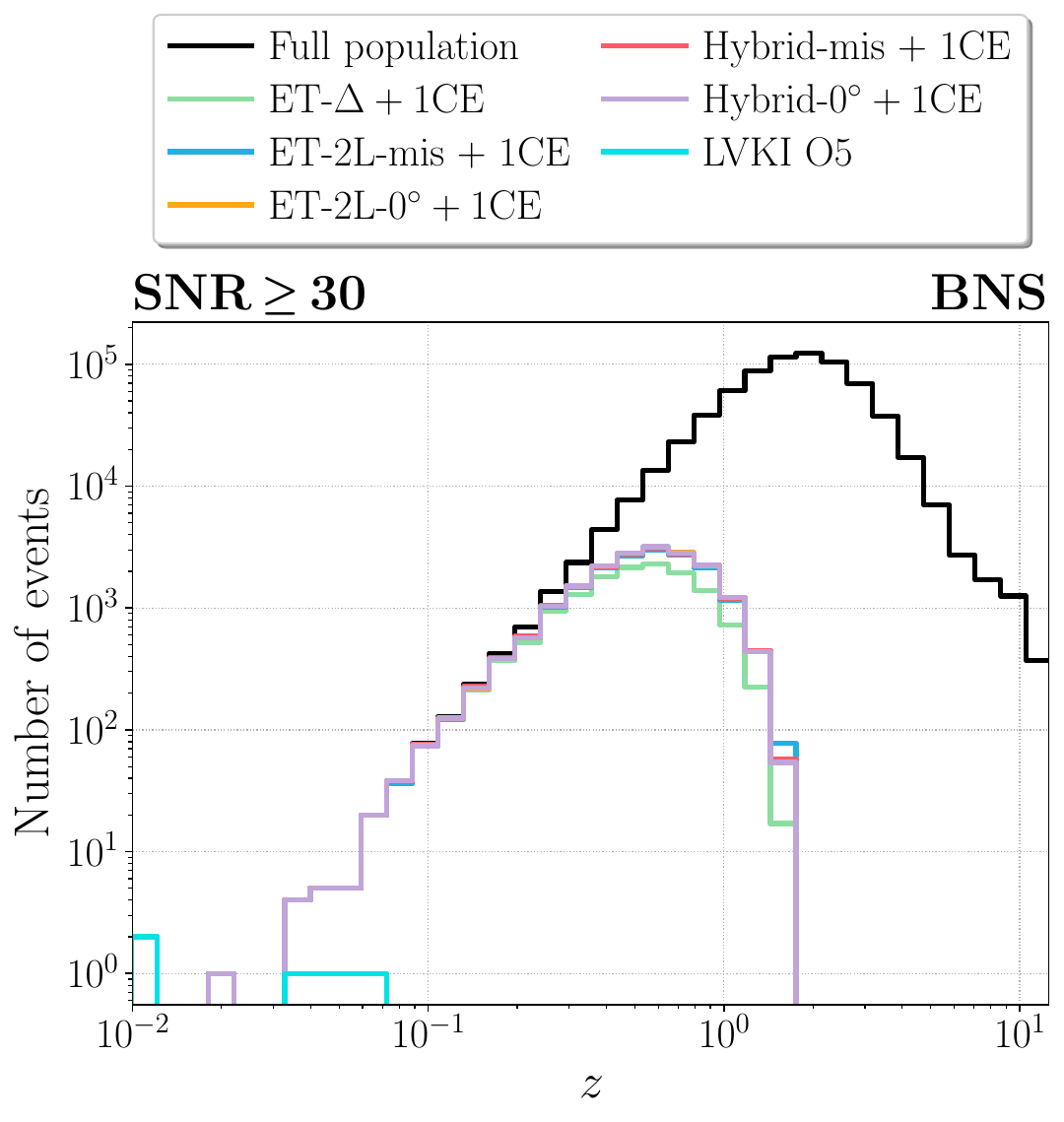}  & \includegraphics[width=.31\textwidth]{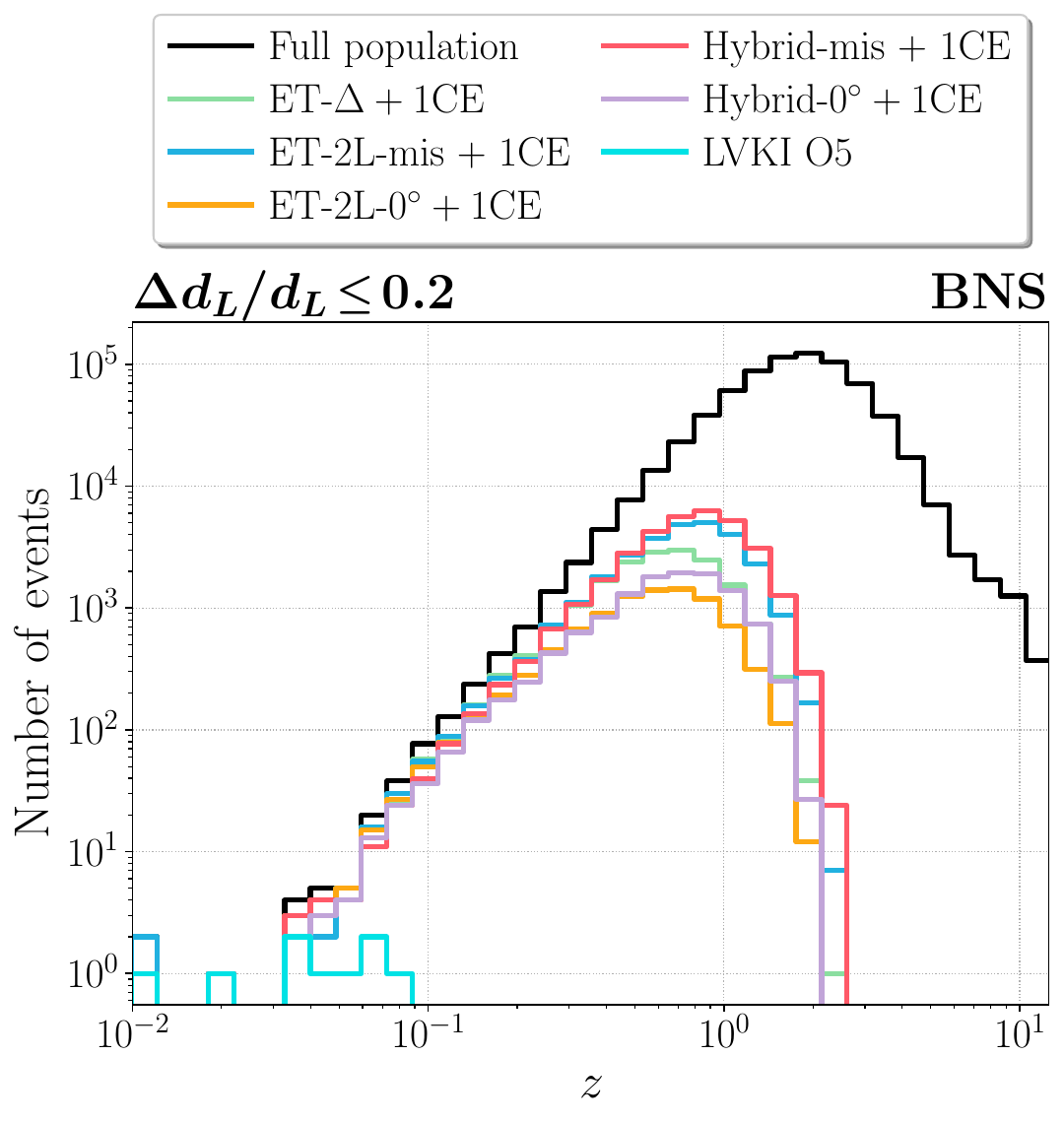} & \includegraphics[width=.31\textwidth]{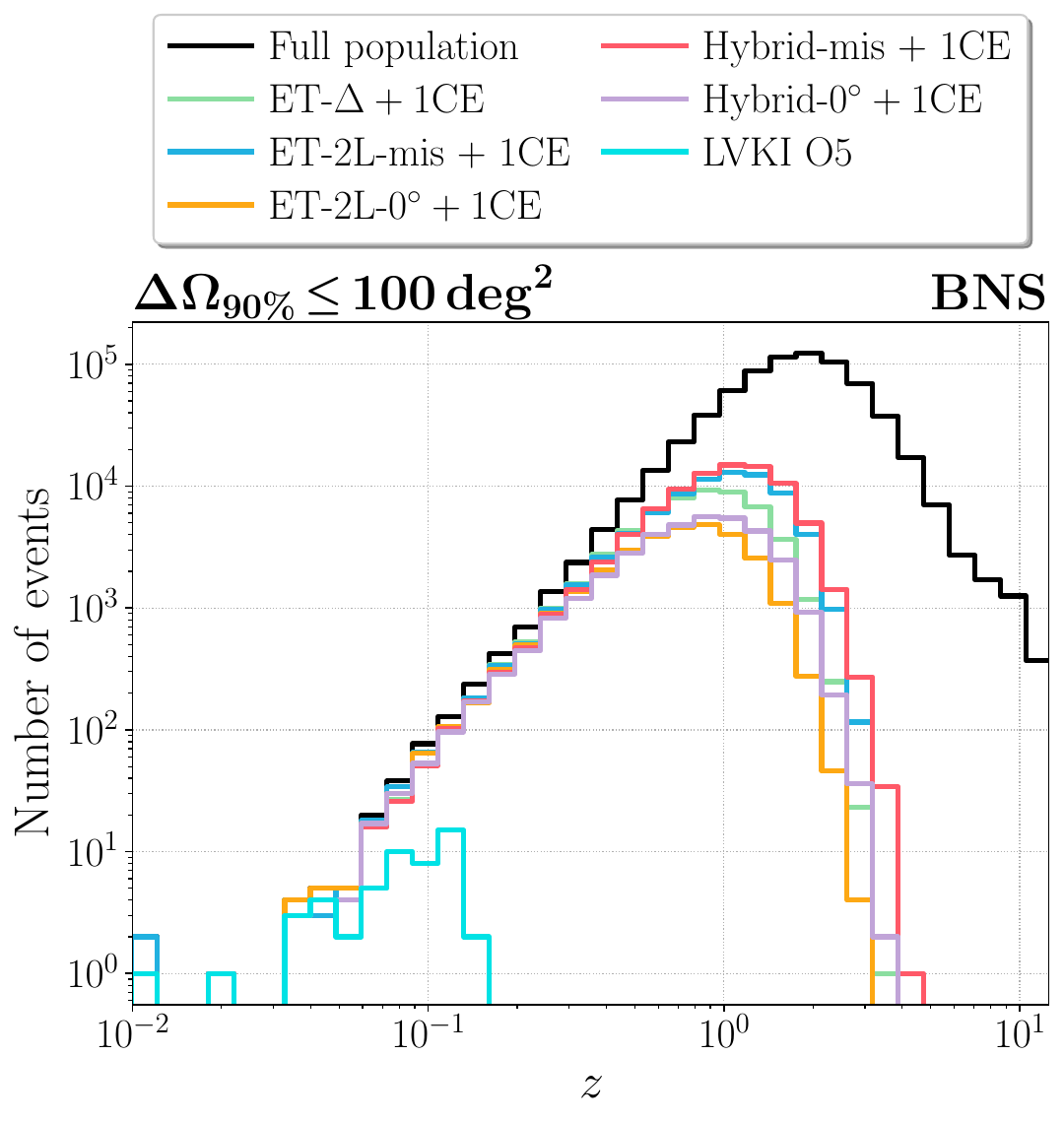} 
    \end{tabular}
    \caption{Redshift distribution of BNSs detected with ${\rm SNR} \geq 30$ (left column), relative error on the luminosity distance $\Delta d_L/d_L \leq 0.2$ (central column), or sky location $\Delta\Omega_{90\%} \leq 100~{\rm deg}^2$ (right column) for the various detector geometries considered, including a single 40-km CE detector in the US.}
    \label{fig:histz_BNS_wCE}
\end{figure}

\begin{table}[!tbp]
    \centering
    \begin{tabular}{!{\vrule width 1pt}m{3.1cm}|S[table-format=4.0]|S[table-format=5.0]|S[table-format=4.0]|S[table-format=4.0]!{\vrule width 1pt}}
        \toprule
        \midrule
        \multicolumn{5}{!{\vrule width 1pt}c!{\vrule width 1pt}}{\textbf{BNS}}\\
        \midrule
        \hfil\multirow{3}{*}{\shortstack[c]{Detector\\configuration}}\hfill & \multicolumn{4}{c!{\vrule width 1pt}}{Detections with} \\
        & \multicolumn{2}{c|}{$\Delta\Omega_{90\%}\leq$} & \multicolumn{2}{c!{\vrule width 1pt}}{$\Delta d_L/d_L \leq$}\\
        \cmidrule(lr){2-5}
        & \multicolumn{1}{c|}{$10~{\rm deg}^2$} & \multicolumn{1}{c|}{$100~{\rm deg}^2$}  & \multicolumn{1}{c|}{$5\times10^{-2}$} & \multicolumn{1}{c!{\vrule width 1pt}}{$10^{-1}$}\\
        \midrule
        \textrm{ET-}$\Delta + {\rm 1CE}$ & 2427 & 54994 & 535 & 4100 \\
        \textrm{ET-2L-mis + 1CE} & 3838 & 75828 & 1040 & 7949 \\
        \textrm{ET-2L-}$0^\circ + {\rm 1CE}$ & 1515 & 29821 & 288 & 2079 \\
        \textrm{Hybrid-mis + 1CE} & 3932 & 85140 & 1043 & 8961 \\
        \textrm{Hybrid-}$0^\circ + {\rm 1CE}$ & 1704 & 35608 & 294 & 2710 \\
        \midrule
        \bottomrule
    \end{tabular}
    \caption{Number of detected BNS sources at the considered networks including a single 40-km CE detector in the US, with different cuts on the sky localization and relative error on the luminosity distance.}
    \label{tab:BNS_numbers_loc_wCE}
\end{table}

\clearpage

\subsection{Pre-merger alerts}\label{sect:premerger}

The low-frequency sensitivity is particularly important for detecting inspiralling binaries early enough, so to be able to send alerts to electromagnetic observatories. From this point of view, two aspects are important. One is how the  SNR accumulates in the different  detector networks, allowing for an early detection.\footnote{Recall that, by SNR, we always mean the SNR of the whole detector network, obtained combining in quadrature the SNRs of the individual detectors.}
The second is how well one can give the angular localization to electromagnetic observatories, sufficiently early before the merger takes place.

\autoref{fig:premerger} shows how the SNR would accumulate in the various 3G  networks  considered, for an event with the characteristics of GW170817, as the  signal  sweeps up in frequency.  The left panel shows the result for
the European networks, while the right panel gives the result when the 40-km CE in the US is added. The lower horizontal axis gives the frequency of the signal, while the upper horizontal axis gives the corresponding time to merger. For this event, which was so close in distance, all network configurations considered reach very rapidly a large SNR. A SNR larger than a detection threshold, say set at ${\rm SNR}=12$, is  reached in all case  about 10 hours prior to merger, and 20 minutes before merger all configurations have SNRs of several hundreds. 

It is interesting to see, in particular, how the SNR  for the Hybrid configurations raises, in the left panel, as the signal sweeps up in frequency. 
The file containing the official ASD of CE starts from a frequency $f= 5$~Hz. Below this frequency, we can assume that CE is essentially blind. In this regime, in the Hybrid network, only the L-shaped detector with the ASD of ET contributes to the accumulation of the SNR. At $f=5$~Hz, the surface detector with the ASD of CE kicks in, and the  red and violet curves in the left panel suddenly change slope. The effect is even more visible in the right panel, where all European networks are taken together with the 40-km CE in the US, that again kicks in at 5~Hz.

Overall, for this specific event, all the networks considered in the left panel have rather similar performances among them, and this is even more the case for the configurations in the  panel when adding CE.   It is quite interesting to see that the accumulation of the SNR in the early part of the pre-merger phase is  similar for the various network configurations, despite the fact that in the Hybrid configurations only one detector can access the low-frequency regime, while in ET-2L-mis there are two detectors that can access it, and in the 10-km triangle all three interferometers can access it. Basically, this is due to the fact that the accumulation of the SNR is dominated by the most sensitive detector in a network.

Of course, GW170817 was a very special event also because of its close distance from us, $d_L\simeq 40$~Mpc. The SNR, however, scales as $1/d_L$, so the result for a source with the same characteristics as GW170817 but at larger distance can be obtained just by rescaling the results in \autoref{fig:premerger}. So, for a source with the same detector-frame masses, at a distance 10 times larger, i.e. about 400~Mpc, at $10^3$ seconds before merger, all these configurations would still reach a SNR between 20 and 60, by itself very well sufficient for reliable detection.

\begin{figure}[tbp]
    \centering
    \begin{tabular}{c c}
    \includegraphics[height=.38\textwidth]{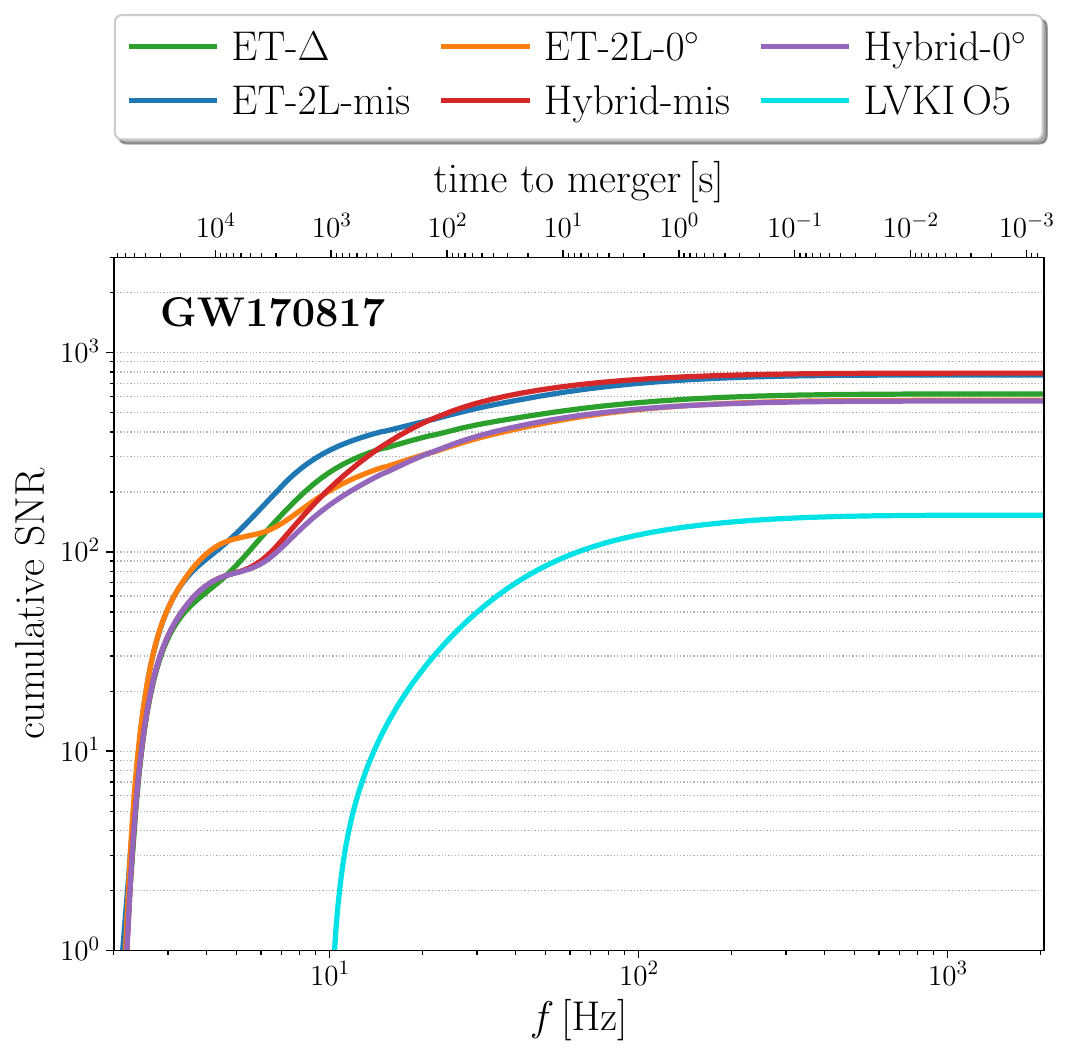} &
       \includegraphics[height=.38\textwidth]{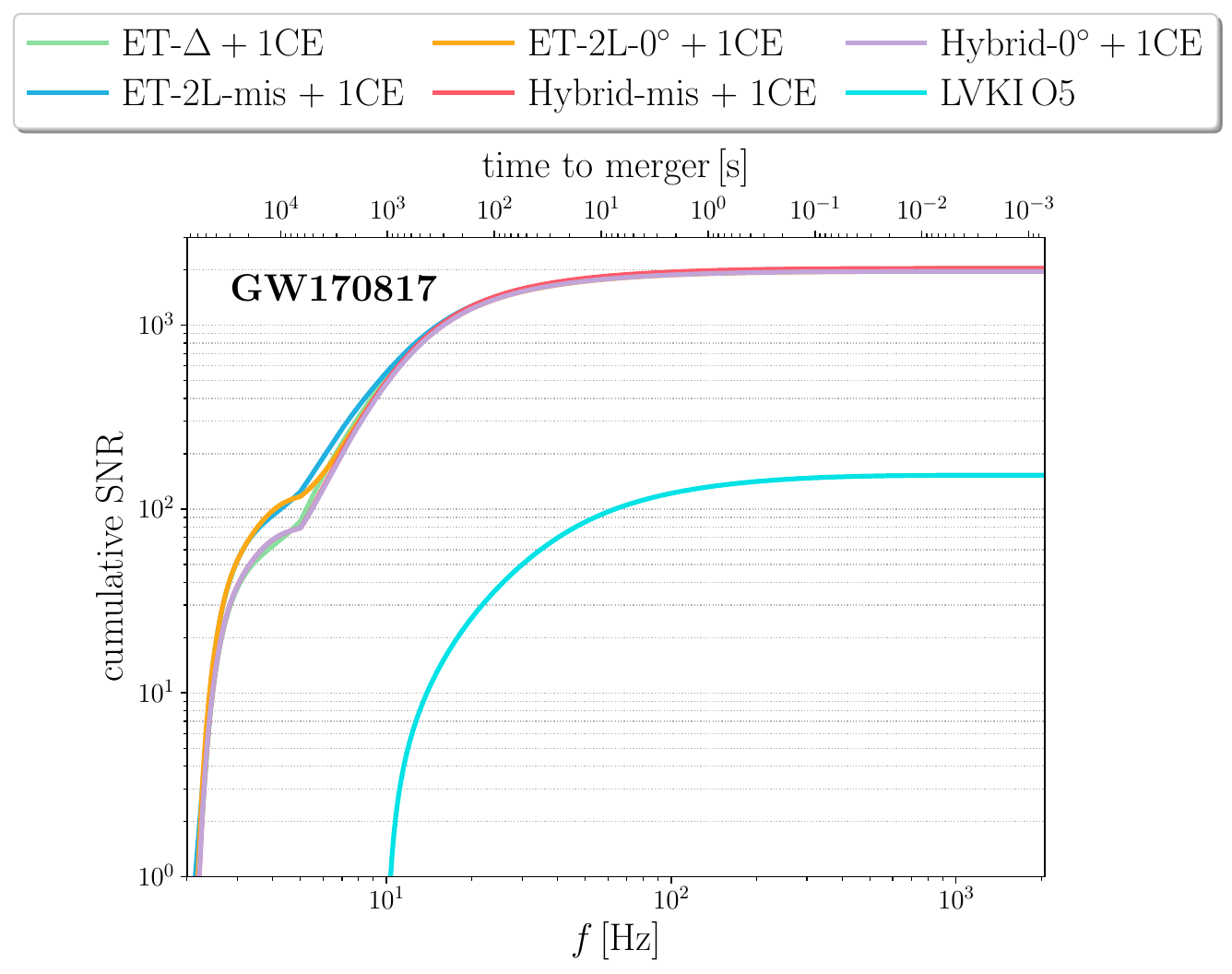}   
    \end{tabular}
    \caption{Accumulation of the SNR in the various 3G network configurations, for a source with the properties of the BNS GW170817, as the signal sweeps up in frequency without (\emph{left panel}) and with (\emph{right panel}) a single 40-km CE detector in the US. The upper horizontal scale gives the corresponding time to merger. Observe that, for this system, the merger takes place at about 2~kHz, see e.g. Figure~2 of \cite{Iacovelli:2022bbs}.}
    \label{fig:premerger}
\end{figure}

Detecting a signal prior to merger is, however, only  part of the story. For multi-messenger studies, we also need to get a good angular resolution before merger, in order to give  to  electromagnetic observatories at least some localization  information, and sufficiently early.
\autoref{tab:premerger_time_angular_res_noCE} shows the
number of BNSs detected with $\rm SNR \geq 12$ and different cuts on the sky localization, at 30~min, 10~min and 1~min prior to merger, for the various European-only configurations studied in this work.\footnote{As in the results of the previous subsections, all these figures are for one year of data taking, with the duty cycles given at the end of \autoref{sect:methods}.}\textsuperscript{,}\footnote{For 
ET-2L-mis  and ET-$\Delta$, these results can be compared to those in Table~3 of \cite{Branchesi:2023mws}. Note however that, there, was used a cut
${\rm SNR}\geq 8$, while here we are using ${\rm SNR}\geq 12$. The rational for a lower threshold, ${\rm SNR}\geq8$, in multi-messenger studies, is that combining the GW detection with the electromagnetic detection increases the statistical significance, so the GW detection can be performed at lower SNR. However, if we are interested in providing an alert to electromagnetic observatories, the electromagnetic detection has not yet been obtained, so it can be safer to stick to ${\rm SNR}\geq 12$, to avoid giving too many false alerts. In any case, the events for which one obtains the best pre-merger localization are those with high SNR, so for these events the precise value of the threshold has limited relevance. Using a higher SNR threshold also reduces the technical problems with the inversion of  ill-conditioned Fisher matrices, see footnote~9 of \cite{Branchesi:2023mws}.}
We show separately the BNS detections where the  orbit has a generic inclination, and  those close to face-on, defined as those such that $\Theta \equiv {\rm min}\{\iota, 180^\circ - \iota\}$ is smaller  than $ 15^\circ$, which can result more likely in coincident $\gamma$-ray burst  (GRB) detections.

To illustrate these results, let us begin by discussing the events with arbitrary orbit inclination, and consider first the detections with angular resolution $\Delta\Omega_{90\%}$  smaller than $10~{\rm deg}^2$. The events that are  well localized have a special importance for multi-messenger astronomy because, unless some telescope devoted just to the follow-up of GW detections will be developed, realistically only a small fraction of the alerts associated to  BNS signals will lead to a follow-up campaign from electromagnetic telescopes, so the best-localized events will be the natural candidates for such a follow-up. In the European-only setting, none of the configurations studied can provide such a localization 30~min or 10~min before merger. However, such a localization will be obtained for a few events 1~min before  merger. In this case (within the statistical fluctuations associated to such small numbers) similar  performances are obtained from
ET-2L-mis, ET-$\Delta$ and Hybrid-mis, with $\{5,4,4\}$ events, respectively; these numbers  become $\{139,79,76\}$ requiring  $\Delta\Omega_{90\%}\leq 100~{\rm deg}^2$ at 1~min before merger; and
$\{53,28,16\}$ for $\Delta\Omega_{90\%}\leq100~{\rm deg}^2$ at 10~min
before merger.

Observe that, for  $\Delta\Omega_{90\%}\leq100~{\rm deg}^2$, 10 min before merger ET-$\Delta$ has more events than  Hybrid-mis, 28 against 16, while 1~min before merger the performances are equivalent, with 79 events against 76. This is due to the fact that the surface detector, in the Hybrid configuration, kicks in later, but then catches up. The same pattern repeats requiring
$\Delta\Omega_{90\%}\leq1000~{\rm deg}^2$, where 10 min before merger
ET-$\Delta$ has localized 125 events, to be compared with 74 for  Hybrid-mis; 1~min before merger, these number raise to 372 and 360, respectively, therefore becoming again very similar.
In any case, even at 10~min before merger, the relative performances of these two networks differ by less than a factor of 2.\footnote{The result for ET-$\Delta$  can also be compared to those in  ref.~\cite{Nitz:2021pbr}, which obtains  larger numbers: 6 and 2 events/yr with
$\Delta\Omega_{90\%}\leq 10\, {\rm deg}^2$
at 5 and 30 min before merger, respectively; by comparison,  for $\Delta\Omega_{90\%}\leq 10\, {\rm deg}^2$ we have 4 events/yr at 1 min and 0 at 30~min. However, the comparison must  take into account some different assumptions. In particular, ref.~\cite{Nitz:2021pbr} uses a duty cycle of 100\%, while we  assume an uncorrelated 85\% duty cycle in  each of the three detectors composing the triangle. Furthermore, ref.~\cite{Nitz:2021pbr} uses the ET-D sensitivity curve, while we use a more up-to-date sensitivity curve, already used in \cite{Branchesi:2023mws}, which is less good at low frequencies (see  Figure~2, left panel,  of  \cite{Branchesi:2023mws}). Other technical differences are that ref.~\cite{Nitz:2021pbr} uses full \texttt{PyCBC Inference}, while we restrict to Fisher matrices. The population model used is also not the same.}

Let us focus next on the events close to face-on, $\Theta \leq  15^\circ$, for which the probability of detecting an associated GRB is higher. None of the network configurations considered, in one year of data taking, can detect such events before merger with a localization accuracy $\Delta\Omega_{90\%}\leq 10~{\rm deg}^2$, and also for $\Delta\Omega_{90\%}\leq 100\, {\rm deg}^2$ there is just a handful of events in each configuration (so that the results are also more sensitive to statistical fluctuations associated to our sample of the catalog, and to random down time associated to the duty cycle).
If, in order to deal with statistically more significant numbers, 
we rather set the cut on angular resolution at $\Delta\Omega_{90\%}\leq 1000\, {\rm deg}^2$, at 10~min before merger the number of BNSs detections (again, in 1 yr of data and with our assumption for the duty cycle) with  $\Theta \leq  15^\circ$ that we  read from 
\autoref{tab:premerger_time_angular_res_noCE} are as follows: ET-2L-mis, 8 events;   ET-$\Delta$ and ET-2L-$0^{\circ}$,  4 events;  
Hybrid-mis and Hybrid-$0^{\circ}$,  2 events.

\begin{table}[!tb]
    \setlength\arrayrulewidth{.5pt}
    \renewcommand{\arraystretch}{1.07}
    \centering
    \begin{tabular}{!{\vrule width 1pt}m{3.1cm}|c|c|S[table-format=1.0]|S[table-format=3.0]|S[table-format=3.0]|S[table-format=5.0]!{\vrule width 1pt}}
        \toprule
        \midrule
        \multicolumn{7}{!{\vrule width 1pt}c!{\vrule width 1pt}}{\textbf{BNS}}\\
        \midrule
        \hfil\multirow{2}{*}{\shortstack[c]{Detector\\configuration}}\hfill & \hfil\multirow{2}{*}{\shortstack[c]{Time before\\merger}}\hfill & \multirow{2}{*}{Orientation} & \multicolumn{4}{c!{\vrule width 1pt}}{Detections with $\Delta\Omega_{90\%}\leq$} \\
        \cmidrule(lr){4-7}
        & & & \multicolumn{1}{c|}{$10~{\rm deg}^2$} & \multicolumn{1}{c|}{$100~{\rm deg}^2$}  & \multicolumn{1}{c|}{$1000~{\rm deg}^2$} & \multicolumn{1}{c!{\vrule width 1pt}}{all sky}\\
        \midrule
        \multirow{6}{*}{\textrm{ET-}$\Delta$} & \multirow{2}{*}{30~min} & All $\Theta$ & 0 & 8 & 39 & 345 \\
        & & $\Theta\leq 15^\circ$ & 0 & 0 & 2 & 31 \\
        \cline{2-7}
        & \multirow{2}{*}{10~min} & All $\Theta$ & 0 & 28 & 125 & 1544 \\
        & & $\Theta\leq 15^\circ$ & 0 & 2 & 4 & 153 \\
        \cline{2-7}
        & \multirow{2}{*}{1~min} & All $\Theta$ & 4 & 79 & 372 & 7599 \\
        & & $\Theta\leq 15^\circ$ & 0 & 3 & 9 & 767 \\
        \midrule
        \multirow{6}{*}{\textrm{ET-2L-mis}} & \multirow{2}{*}{30~min} & All $\Theta$ & 0 & 19 & 83 & 807 \\
        & & $\Theta\leq 15^\circ$ & 0 & 0 & 3 & 78 \\
        \cline{2-7}
        & \multirow{2}{*}{10~min} & All $\Theta$ & 0 & 53 & 288 & 3439 \\
        & & $\Theta\leq 15^\circ$ & 0 & 3 & 8 & 308 \\
        \cline{2-7}
        & \multirow{2}{*}{1~min} & All $\Theta$ & 5 & 139 & 697 & 14765 \\
        & & $\Theta\leq 15^\circ$ & 0 & 5 & 22 & 1562 \\
        \midrule
        \multirow{6}{*}{\textrm{ET-2L-}$0^\circ$} & \multirow{2}{*}{30~min} & All $\Theta$ & 0 & 4 & 50 & 818 \\
        & & $\Theta\leq 15^\circ$ & 0 & 0 & 2 & 78 \\
        \cline{2-7}
        & \multirow{2}{*}{10~min} & All $\Theta$ & 0 & 11 & 117 & 3458 \\
        & & $\Theta\leq 15^\circ$ & 0 & 1 & 4 & 311 \\
        \cline{2-7}
        & \multirow{2}{*}{1~min} & All $\Theta$ & 2 & 25 & 323 & 14991 \\
        & & $\Theta\leq 15^\circ$ & 0 & 0 & 13 & 1593 \\
        \midrule
        \multirow{6}{*}{\textrm{Hybrid-mis}} & \multirow{2}{*}{30~min} & All $\Theta$ & 0 & 2 & 16 & 354 \\
        & & $\Theta\leq 15^\circ$ & 0 & 0 & 0 & 28 \\
        \cline{2-7}
        & \multirow{2}{*}{10~min} & All $\Theta$ & 0 & 16 & 74 & 1686 \\
        & & $\Theta\leq 15^\circ$ & 0 & 0 & 2 & 159 \\
        \cline{2-7}
        & \multirow{2}{*}{1~min} & All $\Theta$ & 4 & 76 & 360 & 12075 \\
        & & $\Theta\leq 15^\circ$ & 0 & 3 & 13 & 1185 \\
        \midrule
        \multirow{6}{*}{\textrm{Hybrid-}$0^\circ$} & \multirow{2}{*}{30~min} & All $\Theta$ & 0 & 0 & 8 & 354 \\
        & & $\Theta\leq 15^\circ$ & 0 & 0 & 0 & 27 \\
        \cline{2-7}
        & \multirow{2}{*}{10~min} & All $\Theta$ & 0 & 4 & 36 & 1704 \\
        & & $\Theta\leq 15^\circ$ & 0 & 0 & 2 & 162 \\
        \cline{2-7}
        & \multirow{2}{*}{1~min} & All $\Theta$ & 1 & 12 & 167 & 12219 \\
        & & $\Theta\leq 15^\circ$ & 0 & 0 & 4 & 1190 \\
        \midrule
        \bottomrule
    \end{tabular}
    \caption{Number of BNS detected with $\rm SNR \geq 12$ and different cuts on the sky localization at 30~min, 10~min and 1~min prior to merger for the various European-only configurations studied in this work. We further report the numbers both for all the sources in the catalog and the sub-sample of sources having an angle $\Theta \leq 15^\circ$, which can result more likely in coincident GRB detections.}
    \label{tab:premerger_time_angular_res_noCE}
\end{table}

\begin{table}[!tbp]
    \setlength\arrayrulewidth{.5pt}
    \renewcommand{\arraystretch}{1.07}
    \centering
    \begin{tabular}{!{\vrule width 1pt}m{3.1cm}|c|c|S[table-format=3.0]|S[table-format=4.0]|S[table-format=5.0]|S[table-format=5.0]!{\vrule width 1pt}}
        \toprule
        \midrule
        \multicolumn{7}{!{\vrule width 1pt}c!{\vrule width 1pt}}{\textbf{BNS}}\\
        \midrule
        \hfil\multirow{2}{*}{\shortstack[c]{Detector\\configuration}}\hfill & \hfil\multirow{2}{*}{\shortstack[c]{Time before\\merger}}\hfill & \multirow{2}{*}{Orientation} & \multicolumn{4}{c!{\vrule width 1pt}}{Detections with $\Delta\Omega_{90\%}\leq$} \\
        \cmidrule(lr){4-7}
        & & & \multicolumn{1}{c|}{$10~{\rm deg}^2$} & \multicolumn{1}{c|}{$100~{\rm deg}^2$}  & \multicolumn{1}{c|}{$1000~{\rm deg}^2$} & \multicolumn{1}{c!{\vrule width 1pt}}{all sky}\\
        \midrule
        \multirow{6}{*}{\textrm{ET-}$\Delta + {\rm 1CE}$} & \multirow{2}{*}{30~min} & All $\Theta$ & 1 & 25 & 229 & 418 \\
        & & $\Theta\leq 15^\circ$ & 0 & 0 & 5 & 37 \\
        \cline{2-7}
        & \multirow{2}{*}{10~min} & All $\Theta$ & 11 & 234 & 1888 & 2493 \\
        & & $\Theta\leq 15^\circ$ & 0 & 9 & 64 & 233 \\
        \cline{2-7}
        & \multirow{2}{*}{1~min} & All $\Theta$ & 77 & 2140 & 22906 & 33042 \\
        & & $\Theta\leq 15^\circ$ & 3 & 76 & 790 & 3136 \\
        \midrule
        \multirow{6}{*}{\textrm{ET-2L-mis + 1CE}} & \multirow{2}{*}{30~min} & All $\Theta$ & 1 & 41 & 307 & 875 \\
        & & $\Theta\leq 15^\circ$ & 0 & 0 & 9 & 82 \\
        \cline{2-7}
        & \multirow{2}{*}{10~min} & All $\Theta$ & 10 & 363 & 2521 & 4542 \\
        & & $\Theta\leq 15^\circ$ & 0 & 10 & 79 & 417 \\
        \cline{2-7}
        & \multirow{2}{*}{1~min} & All $\Theta$ & 101 & 2824 & 27880 & 42804 \\
        & & $\Theta\leq 15^\circ$ & 7 & 82 & 909 & 4057 \\
        \midrule
        \multirow{6}{*}{\textrm{ET-2L-}$0^\circ + {\rm 1CE}$} & \multirow{2}{*}{30~min} & All $\Theta$ & 1 & 19 & 195 & 885 \\
        & & $\Theta\leq 15^\circ$ & 0 & 0 & 3 & 86 \\
        \cline{2-7}
        & \multirow{2}{*}{10~min} & All $\Theta$ & 3 & 100 & 1080 & 4540 \\
        & & $\Theta\leq 15^\circ$ & 0 & 1 & 41 & 426 \\
        \cline{2-7}
        & \multirow{2}{*}{1~min} & All $\Theta$ & 29 & 627 & 6844 & 42977 \\
        & & $\Theta\leq 15^\circ$ & 0 & 21 & 220 & 4063 \\
        \midrule
        \multirow{6}{*}{\textrm{Hybrid-mis + 1CE}} & \multirow{2}{*}{30~min} & All $\Theta$ & 0 & 10 & 149 & 416 \\
        & & $\Theta\leq 15^\circ$ & 0 & 0 & 4 & 36 \\
        \cline{2-7}
        & \multirow{2}{*}{10~min} & All $\Theta$ & 7 & 159 & 1577 & 2655 \\
        & & $\Theta\leq 15^\circ$ & 1 & 6 & 48 & 254 \\
        \cline{2-7}
        & \multirow{2}{*}{1~min} & All $\Theta$ & 103 & 2730 & 25179 & 39062 \\
        & & $\Theta\leq 15^\circ$ & 5 & 85 & 867 & 3591 \\
        \midrule
        \multirow{6}{*}{\textrm{Hybrid-}$0^\circ + {\rm 1CE}$} & \multirow{2}{*}{30~min} & All $\Theta$ & 0 & 7 & 101 & 417 \\
        & & $\Theta\leq 15^\circ$ & 0 & 0 & 0 & 36 \\
        \cline{2-7}
        & \multirow{2}{*}{10~min} & All $\Theta$ & 2 & 56 & 614 & 2681 \\
        & & $\Theta\leq 15^\circ$ & 0 & 5 & 29 & 259 \\
        \cline{2-7}
        & \multirow{2}{*}{1~min} & All $\Theta$ & 24 & 458 & 5177 & 39221 \\
        & & $\Theta\leq 15^\circ$ & 0 & 22 & 196 & 3605 \\
        \midrule
        \bottomrule
    \end{tabular}
    \caption{As in \autoref{tab:premerger_time_angular_res_noCE}, for the networks including a 40-km CE in the US.}
    \label{tab:premerger_time_angular_res_wCE}
\end{table}

We now consider the networks obtained adding also  a 40-km CE in the US. The results are shown in \autoref{tab:premerger_time_angular_res_wCE}. Now, even for $\Delta\Omega_{90\%}\leq 10\, {\rm deg}^2$, at least for generic  orbit inclinations, there are events  localized to such accuracy even 30 or 10 min before mergers. For instance, for (ET-$\Delta$ + 1CE) we get $\{1,11,77\}$ events localized to
$\Delta\Omega_{90\%}\leq 10\, {\rm deg}^2$
at, respectively, 30, 10 and 1 min before merger. For (ET-2L-mis + 1CE) we get $\{1,10,101\}$ and, for
(Hybrid-mis + 1CE),  $\{0,7,103\}$.
Requiring instead a less stringent angular localization, $\Delta\Omega_{90\%}\leq 100\, {\rm deg}^2$, these numbers become, respectively:
$\{25,234,2140\}$ for (ET-$\Delta$ + 1CE);
$\{41,363,2824\}$ for  (ET-2L-mis + 1CE); and
$\{10,159,2730\}$ for (Hybrid-mis + 1CE).

Overall, the conclusion is that the performances of these configurations for pre-merger alert are very comparable for the detections at about 10~min to merger and later, while, requiring a 30~min pre-alert time,
(Hybrid-mis + 1CE) is less performant, by about a factor of 2.5 with respect to (ET-$\Delta$ + 1CE) and a factor of 4 with respect to (ET-2L-mis + 1CE).

\subsection{Stochastic backgrounds}

Finally, we consider the sensitivity to stochastic GW backgrounds (SGWBs) of these configurations. 
Isotropic and Gaussian stochastic backgrounds of GWs  are characterized  by their energy density  per unit logarithmic interval of frequency, 
${\rm d}\rho_{\rm gw}/{\rm d}\log f$~\cite{Allen:1997ad,Maggiore:1999vm}.
To have a dimensionless quantity, this is normalized 
 to the critical density for closing the Universe, $\rho_c=3H_0^2/(8\pi G)$,
defining $\Ogw(f)=(1/\rho_c) {\rm d}\rho_{\rm gw}/{\rm d}\log f$. Writing $H_0=h \times (100~{\rm km\,s}^{-1}{\rm Mpc}^{-1})$, it is then convenient to use  $h^2\,\Ogw(f)$  to characterize the stochastic background, to get rid of the uncertainty in $h$.

\begin{figure}[tbp]
    \centering
    \begin{tabular}{c c}
       \includegraphics[height=.4\textwidth]{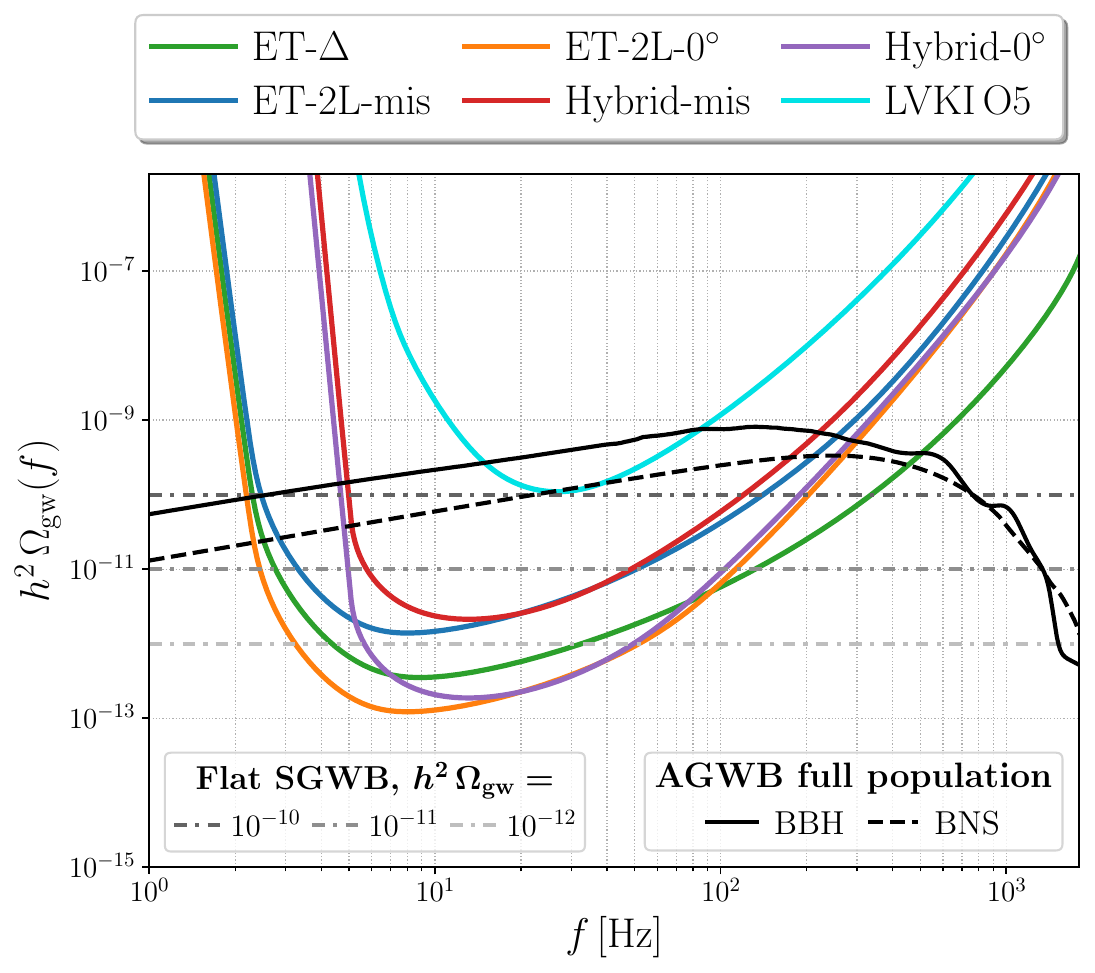}  & \includegraphics[height=.4\textwidth]{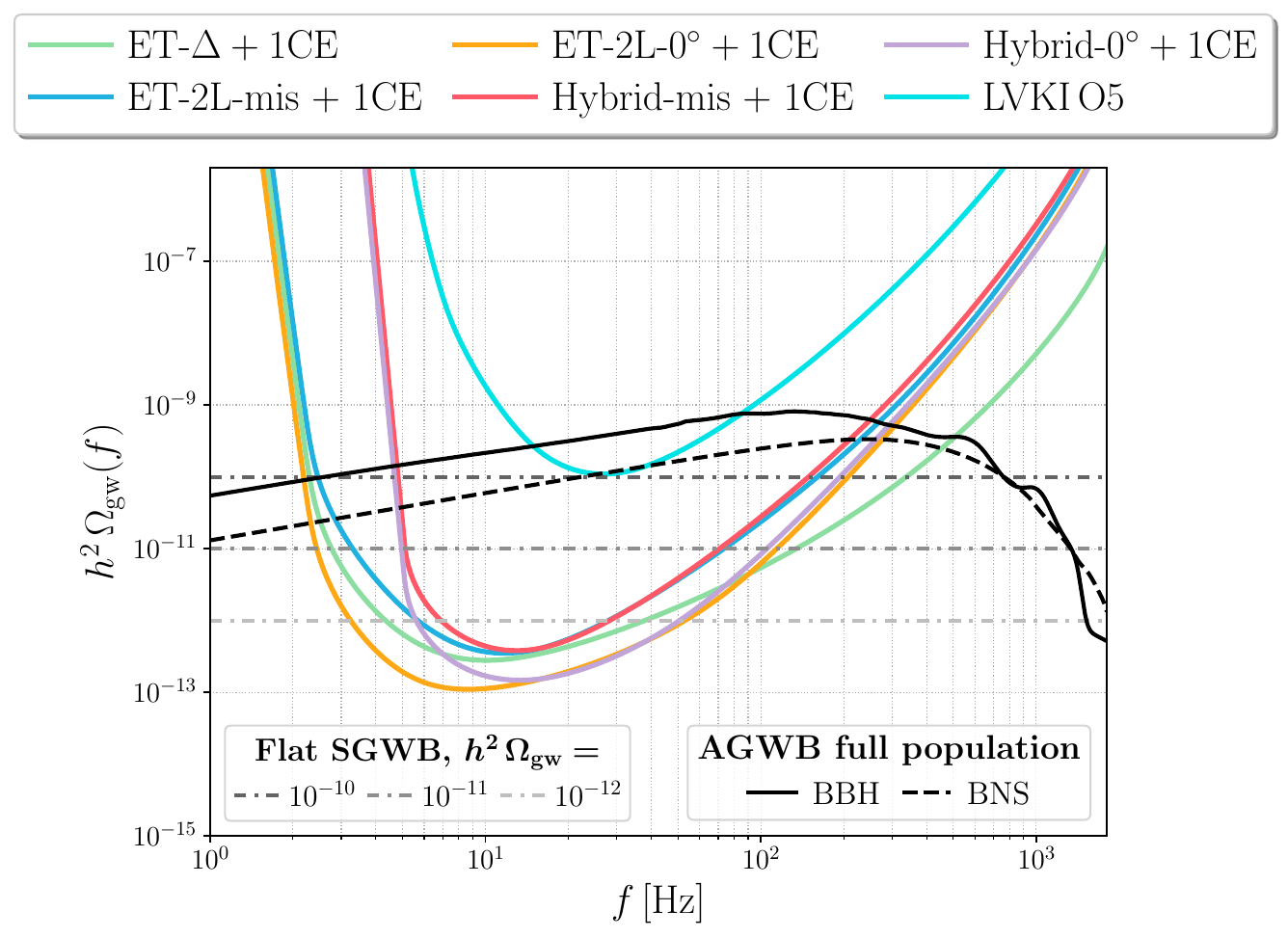} 
    \end{tabular}
    \caption{Power-law integrated sensitivity curves for $h^2\,\Ogw(f)$  for the various configurations studied without (\emph{left panel}) and with (\emph{right panel}) a single 40-km CE detector in the US. These are computed assuming an observational time $T_{\rm obs}=1~{\rm yr}$ and an SNR threshold of 1. We further report some SGWB examples, namely a flat-in-frequency background with different amplitudes and the AGWB generated by our 1~yr catalogs of BBHs and BNSs.}
    \label{fig:PLSs}
\end{figure}

The sensitivity to stochastic background can  be expressed in terms of the power-law integrated sensitivity (PLS) introduced in \cite{Thrane:2013oya} (see also App.~A of ~\cite{Branchesi:2023mws} for conventions and definitions).
\autoref{fig:PLSs} shows the  PLS of the various network configurations studied, expressed in terms of $h^2\,\Ogw(f)$. We see that, at high frequencies (above about 100~Hz),  the 10-km triangle provides the best sensitivity. This is due to the fact that, for colocated detectors, there is almost no  suppression from the  overlap reduction function which instead, for two detectors at a distance $d$, suppresses the correlation when $f d/c\gg 1$ (see e.g. \cite{Maggiore:1999vm}).\footnote{Of course, the reverse of the coin is that, for colocated detectors, there will be a larger correlated noise (particularly at low frequencies), that would simulate a stochastic GW background, see 
\cite{Janssens:2021cta,Janssens:2024jln} and Section~5.4.1 of \cite{Branchesi:2023mws}. The PLS in \autoref{fig:PLSs}, as well as the corresponding SNR for various signals reported below, are computed assuming that the noise in the different detectors are uncorrelated.} We also observe that, going toward
low-frequencies, the Hybrid configurations hit a sensitivity wall earlier, compared to the ET-$\Delta$ or ET-2L. This is due to the fact that stochastic backgrounds are measured by correlating the output of two detectors so, in a two-detector correlation, the result is dominated by the less sensitive detector, and below $\ssim 5$~Hz, where the surface detector with the ASD of CE becomes blind, the whole Hybrid networks also become blind to stochastic backgrounds.\footnote{Note that this is different from what we found in \autoref{sect:premerger} for the accumulation of the SNR in the pre-merger phase. Indeed, in that case the result was dominated by the most sensitive detector in the low-frequency region, so having just one  detector sensitive to the low frequencies was sufficient.}

\begin{table}[tbp]
    \centering
    \begin{tabular}{!{\vrule width 1pt}m{3.1cm}|S[table-format=3.1]|S[table-format=2.1]|S[table-format=1.1]|S[table-format=4.1]|S[table-format=3.1]!{\vrule width 1pt}}
        \toprule
        \midrule
        \hfil\multirow{3}{*}{\shortstack[c]{Detector\\configuration}}\hfill & \multicolumn{5}{c!{\vrule width 1pt}}{SNR for} \\
        & \multicolumn{3}{c|}{flat SGWB, $h^2\,\Omega_{\rm gw}=$} & \multicolumn{2}{c!{\vrule width 1pt}}{AGWB} \\
        \cmidrule(lr){2-4} \cmidrule(lr){5-6}
        & \multicolumn{1}{c|}{$10^{-10}$} & \multicolumn{1}{c|}{$10^{-11}$} & \multicolumn{1}{c|}{$10^{-12}$} & \multicolumn{1}{c|}{all BBH} & \multicolumn{1}{c!{\vrule width 1pt}}{all BNS}\\
        \midrule
        ET-$\Delta$ & 285.2 & 28.5 & 2.9 & 621.6 & 174.5 \\
        ET-2L-mis & 72.1 & 7.2 & 0.7 & 148.6 & 41.3 \\
        ET-2L-$0^\circ$ & 821.4 & 82.1 & 8.2 & 1693.7 & 470.6 \\
        Hybrid-mis & 47.1 & 4.7 & 0.5 & 126.6 & 36.8 \\
        Hybrid-$0^\circ$ & 536.1 & 53.6 & 5.4 & 1439.2 & 417.8 \\
        \midrule
        \bottomrule
    \end{tabular}
    \caption{SNRs at the various European configurations for different stochastic background sources in an observational time $T_{\rm obs}=1~{\rm yr}$. The columns 2, 3 and 4 show the results for a flat-in-frequency SGWB with different amplitudes, the fifth and sixth column for the superposition of all the BBH and BNS signals present in our 1~yr catalogs, respectively.}
    \label{tab:SGWB_numbers_noCE}
\end{table}

\begin{table}[tbp]
    \centering
    \begin{tabular}{!{\vrule width 1pt}m{3.1cm}|S[table-format=3.1]|S[table-format=2.1]|S[table-format=1.1]|S[table-format=4.1]|S[table-format=3.1]!{\vrule width 1pt}}
        \toprule
        \midrule
        \hfil\multirow{3}{*}{\shortstack[c]{Detector\\configuration}}\hfill & \multicolumn{5}{c!{\vrule width 1pt}}{SNR for} \\
        & \multicolumn{3}{c|}{flat SGWB, $h^2\,\Omega_{\rm gw}=$} & \multicolumn{2}{c!{\vrule width 1pt}}{AGWB} \\
        \cmidrule(lr){2-4} \cmidrule(lr){5-6}
        & \multicolumn{1}{c|}{$10^{-10}$} & \multicolumn{1}{c|}{$10^{-11}$} & \multicolumn{1}{c|}{$10^{-12}$} & \multicolumn{1}{c|}{all BBH} & \multicolumn{1}{c!{\vrule width 1pt}}{all BNS}\\
        \midrule
        ET-$\Delta + {\rm 1CE}$ & 358.8 & 35.9 & 3.6 & 828.7 & 234.9 \\
        ET-2L-mis + 1CE & 285.3 & 28.5 & 2.9 & 698.7 & 199.7 \\
        ET-2L-$0^\circ + {\rm 1CE}$ & 909.2 & 90.9 & 9.1 & 1948.3 & 545.3 \\
        Hybrid-mis + 1CE & 263.0 & 26.3 & 2.6 & 677.7 & 195.3 \\
        Hybrid-$0^\circ + {\rm 1CE}$ & 679.4 & 67.9 & 6.8 & 1807.2 & 523.6 \\
        \midrule
        \bottomrule
    \end{tabular}
    \caption{As in \autoref{tab:SGWB_numbers_noCE}, adding to each network a single 40-km CE detector in the US.}
    \label{tab:SGWB_numbers_wCE}
\end{table}

For the same reasons,
adding a CE in the US (right panel) does not improve the sensitivity below 5~Hz, and also has a limited effect at large frequencies, because of the long baseline $d$ between  US and European detectors, and the corresponding suppression due to the overlap reduction function. In general, adding a CE in the US has limited effect in the whole frequency range for 
ET-$\Delta$ and for the Hybrid-$0^{\circ}$ and 
ET-2L-$0^{\circ}$ configurations; this is due to the fact that in any case, because of the suppression by the overlap reduction function,  the correlation between a detector in the US and detectors in Europe is less good, compared to what can be achieved for stochastic backgrounds by well-correlated 3G detectors in Europe.
However, between about 10~Hz and a few hundred Hz, adding a CE in the US improves significantly the performances of the 
Hybrid-mis and ET-2L-mis networks, where the relative orientation between the two European L-shaped detectors  suppresses the sensitivity to stochastic backgrounds, and therefore adding one more detector, not parallel, has a significant effect.

As we see from \autoref{fig:PLSs}, the relative  performances of some configurations depend on the frequency range. For instance,  comparing the PLS of ET-2L-$0^{\circ}$ and ET-$\Delta$, we see that the former is better  for $f$ below about 100~Hz, and the latter is better above 100~Hz. Comparing ET-$\Delta$ to Hybrid-$0^{\circ}$, we see that ET-$\Delta$ is better below about 8; then, from 8 to 80~Hz, is better Hybrid-$0^{\circ}$; and above 80~Hz is better again ET-$\Delta$. The SNR of a given stochastic background signal $h^2\,\Ogw(f)$ is a quantity integrated over the frequencies, so in the end which configuration will perform better depends on the shape of the signal in frequency. To obtain a more quantitative understanding, we consider some examples of possible signals, also shown in \autoref{fig:PLSs}: 
\begin{enumerate}[label=(\arabic*)]
    \item  A stochastic background where $h^2\,\Ogw$ is flat in frequency, and chosen to have the values  $10^{-10}$, $10^{-11}$, or $10^{-12}$. Signals for which $h^2\,\Ogw$ is approximately constant in $f$, at least over the  range of frequencies explored by ground-based GW detectors, emerge rather naturally in cosmology, where typical spectra extend over many more decades in frequency, and the range explored by ground-based detectors is comparatively small. An explicit example is provided by some models for cosmic strings, see e.g. Figure~63 of \cite{Branchesi:2023mws}.\footnote{The stochastic background produced by the amplification  of vacuum fluctuations in single-field slow-roll inflation is also almost flat in $h^2\,\Ogw$, at the frequencies of ground-based detectors. Unfortunately, at the frequencies of 3G ground-based detectors, or of LISA, it gives a prediction for $h^2\,\Ogw$ which is at most $10^{-16}$ (see e.g. Figure~21.9 of \cite{Maggiore:2018sht}) and therefore is not detectable, neither by 3G ground based detectors nor by LISA.}
    \item  The  current best estimate of the astrophysical GW background (AGWB)  obtained from the superposition of all BBH signals, and that obtained from  all BNS signals. We compute it as in Section~5.3 of \cite{Branchesi:2023mws}.\footnote{Observe that, at a given detector network, some of these signals from compact binaries will be resolved, while those below detection threshold, together also with the error in the subtraction of the resolved sources, will form a stochastic background \cite{Cutler:2005qq,Harms:2008xv,Regimbau:2016ike,Sachdev:2020bkk,Zhou:2022nmt}. This residual background is of course dependent on the detector network. For instance, at an ideal detector network that would resolve all sources, and subtract them  with no error, no residual background would be left. To compare the PLS with some example of signals, it is therefore more meaningful to consider the full AGWB generated by the superposition of all sources, resolved or unresolved.}
\end{enumerate}
The  SNR for a given stochastic background  signal is then computed as in Section~7.8.3 of \cite{Maggiore:2007ulw}. The results for the European-only networks are shown in \autoref{tab:SGWB_numbers_noCE}. We see that, for these spectra, the best results come from ET-2L-$0^{\circ}$ followed by Hybrid-$0^{\circ}$. These perform clearly better than ET-$\Delta$ which, in turn, performs clearly better than ET-2L-mis and Hybrid-mis.
\autoref{tab:SGWB_numbers_wCE} shows the results when adding also the 40-km CE in the US. We see that  the hierarchy between the configurations remains the same, and
ET-2L-$0^{\circ}$ and Hybrid-$0^{\circ}$ are still clearly the best configurations. However,
now the performances of ET-2L-mis and Hybrid-mis become quite comparable to that of ET-$\Delta$.

The signals considered in \autoref{fig:PLSs} and 
\hyperref[tab:SGWB_numbers_noCE]{Tables~\ref*{tab:SGWB_numbers_noCE}} and \ref{tab:SGWB_numbers_wCE} are either flat or only mildly varying with frequency, within the detector's bandwidth. 
It is also interesting to compute the SNR for signals with a stronger frequency dependence. We then consider a power-law spectrum 
\be\label{Ogwpower}
\Ogw(f)=\Omega_0\, \(\frac{f}{f_0}\)^{\alpha}\, ,
\ee
where $f_0$ is a reference frequency such that  $\Ogw(f_0)$ has the value $\Omega_0$, and we consider a very steep ``blue" spectrum with $\alpha = 3$ (such spectra could for instance emerge for instance in the context of pre-big-bang models \cite{Brustein:1995ah,Buonanno:1996xc,Ben-Dayan:2024aec}), and a ``red" spectrum with $\alpha=-1$ (this could emerge, in particular,  in the stochastic background generated by bubble collisions at frequencies above the peak of the spectrum; see  Section~22.4.4 of \cite{Maggiore:2018sht}). In both cases we set the reference frequency to the value $f_0=10\, {\rm Hz}$; for $\alpha = 3$ we then consider the cases $h^2\Omega_0=\{10^{-12}, 10^{-13}, 10^{-14} \}$, while for  $\alpha = -1$ we consider the cases $h^2\Omega_0=\{10^{-10}, 10^{-11}, 10^{-12} \}$. These spectra are shown in \autoref{Fig:Omegapower}, together with the PLS of the various configuration. The corresponding values of the SNR are given in 
\hyperref[tab:SGWB_numbers_noCE_alpha]{Tables~\ref*{tab:SGWB_numbers_noCE_alpha}} and \ref{tab:SGWB_numbers_wCE_alpha}. Blue spectra are more sensitive to the PLS at high frequency, and therefore for them the relative performance of the triangle with respect to 2L improves. So, for instance, we saw from \autoref{tab:SGWB_numbers_noCE} that, for a flat background, the SNR of the Hybrid-$0^{\circ}$ configuration is almost a factor of 2  larger than the SNR for ET-$\Delta$ (and that of  ET-2L-$0^{\circ}$ is larger than that of ET-$\Delta$ by a factor about 3); in contrast, from 
 \autoref{tab:SGWB_numbers_noCE_alpha}, we see that for a steep blue spectrum with $\alpha=3$ the SNR of the triangle  is larger  than that of Hybrid-$0^{\circ}$ by a factor of about 3 (and is larger than that of ET-2L-$0^{\circ}$
 by a factor of about 2).

\begin{figure}[tbp]
    \centering
    \begin{tabular}{c c}
       \includegraphics[height=.4\textwidth]{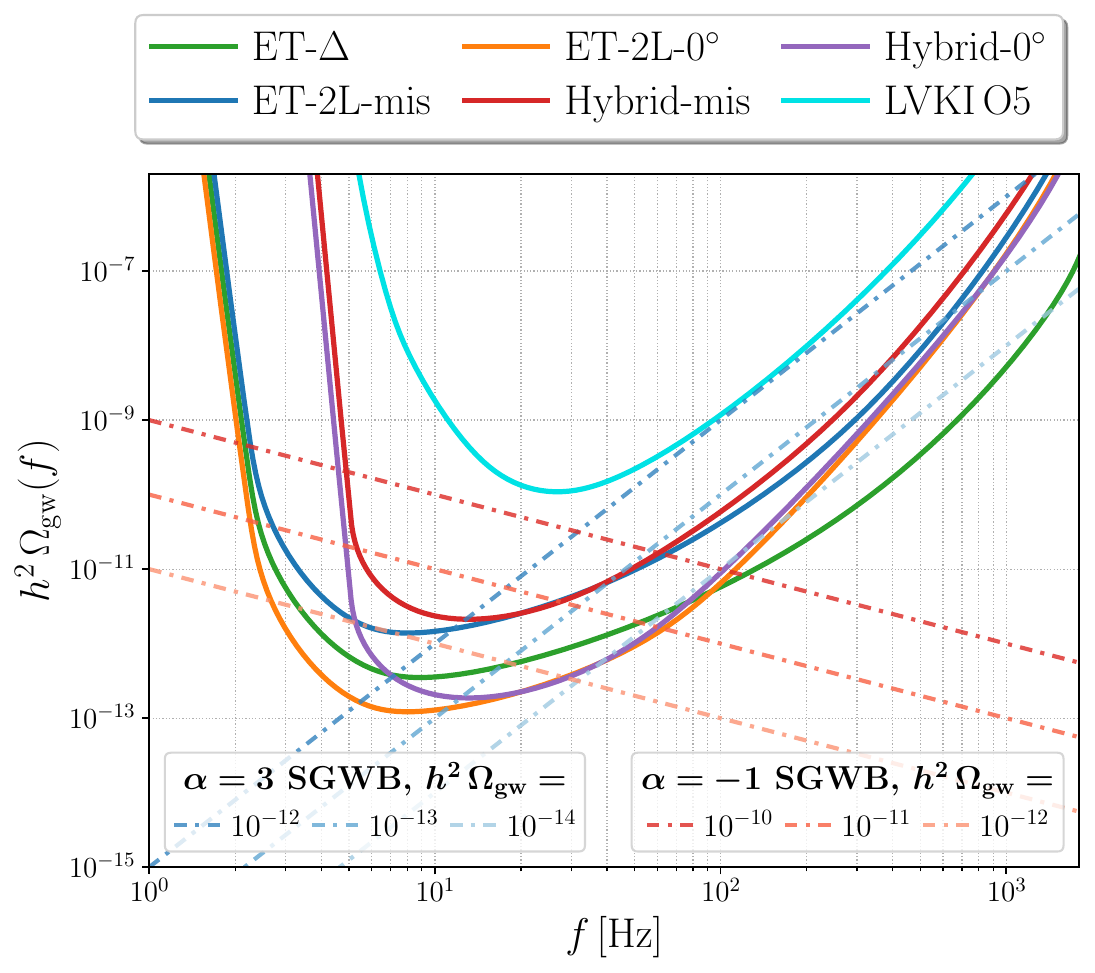}  
        & \includegraphics[height=.4\textwidth]{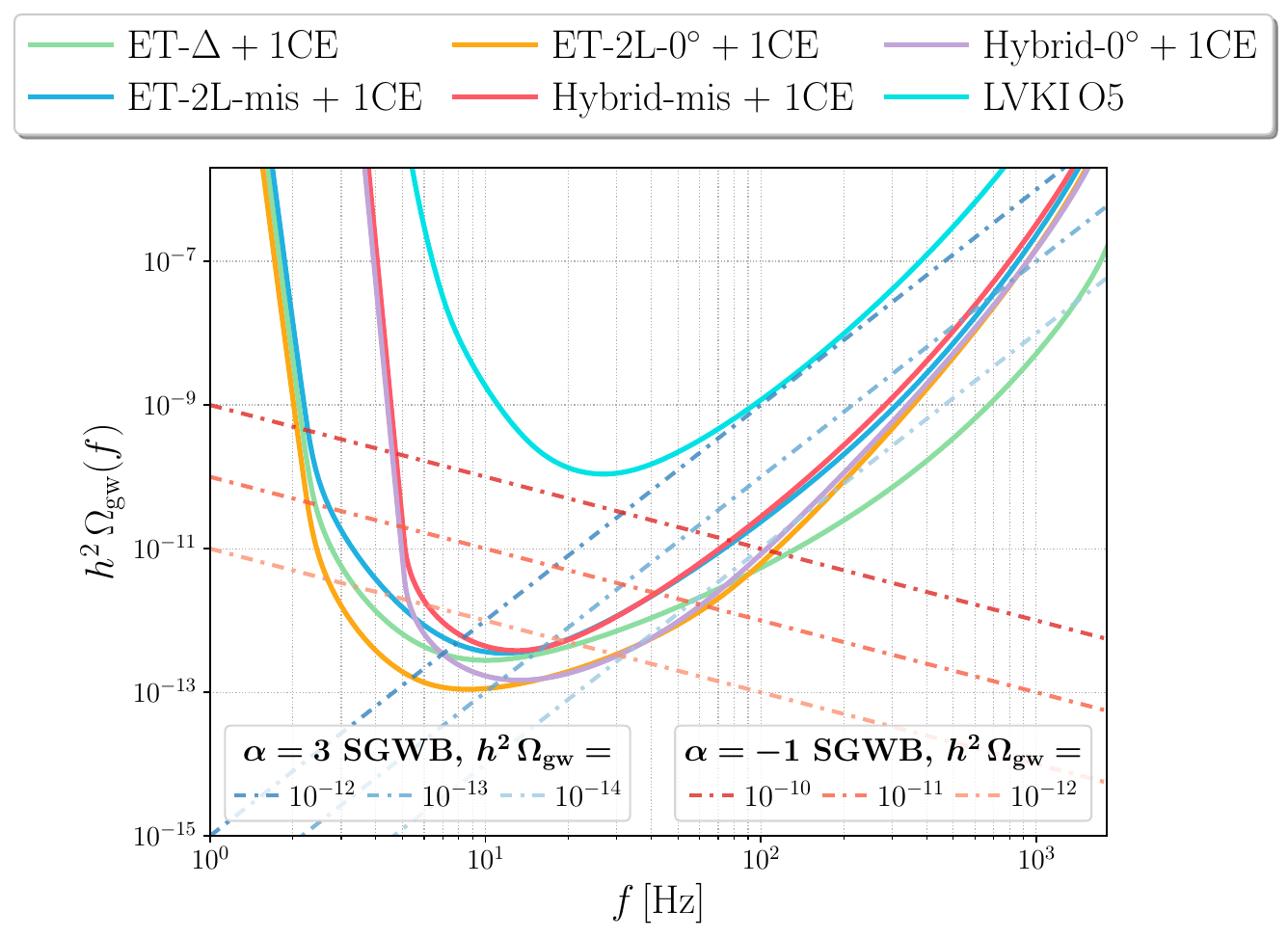} 
    \end{tabular}
    \caption{The same PLS shown in Fig.~\ref{fig:PLSs}, now compared to  power law signals 
    $\Ogw(f)=\Omega_0\, (f/f_0)^{\alpha}$ with $\alpha =3$  and $\alpha=-1$  and different amplitudes.  }
    \label{Fig:Omegapower}
\end{figure}

\begin{table}[tbp]
    \centering
    \begin{tabular}{!{\vrule width 1pt}m{3.1cm}|S[table-format=3.1]|S[table-format=2.1]|S[table-format=1.1]|S[table-format=4.1]|S[table-format=3.1]|S[table-format=2.1]!{\vrule width 1pt}}
        \toprule
        \midrule
        \hfil\multirow{3}{*}{\shortstack[c]{Detector\\configuration}}\hfill & \multicolumn{6}{c!{\vrule width 1pt}}{SNR for} \\
        & \multicolumn{3}{c|}{$\alpha=3$, $h^2\Omega_0=$} &  \multicolumn{3}{c!{\vrule width 1pt}}{$\alpha=-1$, $h^2\Omega_0=$} \\
        \cmidrule(lr){2-4} \cmidrule(lr){5-7}
        & \multicolumn{1}{c|}{$10^{-12}$} & \multicolumn{1}{c|}{$10^{-13}$} & \multicolumn{1}{c|}{$10^{-14}$} & \multicolumn{1}{c|}{$10^{-10}$} & \multicolumn{1}{c|}{$10^{-11}$} & \multicolumn{1}{c!{\vrule width 1pt}}{$10^{-12}$} \\
        \midrule
        ET-$\Delta$ & 385.5 & 38.6 & 3.9 & 374.0 & 37.4 & 3.7 \\
        ET-2L-mis & 29.4 & 2.9 & 0.3 & 103.5 & 10.3 & 1.0 \\
        ET-2L-$0^\circ$ & 166.5 & 16.7 & 1.7 & 1178.1 & 117.8 & 11.8 \\
        Hybrid-mis & 17.7 & 1.8 & 0.2 & 42.7 & 4.3 & 0.4 \\
        Hybrid-$0^\circ$ &130.6 & 13.1 & 1.3 & 486.0 & 48.6 & 4.9 \\
        \midrule
        \bottomrule
    \end{tabular}
    \caption{SNRs at the various European configurations for different stochastic background sources in an observational time $T_{\rm obs}=1~{\rm yr}$, for the power-law spectra given in \eq{Ogwpower}, with $f_0=10\, {\rm Hz}$ and different values of $\Omega_0$ and $\alpha$.}
    \label{tab:SGWB_numbers_noCE_alpha}
\end{table}

\begin{table}[tbp]
    \centering
    \begin{tabular}{!{\vrule width 1pt}m{3.1cm}|S[table-format=3.1]|S[table-format=2.1]|S[table-format=1.1]|S[table-format=4.1]|S[table-format=3.1]|S[table-format=2.1]!{\vrule width 1pt}}
        \toprule
        \midrule
        \hfil\multirow{3}{*}{\shortstack[c]{Detector\\configuration}}\hfill & \multicolumn{6}{c!{\vrule width 1pt}}{SNR for} \\
        & \multicolumn{3}{c|}{$\alpha=3$, $h^2\Omega_0=$} &  \multicolumn{3}{c!{\vrule width 1pt}}{$\alpha=-1$, $h^2\Omega_0=$} \\
        \cmidrule(lr){2-4} \cmidrule(lr){5-7}
        & \multicolumn{1}{c|}{$10^{-12}$} & \multicolumn{1}{c|}{$10^{-13}$} & \multicolumn{1}{c|}{$10^{-14}$} & \multicolumn{1}{c|}{$10^{-10}$} & \multicolumn{1}{c|}{$10^{-11}$} & \multicolumn{1}{c!{\vrule width 1pt}}{$10^{-12}$} \\
        \midrule
        ET-$\Delta + {\rm 1CE}$ & 386.2 & 38.6 & 3.9 & 420.5 & 42.0 & 4.2 \\
        ET-2L-mis + 1CE & 42.8 & 4.3 & 0.4 & 278.9 & 27.9 & 2.8 \\
        ET-2L-$0^\circ + {\rm 1CE}$ & 171.5 & 17.1 & 1.7 & 1233.1 & 123.3 & 12.3 \\
        Hybrid-mis + 1CE & 36.2 & 3.6 & 0.4 & 228.2 & 22.8 & 2.3 \\
        Hybrid-$0^\circ + {\rm 1CE}$ & 136.6 & 13.7 & 1.4 & 591.0 & 59.1 & 5.9 \\
        \midrule
        \bottomrule
    \end{tabular}
    \caption{As in \autoref{tab:SGWB_numbers_noCE_alpha}, adding to each network a single 40-km CE detector in the US.}
    \label{tab:SGWB_numbers_wCE_alpha}
\end{table}

Again, it should be stressed that  these results are obtained assuming that the noise in the different detectors are uncorrelated, an assumption which is more delicate for the colocated interferometers in the ET-$\Delta$ configuration.

Finally, for the configurations involving two L-shaped detectors, it is interesting to understand how the results depend on the relative orientation of the two Ls.
\autoref{fig:PLS_varyangle} shows how the minimum value of the PLS depends on the angle $\beta$ defined in 
\autoref{sect:detconfig} with reference to the great circle connecting the two detectors. As discussed in \autoref{sect:detconfig},  for a network of two L-shaped detectors at a distance $d$, in the limit $d/\lambda\ra 0$ (where $\lambda =c/f$ is the GW wavelength) 
the sensitivity goes to zero for $\beta=45^{\circ}$; indeed, we see from the left panel of \autoref{fig:PLS_varyangle} that, for this angle, the minimum of the PLS becomes large (not strictly infinity, given that the detectors are not colocated). The misaligned configurations that we have studied correspond to the value of $\beta$ marked by the vertical dashed line in the figure. We see that it is quite close to the value where the sensitivity to stochastic background is the worst.
However,  a somewhat larger misalignment angle would not significantly affect the performances for coalescing binaries, while we see from the figure that it would give a non-negligible improvement to the sensitivity to stochastic backgrounds.

Having three L-shaped detectors, as in the (ET-2L + 1CE) and (Hybrid-2L + 1CE) configurations, also allows us to play with all the relative angles, to optimize both the sensitivity to coalescing binaries and to stochastic backgrounds. In particular, one could set the two European L-shaped detectors close to parallel, and the US detector close to $45^{\circ}$ to them. Putting at $45^{\circ}$ the detectors that have a large baseline optimizes the sensitivity to parameter estimation  of coalescing binaries (in particular, angular localization), while setting parallel the two European detectors, that have the shorter baseline, is the best way of optimizing the sensitivity to stochastic backgrounds. Eventually, if one of these configurations should be implemented, dedicated study of the optimization of the science output as a function of the relative orientation angles will be necessary.

\begin{figure}[tbp]
    \centering
    \begin{tabular}{c c}
       \includegraphics[height=.38\textwidth]{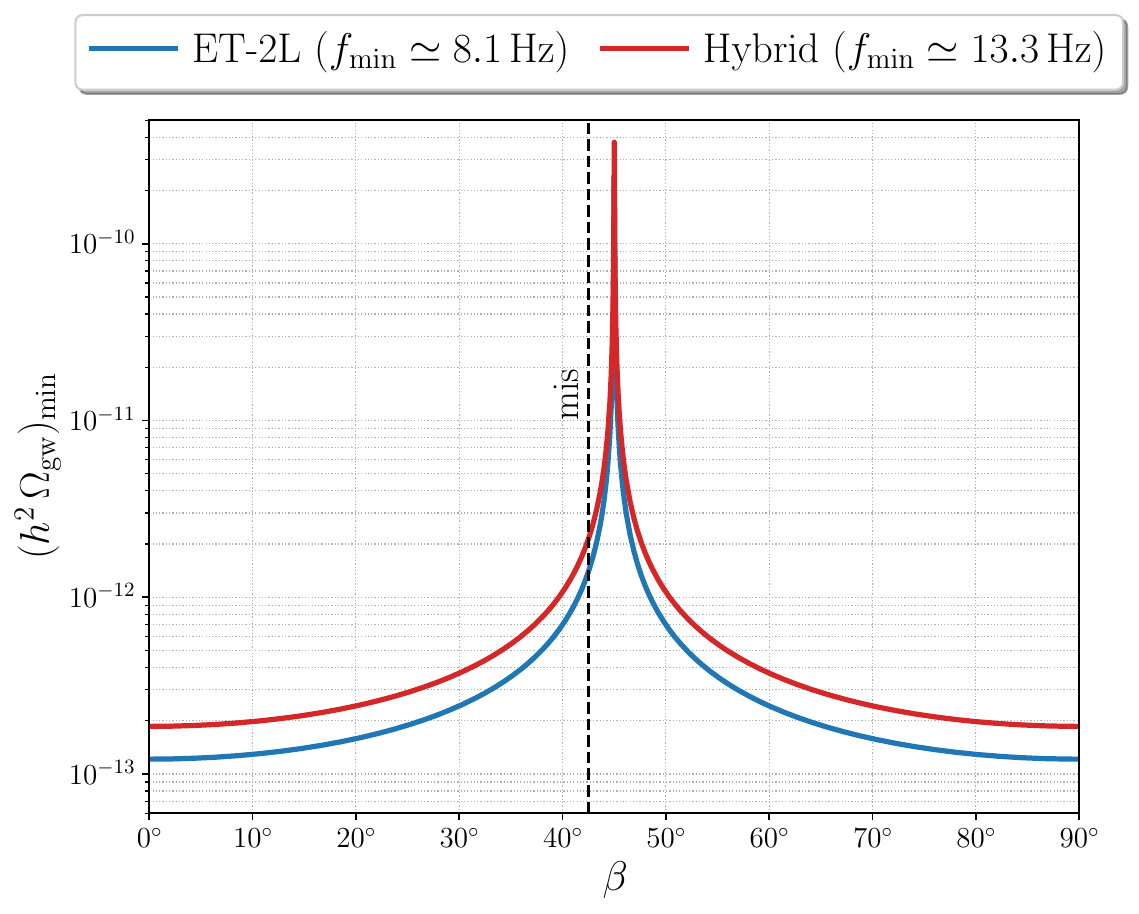}  & \includegraphics[height=.38\textwidth]{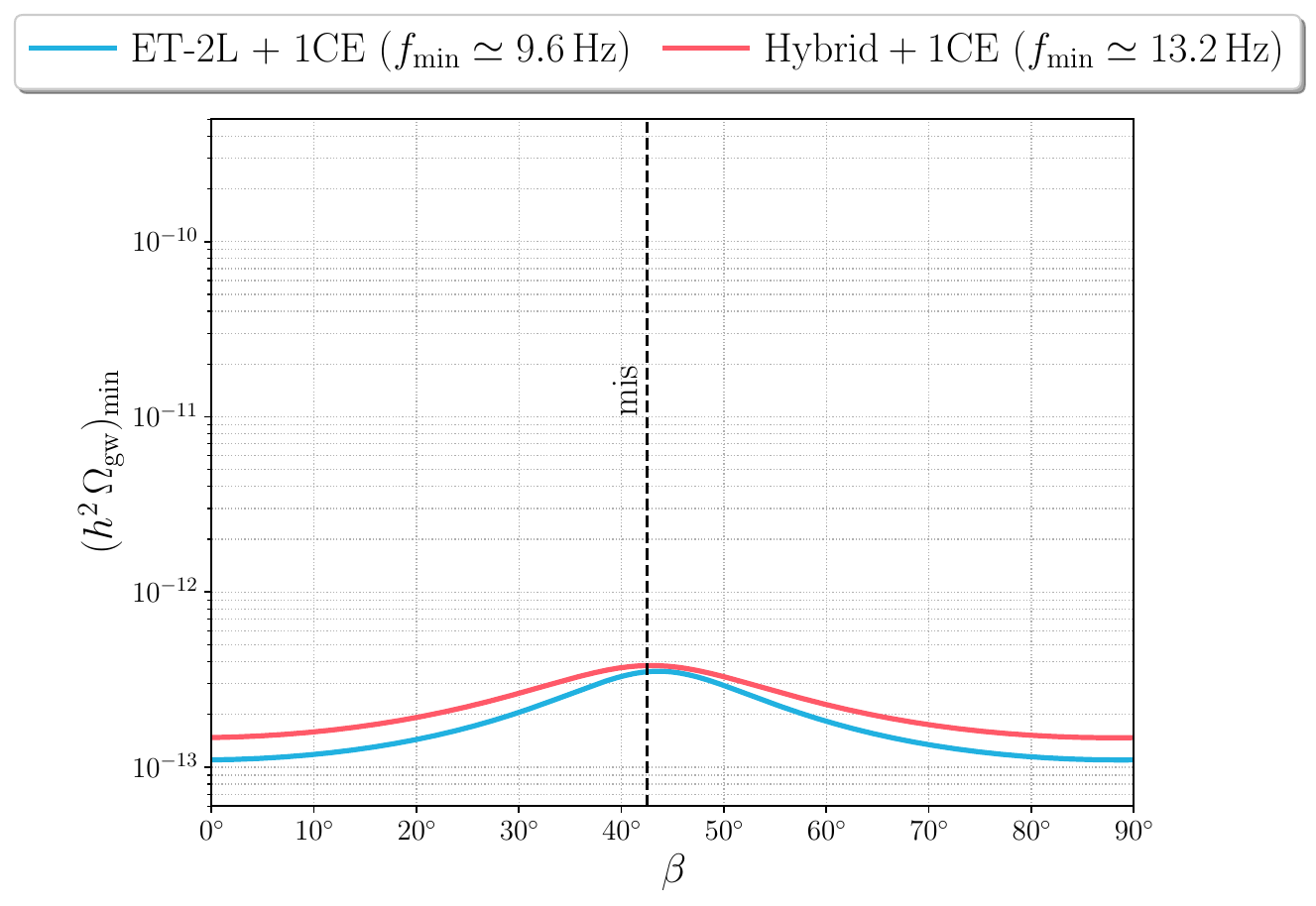} 
    \end{tabular}
    \caption{Minimum value attained by the PLS for $h^2\,\Ogw(f)$,  as a function of the relative angle $\beta$ between the European L-shaped detectors,  for the ET-2L and Hybrid configurations, without (\emph{left panel}) and with (\emph{right panel}) a single 40-km CE detector in the US (whose orientation is kept fixed).  In each panel, the black dashed line denote the alignment corresponding to the ``mis'' configuration considered in the present work. The frequency where the PLS attains its minimum is shown in the legend.}
    \label{fig:PLS_varyangle}
\end{figure}
 
\section{Conclusions}

In this paper we have studied the performance of a hypothetical European detector network made by a 
surface detector and an underground detector. For the underground detector we have considered 
a single L-shaped detector with the ASD of ET (therefore featuring a high-frequency   and a low-frequency interferometer), taken to have  15~km arms. For the detector on-surface  we have ``borrowed''  the ASD of CE with 20~km arms but, as we stressed in the Introduction, this choice is only made for simplicity, since the ASD of CE is the only  publicly available curve for a 3G surface detector.

We have then compared this ``Hybrid'' configuration to the two  options for ET that are currently under active investigation, namely a single-site 10-km triangle, or two 15-km L-shaped underground detectors in different European sites.
We have compared these three networks among them, and we have further compared them when, to each of these networks, is added a 40-km CE detector in the US.

The motivation that inspired this work is that of extending the study of the performance of a third-generation observatory to configurations not previously considered. The results presented here therefore go in the direction of completing the picture of the capabilities of various configurations using 3G detectors whose individual performances have already been presented in literature.
From here the interest to understand what one can do by correlating a single L-shaped interferometer with the characteristics of ET, with a second interferometer built on the Earth's surface rather than underground, and not requiring the ``xylophone'' configuration with a low-frequency cryogenic instrument.

In the context of a European-only network, our main results are as follows:

\begin{itemize}

\item For the detection horizons, above about $30\msun$ all these 3G configurations are very similar, but below   there are significant differences, and ET-2L-mis and Hybrid-mys  are superior to ET-$\Delta$. For instance, for the total mass typical of BNS, $\mtot=2.7\msun$, ET-$\Delta$ has a detection horizon $z_{\rm hor}\simeq 3.3$, while the Hybrid configurations and  the ET-2L configurations both have $z_{\rm hor}\simeq 5.3$;
see the left panel of \autoref{fig:All_horizons} and the left table in
\autoref{tab:horizon_tabs}. Similarly considerations hold  for coalescing binaries with subsolar masses,
that would be a smoking-gun signature of primordial black holes; e.g., for $\mtot=1.0\msun$, ET-$\Delta$ has a detection horizon $z_{\rm hor}\simeq 1.0$, while the Hybrid configurations and  the ET-2L configurations  have $z_{\rm hor}\simeq 1.4$. In terms of comoving volumes, between $z=1.4$ and $z=1.0$ there is a factor $2.0$, which would reflect in the probability of primordial black holes detection.

\item For  BBHs,   the configurations ET-2L-mis and Hybrid-mys are clearly superior to ET-$\Delta$ for SNR distribution, angular localization, and reconstruction of the luminosity distance, see \hyperref[fig:PE_BBH_noCE]{Figures~\ref*{fig:PE_BBH_noCE}} and \ref{fig:histz_BBH_noCE}, and \autoref{tab:BBH_numbers_loc_noCE}.

\item For BNS,  for angular localization the best configuration is ET-2L-mis, followed by ET-2L-$0^{\circ}$ and Hybrid-mys, which in turn are superior to  ET-$\Delta$ and Hybrid-$0^{\circ}$. For luminosity distance, ET-2L-mis is again clearly the best, followed by Hybrid-mys.
See \hyperref[fig:PE_BNS_noCE]{Figures~\ref*{fig:PE_BNS_noCE}} and \ref{fig:histz_BNS_noCE}, and \autoref{tab:BBH_numbers_loc_noCE}.

\item For pre-merger alerts,  the left panel of  \autoref{fig:premerger} shows that, for a GW170817-like event, the SNR  accumulates before merger in a rather similar manner between the different 3G configurations considered. For the angular localization before merger, we see from \autoref{tab:premerger_time_angular_res_noCE} that all these 3G configurations have  comparable performances at 10~min to merger or later; at a finer level,   ET-2L-mis is somewhat better than ET-$\Delta$ and Hybrid-mis, which are very similar among them, within a factor less than 2. At 30~min before merger, however,
Hybrid-mis is  less performant than ET-$\Delta$,   by a factor between  2 and 4 in the number of events with a given localization.

\item The sensitivities to stochastic backgrounds of the various configurations  of European-only networks are shown in the left panel of \autoref{fig:PLSs}, and the corresponding SNR for some examples of stochastic  backgrounds  are given in \autoref{tab:SGWB_numbers_noCE} and  \autoref{tab:SGWB_numbers_noCE_alpha}. As we see from \autoref{tab:SGWB_numbers_noCE}, for flat or almost flat backgrounds
the best configurations in general are ET-2L-$0^{\circ}$ and Hybrid-$0^{\circ}$, which are  better than ET-$\Delta$, which in turn is  better than 
ET-2L-mis and Hybrid-mis. However, for steep blue spectra, the ET-$\Delta$ configurations becomes the best one, while for red spectra the ET-2L-$0^\circ$ is the best one, see \autoref{tab:SGWB_numbers_noCE_alpha}. It should also be observed   that, in the 2L case, 
the results for the misaligned configurations are quite specific to our choice of misalignment angle; the sensitivity to stochastic background of these configurations could be improved by increasing this misalignment angle which, to some extent, can be done without significantly affecting  the performance for parameter estimation of coalescing binaries. 

It should also be stressed that we have computed the  sensitivity to stochastic backgrounds  by correlating the outputs of pairs of detectors, and exploiting the fact that  the GW signal is correlated, while assuming that the noise is uncorrelated. This assumption is never totally correct, but this could be a particularly  serious concern for the colocated interferometers of the triangle configuration, in particular at low frequencies, where seismic and Newtonian noise 
induce correlated noise among
the colocated  interferometers of the triangle.

\end{itemize}

When we put these European configurations in a broader world-wide network with a 40-km CE in the US,  the hierarchy of performances among the configurations remains the same, but the relative differences between them  become much less important. In particular:

\begin{itemize}

\item The detection horizons becomes practically identical, see the right panel of \autoref{fig:All_horizons} and the right panel in
\autoref{tab:horizon_tabs}.

\item The SNR distribution and parameter estimation for BBHs becomes very similar for all configurations, with differences in the number of events with given cuts that, among all configurations, do not exceed a factor of 2, with (ET-2L-mis + 1CE) and  (Hybrid-mis + 1CE) being the best configurations;
see \hyperref[fig:PE_BBH_wCE]{Figures~\ref*{fig:PE_BBH_wCE}} and \ref{fig:histz_BBH_wCE}, and \autoref{tab:BBH_numbers_loc_wCE}.

\item For BNSs the spread in the results between configurations is slightly larger than for BBHs but still, among all configurations, it does not exceed a factor of about 3.5 between the best and the less good configuration; again, (ET-2L-mis + 1CE) is the best configuration and   (Hybrid-mis + 1CE) is the second best;
see \hyperref[fig:PE_BNS_wCE]{Figures~\ref*{fig:PE_BNS_wCE}} and \ref{fig:histz_BNS_wCE}, and \autoref{tab:BNS_numbers_loc_wCE}.

\item For pre-merger alerts, the performances of all configurations are comparable; the best configuration
is again (ET-2L-mis + 1CE), but (ET-$\Delta$ + 1CE) and (ET-2L-$0^{\circ}$ + 1CE) follow very closely;
for alerts at 30~min before merger the (Hybrid + 1CE) configurations are less performant, but they partially catch up for alert at 10~min or less before merger.
For instance, the number of BNS localized to better than $100\, {\rm deg}^2$, 30~min before merger, for (ET-$\Delta$ + 1CE), (ET-2L-mis + 1CE), (ET-2L-$0^{\circ}$ + 1CE), (Hybrid-mis + 1CE) and 
(Hybrid-$0^{\circ}$ + 1CE)
are, respectively, $\{25,41,19,10,7\}$; at 10~min before merger, these numbers become, respectively, $\{234,363,100,159,56\}$;
see the right panel of \autoref{tab:premerger_time_angular_res_wCE}. \autoref{fig:premerger} shows how the SNR accumulates in an event such as GW170817,
and  for such an event the differences between configurations are marginal. 

\item The sensitivities of the various configurations to stochastic backgrounds are shown in the right panel of \autoref{fig:PLSs}, and the corresponding SNR of some example of stochastic  backgrounds  are given in \autoref{tab:SGWB_numbers_wCE}. For instance, for the astrophysical background due to the superposition of BBHs, \autoref{tab:SGWB_numbers_wCE} gives  the SNR values 
$\{829,699,1948,678,1807\}$, respectively for ET-$\Delta$, ET-2L-mis, ET-2L-$0^{\circ}$,  Hybrid-mis and 
Hybrid-$0^{\circ}$ (each one together with 1CE). Similar results hold for the other stochastic backgrounds considered. The best configurations are therefore now (ET-2L-$0^{\circ}$ + 1CE) and (Hybrid-$0^{\circ}$ + 1CE), but the results for the SNR differ by less than a factor of 3 between the best and the less good configuration.
In a direct comparison between (ET-$\Delta$ + 1CE) and 
(Hybrid-mis + 1CE), we see that the SNR for this background are 829 for (ET-$\Delta$ + 1CE) and 678 for
(Hybrid-mis + 1CE), a difference of less than $20\%$. The same holds for the other stochastic backgrounds considered.

\end{itemize}

These results could be a useful input to an analysis of the costs and risks of different detector configurations.
We stress again that the specific choice that we made for the location of a surface detector in Spain does not correspond to any project currently under study from the many points of
view (geological, topographical, financial, political, etc.) that are necessary for determining
viable detector configurations and optimal site selection, and that
the ``hybrid" configuration that we have studied, with an underground  L-shaped detector with the ASD of ET and an on-surface 3G  detector, does not correspond to any project currently endorsed by the ET Collaboration. This work should  be considered as an investigation with the purpose of understanding better what builds up the sensitivity of a GW network.

\section*{Acknowledgements}
 
We thank Eugenio Coccia and Mario Martinez for very valuable discussions that inspired this study.
We also thank Giancarlo Cella, Jerome Degallaix, Ik Siong Heng, Mikhail Korobko and Andrea Maselli for their very useful internal ET review.

The work of F.I., M.Mag. and N.M.   is supported by the
Swiss National Science Foundation (SNSF) grants 200020$\_$191957  and by the SwissMap National Center for Competence in Research.  E.B. and M.Mag. are supported by the SNSF
grant CRSII5$\_$213497. 
M.Mag. thanks for the hospitality the IFAE  in Barcelona, where part of this work was done.
The work of M.Manc. received support from the French government under the France 2030 investment plan, as part of the Initiative d'Excellence d'Aix-Marseille Universit\'e -- A*MIDEX AMX-22-CEI-02.
This work made use of the clusters Yggdrasil and Baobab at the University of Geneva.

\bibliographystyle{utphys}
\bibliography{myrefs.bib}

\end{document}